\documentclass{article}

\usepackage{geometry}
\usepackage{amsmath, amssymb, amsthm, amsfonts, mathrsfs}
\usepackage{graphicx}
\usepackage[numbers]{natbib}

\theoremstyle{plain}
\newtheorem{theorem}{Theorem}
\newtheorem{algorithm}[theorem]{Algorithm}
\theoremstyle{definition}
\newtheorem*{definition}{Definition}
\theoremstyle{remark}

\numberwithin{equation}{section}

\DeclareMathOperator*{\argmax}{argmax}
\DeclareMathOperator*{\argmin}{argmin}
\DeclareMathOperator{\diag}{diag}
\newcommand{\R}{\mathbb{R}}
\newcommand{\E}{\mathbb{E}}
\newcommand{\ud}{\mathrm{d}}
\newcommand{\cL}{\mathcal{L}}
\newcommand{\cN}{\mathcal{N}}
\newcommand{\cQ}{\mathcal{Q}}
\newcommand{\cU}{\mathcal{U}}
\newcommand{\one}{\mathbf{1}}

\usepackage[bookmarks]{hyperref}

\begin{document}

\title{%
  An Expectation-Maximization Algorithm for\\
  Continuous-time Hidden Markov Models
}
\author{%
  Qingcan Wang\thanks{%
    Program in Applied and Computational Mathematics, Princeton University
    (\texttt{qingcanw@princeton.edu}).
  }
  \ and Weinan E\thanks{%
    Department of Mathematics and Program in Applied and Computational
    Mathematics, Princeton University (\texttt{weinan@math.princeton.edu}).
  }
}
\date{}
\maketitle

\begin{abstract}
  We propose a unified framework that extends the inference methods for
  classical hidden Markov models to continuous settings, where both the hidden
  states and observations occur in continuous time. Two different settings are
  analyzed: hidden jump process with a finite state space, and hidden diffusion
  process with a continuous state space. For each setting, we first estimate the
  hidden states given the observations and model parameters, showing that the
  posterior distribution of the hidden states can be described by differential
  equations in continuous time. We then consider the estimation of unknown model
  parameters, deriving the continuous-time formulas for the
  expectation-maximization algorithm. We also propose a Monte Carlo method based
  on the continuous formulation, sampling the posterior distribution of the
  hidden states and updating the parameter estimation.
\end{abstract}

\section{Introduction}

Hidden Markov models (HMM) are widely used for inferring the underlying
dynamical process from observed data. The classical discrete-time hidden Markov
model (DT-HMM) contains a pair of stochastic processes $\{(X_t, Y_t): t = 0, 
\dots, T\}$. The hidden process $\{X_t\}$ is a Markov chain with transition
probability $\Pr\{X_{t+1} | X_t; \theta\}$; the observations $\{Y_t\}$ are
sampled independently at each time $t$ based on the observation probability
$\Pr\{Y_t | X_t; \theta\}$. Here $\theta$ denotes the model parameters. The
following are two basic problems for HMM:\@
\begin{description}
  \item[State estimation:] given the observations $\{Y_t\}$ and the model
    parameters $\theta$, calculate the posterior distribution of the hidden
    states $\Pr\{X_t | Y_0, \dots, Y_T; \theta\}$.
  \item[Parameter estimation:] given the observations $\{Y_t\}$ with unknown
    model parameters $\theta$, learn $\theta$.
\end{description}
For DT-HMM, the most widely used approach to address both of the problems is the
Baum-Welch algorithm \cite{baum1966statistical, baum1970maximization,
rabiner1989tutorial}. It uses the forward-backward equations for state
estimation, and the expectation-maximization (EM) algorithm for parameter
estimation (see Section~\ref{sec:prelim}).

A natural extension of the classical HMM is the continuous-time Hidden Markov
models (CT-HMM), where both the hidden process $\{X_t\}$ and observations
$\{Y_t\}$ occur in continuous time. Different settings of CT-HMM have been
proposed in the literature, along with the corresponding algorithms.  One
setting that has been considered is to take the hidden $\{X_t\}$ as a jump
process with finite state space.  Along this line, \cite{elliott1995hidden}
establishes some theoretical results for the case when the observations
$\{Y_t\}$ are either a non-homogeneous jump process determined by $\{X_t\}$, or
a Brownian motion with drift (see also \cite{ephraim2008algorithm,
wonham1964some, clements1975nonlinear, elliott1993forward}). From a more
practical perspective, different settings like the Markov-modulated Poisson
process \cite{deng1993parameter, ryden1996algorithm}, batch Markovian arrival
process \cite{breuer2002algorithm} and bivariate Markov process
\cite{mark2013algorithm} have been considered (see
Section~\ref{sec:jump_setting} for details). To estimate the parameters, the
approach studied in \cite{deng1993parameter, james1996time} is to discretize the
time and solve it as DT-HMM. Another approach is to use the ideas from  the
estimation of (non-hidden) jump process \cite{metzner2007generator,
brockett2009stochastic, hobolth2011summary}, and calculate the dwell time and
the number of jumps among the states. Another setting considered is when the
hidden $\{X_t\}$ is a diffusion process with continuous state space, which
follows the Stratonovich-Kushner and Zakai equations
\cite{stratonovich1959optimum, kushner1964differential, zakai1969optimal}.
Parameter estimation in this case has taken quite different approaches.  In
\cite{ching2006bayesian, campillo2009convolution}, the authors regard the
parameters as being part of the hidden state. Parameter estimation can then be
solved using state estimation. In \cite{zhao2013parameter, feng2016expectation},
the authors propose to discretize the time first, and then use the discrete EM
formulation.

Given the different settings and algorithms in CT-HMM, our goal is to propose a
unified framework for parameter estimation in CT-HMM. Our basic idea is to study
the continuous-time limit of the classical discrete-time Baum-Welch algorithm.
We will apply the Baum-Welch framework to both hidden jump process and hidden
diffusion process, and show that the estimation problem can be solved using a
similar approach even though the settings are quite different.

Section~\ref{sec:hmm_jump} considers CT-HMM for hidden jump process, where both
the hidden states and observations take a finite set of values. Consider a
system where the underlying states can switch between several discrete values,
and we have a sensor that monitors the system in continuous time, whose output
also takes discrete values. It is reasonable to model the hidden states
$\{X_t\}$ as a jump process, and assume that the sensor output $\{Y_t\}$ changes
every time $\{X_t\}$ jumps. Specifically, for each jumping time $\tau$ of
$\{X_t\}$, resample $Y_\tau$ from the observation probability condition on
$X_\tau$, and $X_t$, $Y_t$ remain constant during the holding period $[\tau,
\tau')$, where $\tau'$ is the next jumping time of $\{X_t\}$. Here the holding
time $\tau' - \tau$ is an exponentially distributed random variable, therefore
this can be regarded as an extension of DT-HMM whose holding time is always 1.
Applying the Baum-Welch framework to this continuous setting, the state
estimation, i.e., the posterior distribution $\Pr\{X_t | Y_s, s \in [0, T];
\theta\}$ can be described by forward and backward piecewise ODEs. Parameter
estimation can be done using the EM algorithm, where the parameter update
formula consists of summing terms that correspond to the jumping times, as well
as the integral terms for the holding periods.

Section~\ref{sec:hmm_diffusion} shows that the framework can also be applied to
CT-HMM with hidden diffusion process, where the hidden states and observations
take continuous values. Here we analyze the classical setting in the
Stratonovich-Kushner and Zakai equations. The Zakai equation solves the optimal
nonlinear filtering problem, which is the forward part in our framework. We give
the corresponding SPDE for the backward part. By combining those we can solve
the smoothing problem for state estimation. We then derive the continuous-time
formula for the EM algorithm in parameter estimation. For linear Gaussian
problem, we show that our results are consistent with the discrete-time Kalman
filter and EM algorithm under the limit $\Delta t \to 0$. For general nonlinear
problems, we propose a Monte Carlo method based on particle filter and smoother
to sample the posterior distribution of hidden states, and then update
parameters from the samples in the EM algorithm.

While this work mainly focuses on the continuous-time states and observations,
the Baum-Welch framework can also be applied to the setting where the
observations occur at some different set of discrete times, and we show the
corresponding results in Appendix~\ref{sec:hmm_disc_obs}.

\section{Preliminaries}%
\label{sec:prelim}

\subsection{Notations}

In this paper, we use uppercase $X_t$, $Y_t$ to denote random variables,
lowercase $x_t$, $y_t$ to denote the particular values of the random variables,
and $\{X_t\}$, $\{Y_t\}$ to denote the stochastic processes. Let $\Pr\{\cdot\}$
be the probability distribution and $p(\cdot)$ be the probability density. Let
$\Pr\{\cdot; \theta\}$ be the probability distribution corresponding to the
parameters $\theta$. Without ambiguity, we may simply denote the conditional
probability $\Pr\{X_t = x_t | Y_t = y_t\}$ as $\Pr\{x_t | y_t\}$.

In continuous time, the subscript $x_{0:t}$ denotes all the $x_s$ for $0 \le s
\le t$. In some parts of the paper we may need discretization of the time space
with step $\Delta t$. In this case we only consider $x_t$ on the grid, i.e., $t
= k \Delta t$ for some integer $k$, then $x_{0:t}$ denotes all the $x_{l \Delta
t}$ for $l = 0, 1, \dots, k$, and $\Delta x_t = x_{t + \Delta t} - x_t$. Let
$\one_{\{\cdot\}}$ be the indicator function. Let $\diag(a_1, \dots, a_n)$ be
the $n \times n$ diagonal matrix with $a_1, \dots, a_n$ on the main diagonal.

\subsection{Expectation-maximization algorithm}

The expectation-maximization (EM) algorithm is an iterative method to estimate
parameters in statistical models. Let $\theta$ be the unknown parameters, and
the model generates a set of unobservable latent variables $X$ and a set of
observed data $Y$. The goal is to calculate the maximum likelihood estimation of
the parameters $\theta$ given the observations $Y = y$, i.e., $\theta^\star =
\argmax_\theta L(\theta)$ where
\begin{equation}
  L(\theta) = \log p(Y = y; \theta) = \log \int p(X = x, Y = y; \theta) \ud x.
\end{equation}

In the case where the integral cannot be calculated directly, the EM algorithm
takes an iterative approach as following. Assume that we can find a function
$\cQ(\theta, \theta')$ such that $L(\theta) \ge \cQ(\theta, \theta')$ and
$L(\theta) = \cQ(\theta, \theta)$ for all $\theta$ and $\theta'$. For iterations
$k = 0, 1, 2, \dots$, let $\theta^{k+1} = \argmax_\theta \cQ(\theta, \theta^k)$,
then
\[
  L(\theta^k) = \cQ(\theta^k, \theta^k)
  \le \cQ(\theta^{k+1}, \theta^k) \le L(\theta^{k+1}), 
\]
i.e., $L(\theta^k)$ is nondecreasing during the iterations. If $L$ is upper
bounded, the EM algorithm will converge to a local maximum.

Consider the following construction of $\cQ$:
\begin{equation}
 \cQ(\theta, \theta')
 = \int p(x|y; \theta') \log \frac{p(x, y; \theta)}{p(x|y; \theta')} \ud x.
 \label{eq:em_q0}
\end{equation}
Since
\[
  L(\theta) = \log \int p(x, y; \theta) \ud x
  = \log \int p(x|y; \theta') \frac{p(x, y; \theta)}{p(x|y; \theta')} \ud x
\]
and $\log(\cdot)$ is concave, we have $\cQ(\theta, \theta') \le L(\theta)$.
Meanwhile,
\begin{align*}
  \cQ(\theta, \theta)
  & = \int p(x|y; \theta) \log \frac{p(x, y; \theta)}{p(x|y; \theta)} \ud x \\
  & = \int p(x|y; \theta) \log p(y; \theta) \ud x \\
  & = \log p(y; \theta) = L(\theta), 
\end{align*}
thus $\cQ$ is a qualified construction.

The EM algorithm repeats the following two steps until convergence. In the
\emph{expectation step (E-step)}, calculate the conditional probability $p(X = x
| Y = y; \theta^k)$; in the \emph{maximization step (M-step)}, update the
parameter as $\theta^{k+1} = \argmax_\theta \cQ(\theta, \theta^k)$. In practice
we may use
\begin{equation}
 \cQ(\theta, \theta')
 = \int p(X = x | Y = y; \theta') \log p(X = x, Y = y; \theta)
 \label{eq:em_q}
\end{equation}
instead of (\ref{eq:em_q0}), since they give the same result in the M-step.

\subsection{Baum-Welch algorithm}

The Baum-Welch algorithm solves both the state estimation and parameter
estimation for the classical DT-HMM.\@ In a DT-HMM $\{(X_t, Y_t): t = 0, \dots,
T\}$, the hidden states $X_t \in \{1, \dots, n\}$ is a Markov chain with
transition probability $P$, and the observations $Y_t \in \{1, \dots, m\}$ are
sampled independently at each $t$ according to the observation probability $r$,
i.e., 
\begin{align}
  P_{ij} & = \Pr\{X_{t+1} = j | X_t = i\}, \nonumber \\
  r_i(y) & = \Pr\{Y_t = y | X_t = i\}, 
\end{align}
for $i, j = 1, \dots, n$ and $y = 1, \dots, m$.

\paragraph{State estimation}
Assume that the probabilities $P$ and $r$ and the initial distribution $\pi_0(i)
= \Pr\{X_0 = i\}$ are given, and we have observations $Y_t = y_t$, $t = 0,
\dots, T$. The goal is to estimate the posterior distribution $\Pr\{X_t |
y_{0:T}\}$. The Baum-Welch algorithm introduces two sets of probabilities
\begin{equation}
  \alpha_t(i) = \Pr\{X_t = i, y_{0:t}\}, \quad
  \beta_t(i) = \Pr\{y_{t+1:T} | X_t = i\}, 
\end{equation}
that can be solved by the forward and backward equations inductively:
\begin{gather}
  \alpha_0(i) = \pi_0(i) r_i(y_0), \quad
  \alpha_t(i) = \sum_{j=1}^n \alpha_{t-1}(j) P_{ji} r_i(y_t), \nonumber \\
  \beta_T(i) = 1, \quad
  \beta_t(i) = \sum_{j=1}^n P_{ij} \beta_{t+1}(j) r_j(y_{t+1}).
\end{gather}
Since
\[
  \Pr\{X_t = i, y_{0:T}\}
  = \Pr\{y_{t+1:T} | X_t = i, y_{0:t}\} \Pr\{X_t = i, y_{0:t}\}
  = \alpha_t(i) \beta_t(i), 
\]
the posterior distribution is given by
\begin{equation}
  \rho_t(i) = \Pr\{X_t = i | y_{0:T}\}
  = \frac{\alpha_t(i) \beta_t(i)}{\sum_{i'=1}^n \alpha_t(i') \beta_t(i')}.
\end{equation}
Here we also calculate
\begin{equation}
  \xi_t(i, j) = \Pr\{X_{t-1} = i, X_t = j | y_{0:T}\}
  = \frac{\alpha_{t-1}(i) P_{ij} \beta_t(j) r_j(y_t)}%
  {\sum_{i'=1}^n \alpha_t(i') \beta_t(i')}.
\end{equation}

\paragraph{Parameter estimation}
Now assume that $\theta = (P, r)$ are the unknown parameters to be estimated.
Let $\rho$ and $\xi$ be the state estimation based on $\theta = \theta^k$. From
the EM algorithm (\ref{eq:em_q}), the parameter update is given by $\theta^{k+1}
= \argmax_\theta \cQ(\theta, \theta^k)$, and
\begin{align*}
  \cQ(\theta, \theta^k)
  & = \sum_{x_{0:T}} \Pr\{x_{0:T} | y_{0:T}; \theta^k\}
  \log \Pr\{x_{0:T}, y_{0:T}; \theta\} \\
  & = \sum_{x_{0:T}} \Pr\{x_{0:T} | y_{0:T}; \theta^k\}
  \left[\log \Pr\{x_0\} + \sum_{t=1}^T \log P_{x_{t-1}, x_t}
  + \sum_{t=0}^T \log r_{x_t}(y_t)\right] \\
  & = C + \sum_{x_{0:T}} \left[
  \sum_{t=1}^T \Pr\{x_{t-1}, x_t | y_{0:T}; \theta^k\} \log P_{x_{t-1}, x_t}
  + \sum_{t=0}^T \Pr\{x_t | y_{0:T}; \theta^k\} \log r_{x_t}(y_t)\right] \\
  & = C + \sum_{i, j} \sum_{t=1}^T \xi_t(i, j) \log P_{ij}
  +\sum_i \sum_{t=0}^T \rho_t(i) \log r_i(y_t), 
\end{align*}
where $C$ is a constant not depending on $(P, r)$. Solving the optimization
problem, the transition probability is updated as
\begin{equation}
  P_{ij}^{k+1}
  = \frac{\sum_{t=1}^T \xi_t(i, j)}{\sum_{j'} \sum_{t=1}^T \xi_t(i, j')}
  = \frac{\sum_{t=1}^T \xi_t(i, j)}{\sum_{t=0}^{T-1} \rho_t(i)}, 
\end{equation}
and the observation probability is given by
\begin{equation}
  r_i^{k+1}(y)
  = \frac{\sum_{t=0}^T \rho_t(i) \one_{y_t = y}}{\sum_{t=0}^T \rho_t(i)}.
\end{equation}

\subsection{Filtering problem}
\label{sec:filtering}

The goal of the filtering problem is to estimate the state of a stochastic
dynamical system given some noisy measurements of the system. The earliest
results are the Stratonovich-Kushner and Zakai equations
\cite{stratonovich1959optimum, kushner1964differential, kushner1967dynamical,
zakai1969optimal} that solve the optimal nonlinear filtering problems, and this
line of work focus more on the theoretical derivation (see
Section~\ref{sec:dif_state}). From the application prospective, most previous
work considers the discrete-time dynamics
\begin{align*}
  X_{t+1} & = f(X_t) + W_t, \\
  Y_t & = h(X_t) + V_t.
\end{align*}
(For real world problems in continuous time, one may discretize time first.) Two
types of algorithms are widely used: Kalman filter and particle filter.

\paragraph{Kalman filter}
The Kalman filter \cite{kalman1960new} considers the linear case where the
optimal filter has explicit solution. Assume that both $f$ and $h$ are linear
and the initial distribution of $X_0$ is Gaussian, then the conditional
distribution of the states is always Gaussian: $X_t | Y_{0:t} \sim \cN(\mu_t,
P_t)$. The Kalman filter gives the update rule from $(\mu_t, P_t)$ to
$(\mu_{t+1}, P_{t+1})$. \cite{rauch1965maximum} gives the formula of
corresponding smoother (RTS). The extended Kalman filter (EKF) is a finite
dimensional approximation of the nonlinear dynamics, which assumes that the
conditional distribution of $X_t$ is approximately Gaussian.
\cite{brown1992introduction} analyzes the EKF in continuous time.

\paragraph{Particle filter}
The particle filter \cite{gordon1993novel} applies sequential Monte Carlo
methods to generate a set of samples (particles) to represent the conditional
distribution of the states. For instance, later in (\ref{eq:particle}) we
approximate $\tilde\pi_t$ by $\hat\pi_t(x) = \frac{1}{N} \sum_{i=1}^N
\delta_{\xi_t^i}(x)$ with samples $\left\{\xi_t^i\right\}$. Here the dynamics
$f$ and $h$ can be nonlinear and the noise $\{W_t\}$ and $\{V_t\}$ can be
non-Gaussian. \cite{crisan1999particle, kushner2000nonlinear,
kushner2008numerical, kushner2013numerical} apply the particle filters to solve
the Kushner equation. Extensions of particle filter include auxiliary particle
filters \cite{pitt1999filtering}, Gaussian sum particle filters
\cite{kotecha2003gaussian}, and Rao-Blackwellised particle filtering
\cite{cappe2006inference, doucet2013rao}. The particle smoother is based on the
samples in the filter with updated reweighting. The discrete-time particle
smoother is proposed by \cite{doucet2000sequential}, which is based on the
update formula in \cite{kitagawa1987non}.

\section{CT-HMM with jump process}%
\label{sec:hmm_jump}

\subsection{Problem settings}%
\label{sec:jump_setting}

\begin{definition}
  Denote a CT-HMM $\{(X_t, Y_t): t \in [0, T]\}$ as $(Q, r, \pi_0)$ if the
  hidden states $X_t \in \{1, \dots, n\}$ and the observations $Y_t \in \{1, 
  \dots, m\}$ are generated as following:
  \begin{description}
    \item[Hidden state] $\{X_t\}$ is a jump process given by the initial
      probability $\pi_0(i) = \Pr\{X_0 = i\}$ and \emph{generator} $Q \in \R^{n
      \times n}$, i.e.
      \begin{equation}
        Q_{ij} = \lim_{\Delta t \downarrow 0}
        \frac{\Pr\{X_{t + \Delta t} = j | X_t = i\}}{\Delta t}, 
        \quad i, j \in \{1, \dots, n\},\ i \ne j,
      \end{equation}
      and $Q_{ii} = -\sum_{j \ne i} Q_{ij}$.
    \item[Observation] Let $0 = \tilde\tau_0 < \tilde\tau_1 < \tilde\tau_2 <
      \cdots < \tilde\tau_{\tilde S} \le T$ be the jumping time of $\{X_t\}$. At
      each $\tilde\tau_s$, the observation $Y_{\tilde\tau_s}$ is generated from
      \begin{equation}
        r_{i}(y) = \Pr\{Y_{\tilde\tau_s} = y | X_{\tilde\tau_s} = i\}, 
      \end{equation}
      where $i = 1, \dots, n$, $y = 1, \dots, m$, and $Y_t = Y_{\tilde\tau_s}$
      for $t \in [\tilde\tau_s, \tilde\tau_{s+1})$.
  \end{description}
\end{definition}

According to the definition above, both $\{X_t\}$ and $\{Y_t\}$ are piecewise
constant functions in $t$:
\begin{equation}
  X_t = \sum_{s=0}^{\tilde S}
  X_{\tilde\tau_s} \one_{[\tilde\tau_s, \tilde\tau_{s+1})}(t), \quad
  Y_t = \sum_{s=0}^S Y_{\tau_s} \one_{[\tau_s, \tau_{s+1})}(t), 
\end{equation}
where $\{\tilde\tau_s\}_{s=0}^{\tilde S}$ are the jumping time of $\{X_t\}$, and
$\{\tau_s\}_{s=0}^S$ are the discontinuities of $\{Y_t\}$. A key point is that
$\{\tau_s\}_{s=0}^S \subseteq \{\tilde\tau_s\}_{s=0}^{\tilde S}$ since $\{X_t\}$
may jump at $\tilde\tau_s$ but generate the same $Y_{\tilde\tau_s} =
Y_{\tilde\tau_{s-1}}$. So the \emph{embedded Markov chain} $\{(X_{\tilde\tau_s},
Y_{\tilde\tau_s})\}_{s=0}^{\tilde S}$ is a DT-HMM, but $\{(X_{\tau_s},
Y_{\tau_s})\}_{s=0}^S$ is not. Since we can only observe $\{Y_t\}$ and
$\{\tau_s\}_{s=0}^S$ instead of $\{\tilde\tau_s\}_{s=0}^{\tilde S}$, we cannot
apply the DT-HMM methods directly.

\paragraph{Comparison with related work}
Before proceeding with the detailed calculation, we compare our setting with
previous work where CT-HMM with hidden jump process are modeled under different
assumptions.

The first line of previous work assumes that $\{Y_t\}$ can only be observed at
some discrete time points \cite{bureau2003applications, jackson2011multi,
leiva2011visualization, liu2015efficient}. One of the applications is modeling
disease progression, where the disease states can be described as a jump
process, and the observations of patients are noisy and arrive irregularly in
time. Under this setting, it is reasonable to assume that the observations are
conditionally independent and may not be synchronized with the jumps. However,
the discrete-observation setting cannot be extended to the
continuous-observation setting by simply taking the continuous limit. Let $N$ be
the number of observations. If we keep increasing $N$, the observations
$\{Y_t\}$ cannot be continuous in $t$ if they are conditionally independent;
meanwhile, the state estimation will be more and more accurate, and we will get
perfect estimation $X_t = x_t$ almost surely as $N \to \infty$. Note that the
discrete-observation setting can also be solved by the Baum-Welch framework, and
we list the corresponding results in Appendix~\ref{sec:hmm_disc_obs}.

The second line of work assumes continuous observations $\{Y_t\}$ that are not
conditionally independent. Although here the number of observations $N =
\infty$, the state estimation still has a proper posterior distribution.
\cite{fischer1993markov, ryden1994parameter, ryden1996algorithm,
roberts2006ryde} consider the Markov-modulated Poisson processes where $\{Y_t\}$
is a Poisson process whose rate is determined by $\{X_t\}$, and
\cite{breuer2002algorithm, klemm2003modeling} consider a more general batch
Markovian arrival process. \cite{elliott1995hidden, ephraim2008algorithm}
considers the Markov-modulated Markov processes where $\{Y_t\}$ is a
non-homogeneous jump process whose generator is determined by $\{X_t\}$.
\cite{mark2013algorithm} further considers the bivariate Markov chain where
$\{(X_t, Y_t)\}$ together is a jump process. \cite{leiva2011visualization} has
similar settings as ours except that they need to know the underlying jumping
time $\{\tilde\tau_s\}$, while we only require $\{\tau_s\}$ from $\{Y_t\}$. In
addition, our setting can be regarded as a natural extension of the classical
case since the embedded chain $\{(X_{\tilde\tau_s},
Y_{\tilde\tau_s})\}_{s=0}^{\tilde S}$ is a DT-HMM\@. Although a DT-HMM $\{(X_t,
Y_t)\}$ is also a bivariate Markov chain, people usually further assume that
$\{X_t\}$ itself is a Markov chain, and $Y_t$ depends on $X_t$ only, which is
more similar to our setting.

From the methodology perspective, previous derivations usually follow the
standard approach of parameter estimation for (non-hidden) jump process
\cite{metzner2007generator, hobolth2011summary}, calculating the dwell time of
each hidden state and the number of jumps between two states. Here we take a
more straightforward approach based on the Baum-Welch framework. See the end of
Section~\ref{sec:jump_param} for a detailed comparison. The unified framework
can be applied to all the previous setting easily, and
Section~\ref{sec:hmm_diffusion} shows that it also works for hidden diffusion
process.

\subsection{State estimation}

For state estimation, we assume that the model $(Q, r, \pi_0)$ are given, and we
have continuous-time observations $Y_t = y_t$, $t \in [0, T]$. For parameter
estimation, we assume that the generator $Q$ is unknown and needs to be
calculated by the EM algorithm.

\begin{theorem}\label{thm:jump}
  For a CT-HMM $(Q, r, \pi_0)$ defined above, assume that the observations $Y_t
  = y_t$, $t \in [0, T]$. Let the row vector $\alpha_t = [\alpha_t(1)\ \cdots\
  \alpha_t(n)] \in \R^{1 \times n}$ and the column vector $\beta_t =
  [\beta_t(1)\ \cdots\ \beta_t(n)]^\intercal \in \R^{n \times1}$ be the
  solutions of the forward and backward piecewise ODEs respectively
  \begin{align}
    & \begin{cases}
      \alpha_0 = \pi_0 R(y_0), \\
      \dot\alpha_t = \alpha_t [D + (Q - D) R(y_{\tau_s})],
      & t \in [\tau_s, \tau_{s+1}), \\
      \alpha_{\tau_s} = \alpha_{\tau_s^-} (Q - D) R(y_{\tau_s}),
    \end{cases}
    \label{eq:jump_fwd} \\
    & \begin{cases}
      \beta_T = 1, \\
      \dot\beta_t = -[D + (Q - D) R(y_{\tau_s})] \beta_t,
      & t \in [\tau_s, \tau_{s+1}), \\
      \beta_{\tau_s^-} = (Q - D) R(y_{\tau_s}) \beta_{\tau_s}, 
    \end{cases}
    \label{eq:jump_bwd}
  \end{align}
  where $D = \diag(Q_{11}, \dots, Q_{nn})$, $R(y) = \diag(r_{1}(y), \dots,
  r_{n}(y))$ and the left limit $\alpha_{\tau_s^-} = \lim_{t \uparrow \tau_s}
  \alpha_t$. Then the posterior distribution of the hidden states satisfies
  \begin{equation}
    \rho_t(i) = \Pr\{X_t = i | Y_s = y_s, s \in [0, T]\}
    = \frac{\alpha_t(i) \beta_t(i)}{\alpha_0 \cdot \beta_0},
    \label{eq:jump_state}
  \end{equation}
  where the inner product $\alpha_0 \cdot \beta_0 = \sum_{i=1}^n \alpha_0(i)
  \beta_0(i)$.
\end{theorem}

To prove the theorem, we first discretize $[0, T]$ with time step $\Delta t$ and
then take limit $\Delta t \to 0$. Note that in the following derivation the
convergence of the stochastic process under the limit is nontrivial and requires
more careful stochastic analysis. Here we only focus on illustrating the
main idea and omit the detailed technical analysis.

An important property is that $\{(X_t, Y_t)\}$ together is a Markov process with
transition probability
\begin{equation}
  \Pr\{X_{t + \Delta t} = j, y_{t + \Delta t} | X_t = i, y_t\}
  = \begin{cases}
    (1 + Q_{ii} \Delta t) \one_{y_{t + \Delta t} = y_t}, & j = i, \\
    Q_{ij} r_j(y_{t + \Delta t}) \Delta t, & j \ne i.
  \end{cases}
  \label{eq:jump_markov}
\end{equation}
Here we omit the $o(\Delta t)$ terms on the right hand side. Then the Baum-Welch
forward-backward algorithm for DT-HMM can be modified as follows.

In the forward equation, let
\[
  \alpha_t(i) = \Pr\{X_t = i, y_{0:t}\} \Delta t^{-s},
  \quad t \in [\tau_s, \tau_{s+1}),
\]
(Later we will show that the limit of the right hand side exists when taking
$\Delta t \to 0$.) The initial probability $\alpha_0(i) = \pi_0(i) r_i(y_0)$.
Since
\begin{align*}
  \Pr & \{X_{t + \Delta t} = i, y_{0 : t + \Delta t}\}
  = \sum_{j=1}^n \Pr\{X_{t + \Delta t} = i, X_t = j, y_{0 : t + \Delta t}\} \\
  & = \sum_{j=1}^n \Pr\{X_{t + \Delta t} = i, y_{t + \Delta t} | X_t = j, y_t\}
  \Pr\{X_t = j, y_{0:t}\} \\
  & = (1 + Q_{ii} \Delta t) \one_{y_{t + \Delta t} = y_t}
  \Pr\{X_t = i, y_{0:t}\}
  + \sum_{j \ne i} Q_{ji} r_i(y_{t + \Delta t}) \Delta t \Pr\{X_t = j, y_{0:t}\},
\end{align*}
we have
\[
  \alpha_{t + \Delta t}(i) =
  \begin{cases}
    \alpha_t(i) + \left[\alpha_t(i) Q_{ii}
    + \sum_{j \ne i} \alpha_t(j) Q_{ji} r_i(y_t)\right] \Delta t,
    & y_{t + \Delta t} = y_t, \\
    \sum_{j \ne i} \alpha_t(j) Q_{ji} r_i(y_{t + \Delta t}),
    & y_{t + \Delta t} \ne y_t.
  \end{cases}
\]
Taking $\Delta t \to 0$, we get the forward piecewise ODE (\ref{eq:jump_fwd}).

In the backward equation, let
\[
  \beta_t(i) = \Pr\{y_{t + \Delta t : T} | X_t = i, y_t\} \Delta t^{-(S-s)},
  \quad t \in [\tau_s, \tau_{s+1}).
\]
Then the boundary condition $\beta_T(i) = 1$. Since
\begin{align*}
  \Pr & \{y_{t:T} | X_{t - \Delta t} = i, y_{t - \Delta t}\}
  = \sum_{j=1}^n
  \Pr\{X_t = j, y_{t:T} | X_{t - \Delta t} = i, y_{t - \Delta t}\} \\
  & = \sum_{j=1}^n \Pr\{y_{t + \Delta t : T} | X_t = j, y_t\}
  \Pr\{X_t = j, y_t | X_{t - \Delta t} = i, y_{t - \Delta t}\} \\
  & = \Pr\{y_{t + \Delta t : T} | X_t = i, y_t\}
  (1 + Q_{ii} \Delta t) \one_{y_{t - \Delta t} = y_t} \\
  & \qquad + \sum_{j \ne i} \Pr\{y_{t + \Delta t : T} | X_t = j, y_t\}
  Q_{ij} r_j(y_t) \Delta t,
\end{align*}
we have
\[
  \beta_{t - \Delta t}(i) =
  \begin{cases}
    \beta_t(i) + \left[Q_{ii} \beta_t(i)
    + \sum_{j \ne i} Q_{ij} r_j(y_t) \beta_t(j)\right] \Delta t,
    & y_{t - \Delta t} = y_t, \\
    \sum_{j \ne i} Q_{ij} r_j(y_t) \beta_t(j),
    & y_{t - \Delta t} \ne y_t.
  \end{cases}
\]
Taking $\Delta t\to0$, we get the backward piecewise ODE (\ref{eq:jump_bwd}).

Now calculate $\rho_t = \Pr\{X_t = i | y_{0:T}\}$ in (\ref{eq:jump_state}) from
$\alpha_t$ and $\beta_t$. Since
\[
  \Pr\{X_t = i, y_{0:T}\}
  = \Pr\{y_{t + \Delta t : T} | X_t = i, y_t\} \Pr\{X_t = i, y_{0:t}\}
  = \alpha_t(i) \beta_t(i) \Delta t^S,
\]
and
\[
  \Pr\{y_{0:T}\} = \sum_i \Pr\{X_t = i, y_{0:T}\}
  = \sum_i \alpha_t(i) \beta_t(i) \Delta t^S
  = \alpha_t \cdot \beta_t \Delta t^S,
\]
the inner product $\alpha_t \cdot \beta_t$ is constant for all
$t$. Therefore, 
\[
  \rho_t(i) = \frac{\Pr\{X_t = i, y_{0:T}\}}{\Pr\{y_{0:T}\}}
  = \frac{\alpha_t(i) \beta_t(i)}{\alpha_0 \cdot \beta_0}.
\]

\subsection{Parameter estimation}%
\label{sec:jump_param}

In case when the generator $Q$ is an unknown parameter in the model, it can be
estimated by the following EM algorithm.

\begin{algorithm}\label{alg:jump}
  For a CT-HMM $(Q, r, \pi_0)$ defined above, assume that the observations $Y_t
  = y_t$, $t \in [0, T]$, and $Q$ is the unknown parameter. Let $Q^0$ be the
  initialization of the generator, then repeat the following E-step and M-step
  to update $Q^k$, $k = 0, 1, \dots$ until convergence.
  \begin{description}
    \item[E-step:] under the current estimate $Q = Q^k$, solve the forward and
      backward piecewise ODEs (\ref{eq:jump_fwd}) and (\ref{eq:jump_bwd}) for
      $\alpha_t$ and $\beta_t$, and calculate the posterior distribution
      $\rho_t$ from (\ref{eq:jump_state}).
    \item[M-step:] update the generator from $Q^k$ to $Q^{k+1}$ using
      \begin{equation}
        Q_{ij}^{k+1}
        = \frac{1}{(\alpha_0 \cdot \beta_0) \int_0^T \rho_t(i) \ud t}
        \bigg[\sum_{s=0}^S \int_{\tau_s}^{\tau_{s+1}}
        \alpha_t(i) Q_{ij}^k r_j(y_{\tau_s}) \beta_t(j) \ud t
        + \sum_{s=1}^S
        \alpha_{\tau_s^-}(i) Q_{ij}^k r_{j}(y_{\tau_s}) \beta_{\tau_s}(j)\bigg]
        \label{eq:jump_q}
      \end{equation}
    for $j \ne i$, and $Q_{ii}^{k+1} = -\sum_{j \ne i}Q_{ij}^{k+1}$.
  \end{description}
\end{algorithm}

The goal of the E-step is to calculate the posterior distribution $\Pr\{X_t |
y_{0:T}; Q^k\}$ under the current generator $Q^k$. This is equivalent to the
previous state estimation problem if we replace the true generator $Q$ by $Q^k$.
Besides $\alpha_t$, $\beta_t$ and $\rho_t$, here we also need to calculate
\[
  \xi_t(i, j) = \Pr\{X_{t - \Delta t} = i, X_t = j | y_{0:T}\}.
\]
under the time discretization. Since
\begin{multline*}
  \Pr\{X_{t - \Delta t} = i, X_t = j, y_{0:T}\} \\
  = \Pr\{y_{t + \Delta t:T} | X_t = j, y_t\}
  \Pr\{X_t = j, y_t | X_{t - \Delta t} = i, y_{t - \Delta t}\}
  \Pr\{X_{t - \Delta t} = i, y_{0 : t - \Delta t}\},
\end{multline*}
we can show that
\[
  (\alpha_0 \cdot \beta_0) \xi_t(i, j) =
  \begin{cases}
    \alpha_{t - \Delta t}(i) (1 + Q_{ii} \Delta t) \beta_t(i),
    & y_{t - \Delta t} = y_t,\ j = i, \\
    \alpha_{t - \Delta t}(i) Q_{ij} r_j(y_t) \beta_t(j) \Delta t,
    & y_{t - \Delta t} = y_t,\ j \ne i,\\
    0, & y_{t - \Delta t} \ne y_t,\ j = i,\\
    \alpha_{t - \Delta t}(i) Q_{ij} r_j(y_t) \beta_t(j),
    & y_{t - \Delta t} \ne y_t,\ j \ne i.
  \end{cases}
\]

The goal of the M-step is to calculate the maximum-likelihood estimation of the
parameter $Q$ given the current estimation of $\{(X_t, Y_t)\}$. Under the time
discretization, the update formula of the EM algorithm (\ref{eq:em_q}) gives
$Q^{k+1} = \argmax_Q \cQ(Q, Q^k)$, where
\[
  \cQ(Q, Q^k)
  = \sum_{x_{0:T}} \Pr\{x_{0:T} | y_{0:T}; Q^k\} \log\Pr\{x_{0:T}, y_{0:T}; Q\}.
\]

Let $\alpha_t$, $\beta_t$, $\rho_t$ and $\xi_t$ are calculated in the E-step
with respect to $Q^k$. Since $\{(X_t, Y_t)\}$ together is a Markov process with
transition probabilities (\ref{eq:jump_markov}), 
\[
  \Pr\{x_{0:T}, y_{0:T}; Q\} =
  \Pr\{x_0, y_0\} \prod_{t = \Delta t}^T
  \Pr\{x_t, y_t | x_{t - \Delta t}, y_{t - \Delta t}; Q\}.
\]
then
\begin{align*}
  \cQ(Q, Q^k)
  & = \sum_{x_{0:T}} \Pr\{x_{0:T} | y_{0:T}; Q^k\}
  \left[\log\Pr\{x_0, y_0\} + \sum_{t = \Delta t}^T
  \log\Pr\{x_t, y_t | x_{t - \Delta t}, y_{t - \Delta t}; Q\}\right] \\
  & = \sum_{x_0} \Pr\{x_0 | y_0; Q^k\} \log\Pr\{x_0, y_0\} \\
  & \qquad + \sum_t \sum_{x_{t - \Delta t}, x_t}
  \Pr\{x_{t - \Delta t}, x_t | y_{0:T}; Q^k\}
  \log\Pr\{x_t, y_t | x_{t - \Delta t}, y_{t - \Delta t}; Q\}.
\end{align*}
Since $\Pr\{x_0, y_0\}$ in the first term does not depend on $Q$, we only need
to maximize the second term
\begin{align*}
  \sum_t & \sum_{x_{t - \Delta t}, x_t}
  \Pr\{x_{t - \Delta t}, x_t | y_{0:T}; Q^k\}
  \log\Pr\{x_t, y_t | x_{t - \Delta t}, y_{t - \Delta t}; Q\} \\
  & = \sum_{i, j} \sum_t \xi_t(i, j)
  \log\Pr\{X_t = j, y_t | X_{t - \Delta t} = i, y_{t - \Delta t}; Q\} \\
  & = \sum_i \sum_t \xi_t(i,i) \log(1 + Q_{ii} \Delta t)
  + \sum_{j \ne i} \sum_t \xi_t(i, j) [\log(Q_{ij} \Delta t) + \log r_j(y_t)].
\end{align*}
Notice that $(1 + Q_{ii} \Delta t) + \sum_{j \ne i} (Q_{ij} \Delta t) = 1$ and
$\sum_{j=1}^n \xi_t(i, j) = \rho_{t - \Delta t}(i)$, then
\begin{multline*}
  Q_{ij}^{k+1}
  = \frac{\sum_t \xi_t(i, j)}{\sum_{j'=1}^n \sum_t \xi_t(i, j') \Delta t}
  = \frac{1}{(\alpha_0 \cdot \beta_0) \sum_t\rho_{t - \Delta t}(i) \Delta t} \\
  \cdot \left[\sum_{y_{t - \Delta t} = y_t}
  \alpha_{t - \Delta t}(i) Q_{ij}^k r_j(y_t) \beta_t(j) \Delta t
  + \sum_{y_{t - \Delta t} \ne y_t}
  \alpha_{t - \Delta t}(i) Q_{ij}^k r_j(y_t) \beta_t(j)\right]
\end{multline*}
for $j \ne i$. Taking $\Delta t \to 0$, we will have the update
(\ref{eq:jump_q}) in Algorithm~\ref{alg:jump}.

\paragraph{Comparison with the bivariate Markov process}
The bivariate Markov process \cite{mark2013algorithm} assumes that $\{(X_t,
Y_t)\}$ together is a jump process. This is closely related to our setting. In
our setting $\{(X_t, Y_t)\}$ is also a bivariate Markov process with special
structure. On the other hand, in the bivariate Markov process if we rewrite
$\tilde X_t = (X_t, Y_t)$ together as the hidden state and keep $\tilde Y_t =
Y_t$ as the observation, then $\{(\tilde X_t, \tilde Y_t)\}$ fits in our
setting. In this case $\tilde Y_t$ is a \emph{partial observation} of $\tilde
X_t$, while our setting allows that $\tilde Y_t$ is not necessarily part of
$\tilde X_t$, but contains \emph{partial information} of $\tilde X_t$.

The following is a comparison of the results. For the state estimation,
\cite{mark2013algorithm} derives the forward and backward recursion similar to
our equations for $\alpha_t$ and $\beta_t$. They are defined on the jumping time
$t = \tau_s$ only, but can be generalized to any $t \in [\tau_s, \tau_{s+1})$.
In addition, in our setting we need to calculate the jump of $\alpha_t$ and
$\beta_t$ from $\tau_s^-$ to $\tau_s$.

For the parameter estimation, \cite{mark2013algorithm} considers the dwell time
$D(i)$ and the number of jumps $m(i, j)$ as
\[
  D(i) = \int_0^T \one_{\{\tilde X_t = i\}} \ud t, \quad
  m(i, j) = \sum_{s=0}^{\tilde S}
  \one_{\left\{\tilde X_{\tilde\tau_s^-} = i,\
  \tilde X_{\tilde\tau_s} = j\right\}},
\]
where $\tilde X_t = (X_t, Y_t)$ and $\tilde\tau_s$ is the jumping time of
$\{\tilde X_t\}$ (instead of $\{Y_t\}$). The EM algorithm updates the generator
as
\[
  Q_{ij}^{k+1} =
  \frac{\E[m(i,j) | y_t, t \in [0, T]; Q^k]}{\E[D(i) | y_t, t \in [0, T]; Q^k]}.
\]
Note that \cite{mark2013algorithm} uses the update formula without proof (can be
derived using ideas similar to \cite{ryden1996algorithm}), and then calculates
the conditional expectation of $D(i)$ and $m(i, j)$. In our setting, they can be
calculated in the same way by restricting the generator such that $\{Y_t\}$
jumps simultaneously with $\{X_t\}$. However, we get a more concise formula
(\ref{eq:jump_q}) than the one above using conditional expectations, and their
results can also be derived from. In addition, the bivariate Markov process
might be hard to calculate when the sizes of the state and observation spaces
$m$ and $n$ are large (\cite{mark2013algorithm} only runs a numerical experiment
for $m = n = 2$).

\section{CT-HMM with diffusion process}%
\label{sec:hmm_diffusion}

\subsection{Problem settings}

In this section, we analyze another type of CT-HMM where the hidden process is a
diffusion process, and the hidden states and observations take continuous
values. We consider the same setting as the Stratonovich-Kushner and Zakai
equations. Assume that the hidden process $X_t \in \R^n$ and the observations
$Y_t \in \R^m$ are given by SDEs
\begin{align}
  \ud X_t & = f(X_t)\ud t + \sigma \ud W_t, \nonumber \\
  \ud Y_t & = h(X_t)\ud t + \eta \ud B_t,
  \label{eq:dif_model}
\end{align}
where the drift terms $f: \R^n \to \R^n$, $h: \R^n \to \R^m$; the noise terms
$\{W_t\}$ and $\{B_t\}$ are independent $n$-dimensional and $m$-dimensional
Brownian motions respectively. Here we assume that $\sigma$ and $\eta$ are
scalars for brevity, and all the results can be easily extended to the case when
they are matrices. Let the initial distribution of $X_0$ be $\pi_0(x) = p(X_0 =
x)$, and the time period $t \in [0, T]$.

\subsection{State estimation}
\label{sec:dif_state}

State estimation is closely related to the filtering and smoothing problem (see
Section~\ref{sec:filtering}). The optimal nonlinear filtering of the diffusion
model (\ref{eq:dif_model}) is solved by the \emph{Stratonovich-Kushner equation}
\cite{stratonovich1959optimum, kushner1964differential, kushner1967dynamical}
that describes the dynamics of the density of the states condition on the
\emph{previous} observations $\tilde\pi_t(x) = p(X_t = x | Y_{0:t})$.
\begin{equation}
  \ud \tilde\pi_t(x) = \cL^* \tilde\pi_t(x) +
  \frac{1}{\eta^2} \tilde\pi_t(x) [h(x) -\E_{X \sim \tilde\pi_t} h(X)]^\intercal 
  [\ud Y_t - \E_{X \sim \tilde\pi_t} h(X) \ud t].
\end{equation}
(See Theorem~\ref{thm:diffusion} for the definition of the operators $\cL$ and
$\cL^*$.) The \emph{Zakai equation} \cite{zakai1969optimal} introduces a
simplified dynamics for the unnormalized conditional distribution $\pi_t(x) =
\tilde\pi_t(x) Z_t$ with some constant $Z_t$, and we take it as the forward part
(\ref{eq:dif_fwd}). The corresponding smoothing problem considers the density of
the states condition on the \emph{whole} observation process $\rho_t(x) = p(X_t
= x | Y_{0:T})$. \cite{anderson1972fixed} derives the differential equation
\begin{equation}
  \frac{\ud \rho_t(x)}{\ud t}
  = \frac{\rho_t(x)}{\tilde\pi_t(x)} \cL^* \tilde\pi_t(x)
  - \tilde\pi_t(x) \cL \left[\frac{\rho_t(x)}{\tilde\pi_t(x)}\right].
  \label{eq:dif_state_mix}
\end{equation}
Unlike (\ref{eq:dif_state_mix}) where the equation of the smoother $\rho_{t}$
contains the filtered $\tilde\pi_t$, here we follow the approach in
\cite{pardouxt1980stochastic, campillo1989mle, papanicolaou2014stochastic},
writing the smoother as the product of two terms that are the solutions of the
forward and backward SPDEs respectively, which goes in line with the Baum-Welch
framework.

\begin{theorem}\label{thm:diffusion}
  For a CT-HMM with hidden diffusion process defined above, let $\pi_t$ and
  $\beta_t$ be the solutions of the forward and backward SPDEs respectively
  \begin{gather}
    \pi_0(x) = p\left(X_0 = x\right), \quad
    \ud\pi_t(x) = \cL^*\pi_t(x) \ud t
    + \eta^{-2} \pi_t(x) h^\intercal(x) \ud Y_t,
    \label{eq:dif_fwd} \\
    \beta_T(x) = 1, \quad
    \ud\beta_t(x) = -\cL\beta_t(x) \ud t
    - \eta^{-2} \beta_t(x) h^\intercal(x) [\ud Y_t - h(x) \ud t],
    \label{eq:dif_bwd}
  \end{gather}
  where
  \begin{align*}
    \cL^*\pi_t(x) & = -\nabla \cdot [f(x) \pi_t(x)]
    + \frac{\sigma^2}{2} \nabla^2\pi_t(x), \\
    \cL\beta_t(x) & = f(x) \cdot \nabla\beta_t(x)
    + \frac{\sigma^2}{2} \nabla^2\beta_t(x);
  \end{align*}
  the gradient operator $\nabla = \left[\frac{\partial}{\partial x_1}, \dots,
  \frac{\partial}{\partial x_n}\right]$, and the Laplace operator $\nabla^2 =
  \nabla \cdot \nabla = \sum_{i=1}^n \frac{\partial^2}{\partial x_i^2}$. Then
  the posterior distribution of the hidden states is given by
  \begin{equation}
    \rho_t(x) = p(X_t = x | Y_s = y_s, s \in [0, T])
    = \frac{1}{Z_T} \pi_t(x) \beta_t(x)
    \label{eq:dif_state}
  \end{equation}
  for some normalization factor $Z_T$.
\end{theorem}

The proof is based on the derivation of the Zakai equation and we give the
sketch here. One can follow \cite{papanicolaou2014stochastic} for the complete
proof.

Since $\ud Y_t = h(X_t) \ud t + \eta \ud B_t$, from Girsanov theorem, define an
equivalent measure $Q$ by
\[
  \ud Q = M_T^{-1} \ud P, \quad
  M_t = \exp\left[-\frac{1}{2 \eta^2} \int_0^t h^\intercal(X_s) h(X_s) \ud s
  + \frac{1}{\eta^2} \int_0^t h^\intercal(X_s) \ud Y_s\right].
\]
Then under measure $Q$, $\{Y_t / \eta\}$ is a Brownian motion and $\{M_t\}$ is a
martingale, while $\{X_t\}$ and $\{Y_t\}$ are independent. For any function $g$,
let 
\[
  \psi_t[g] = \E_P[g(X_t) | Y_{0:t}], \quad
  \varphi_t[g] = \E_Q[g(X_t) M_t | Y_{0:t}], \quad
  Z_t = \varphi_t[1] = \E_Q[M_t|Y_{0:t}], 
\]
one can show that 
\[
  \psi_t[g] = \varphi_t[g] / Z_t.
\]
Since $\{X_t\}$ and $\{Y_t\}$ are independent under $Q$, we can derive the
dynamics of
$\varphi_t[g]$ as
\begin{equation}
  \ud\varphi_t[g] = \varphi_t[\cL g] \ud t
  + \varphi_t\left[g \frac{h^\intercal}{\eta^2}\right] \ud Y_t.
  \label{eq:zakai}
\end{equation}
Assume that we can write $\varphi_t[g] = \int_{\R^n} g(x) \pi_t(x) \ud x$, then
$\pi_t$ is an unnormalized density $\pi_t(x) = p(X_t = x | Y_{0:t}) Z_t$, and
satisfies the adjoint of (\ref{eq:zakai}), i.e., the Zakai equation
(\ref{eq:dif_fwd}).

Furthermore, let the smoothing component be
\[
  \beta_t(x) = \E_Q[M_T / M_t | X_t = x, Y_{0:T}], 
\]
then one can show that $\rho_t(x) = \pi_t(x) \beta_t(x) / Z_T$, and $\beta_t$
satisfies the backward SPDE (\ref{eq:dif_bwd}). We can also calculate that
\begin{equation}
  \frac{\ud\rho_t(x)}{\ud t}
  = \frac{1}{Z_T} [\beta_t(x) \cL^*\pi_t(x) - \pi_t(x) \cL\beta_t(x)], 
\end{equation}
which is equivalent to (\ref{eq:dif_state_mix}).

\subsection{Parameter estimation}

For simplicity, we assume that only $f$ contains unknown parameters $f = f(x;
\theta)$. The following is the EM algorithm.

\begin{algorithm}\label{alg:diffusion}
  For a CT-HMM with hidden diffusion process defined above, assume that the
  observations $Y_t = y_t$, $t \in [0, T]$, and $f = f(x; \theta)$ where
  $\theta$ is the unknown parameter. Let $\theta^0$ be the initialization of
  $\theta$, then repeat the following E-step and M-step to update $\theta^k$, $k
  = 0, 1, \dots$ until convergence.
  \begin{description}
    \item[E-step:] using the current value of the estimated $\theta = \theta^k$,
      solve the forward and backward SPDEs (\ref{eq:dif_fwd}) and
      (\ref{eq:dif_bwd}) for $\pi_t$ and $\beta_t$ respectively, and calculate
      the posterior distribution $\rho_t$ from (\ref{eq:dif_state}).
    \item[M-step:] update the parameter by $\theta^{k+1} =
      \argmin_\theta\tilde\cQ(\theta, \theta^k)$ where
      \begin{equation}
        \tilde\cQ(\theta, \theta^k) = \int_0^T\int_{\R^n} \rho_t(x)
        \left[\|f(x; \theta)\|_2^2 - 2 f^\intercal(x; \theta) \left[
        f(x; \theta^k) + \frac{\sigma^2 \nabla\beta_t(x)}{\beta_t(x)}
        \right]\right] \ud x \ud t.
        \label{eq:dif_q}
      \end{equation}
  \end{description}
\end{algorithm}

The E-step is equivalent to the state estimation discussed above. So we will
focus on the M-step. We still discretize $[0, T]$ with time step $\Delta t$ and
then take limit $\Delta t \to 0$. The parameter update formula (\ref{eq:em_q})
gives $\theta^{k+1} = \argmax_\theta \cQ(\theta, \theta^k)$ where
\begin{equation}
  \cQ(\theta, \theta^k) = \int p(x_{0:T}|y_{0:T}; \theta^k)
  \log p(x_{0:T}, y_{0:T}; \theta) \ud x_{0:T}
  \label{eq:dif_q0}
\end{equation}
(the integral takes over $\ud x_{0:T} = \ud x_0 \ud x_{\Delta t} \ud x_{2 \Delta
t} \cdots \ud x_T$). Notice that
\[
  p(x_{0:T}, y_{0:T}; \theta)
  = p(x_0) \prod_{t=0}^{T - \Delta t} p(\Delta x_t | x_t; \theta)
  \prod_{t=0}^T p(\Delta y_t | x_t), 
\]
where
\[
  p(\Delta x_t | x_t; \theta) = \frac{1}{Z_f} \exp\left[
  -\frac{\|\Delta x_t - f(x_t; \theta) \Delta t\|_2^2}{2 \sigma^2 \Delta t}
  \right]
\]
for $Z_f = \left(2 \pi \sigma^2\right)^{n/2}$, and $p(x_0)$, $p(\Delta y_t |
x_t)$ do not depend on $\theta$. So we have
\begin{align*}
  \cQ & (\theta, \theta^k)
  = \int p(x_{0:T} | y_{0:T}; \theta^k)
  \bigg[\log p(x_0) + \sum_{t=0}^{T - \Delta t} \log p(\Delta x_t | x_t; \theta)
  + \sum_{t=0}^T \log p(\Delta y_t | x_t)\bigg] \ud x_{0:T} \\
  & = \int p(x_0 | y_{0:T}; \theta^k) \log p(x_0) \ud x_0
  + \sum_{t=0}^{T - \Delta t} \int p(x_t, \Delta x_t | y_{0:T}; \theta^k)
  \log p(\Delta x_t | x_t; \theta) \ud x_t \ud x_{t + \Delta t} \\
  & \qquad + \sum_{t=0}^T \int
  p(x_t | y_{0:T}; \theta^k) \log p(\Delta y_t | x_t) \ud x_t.
\end{align*}
Only the second summation term depends on $\theta$, whose integral becomes
\begin{align*}
  \int p & (x_t, \Delta x_t | y_{0:T}; \theta^k)
  \log p(\Delta x_t | x_t; \theta) \ud x_t \ud x_{t + \Delta t} \\
  & = -\log Z_f - \int \rho_t(x_t) p(\Delta x_t | x_t, y_{0:T}; \theta^k)
  \frac{\|\Delta x_t - f(x_t; \theta) \Delta t\|_2^2}{2 \sigma^2 \Delta t}
  \ud x_t \ud x_{t + \Delta t} \\
  & = C - \frac{1}{2 \sigma^2} \int \rho_t(x_t)
  \left[\|f(x_t; \theta)\|_2^2 \Delta t - 2 f^\intercal(x_t; \theta)
  \E[\Delta x_t | x_t, y_{0:T}; \theta^k]\right] \ud x_t \\
  & = C - \frac{1}{2 \sigma^2} \int \rho_t(x)
  \left[\|f(x; \theta)\|_2^2 \Delta t - 2 f^\intercal(x; \theta)
  \E[\Delta x | X_t = x, y_{0:T}; \theta^k]\right] \ud x, 
\end{align*}
where $C$ is a constant and not depending on $\theta$. Therefore, $\theta^{k+1}
= \argmin_\theta\tilde\cQ(\theta, \theta^k)$ where
\[
  \tilde\cQ(\theta, \theta^k) = \int_{\R^n} \sum_{t=0}^{T - \Delta t} \rho_t(x)
  \left[\|f(x; \theta)\|_2^2 \Delta t - 2 f^\intercal(x; \theta)
  \E[\Delta x | X_t = x, y_{0:T}; \theta^k]\right] \ud x.
\]
Now take $\Delta t \to 0$, we have
\[
  \tilde\cQ(\theta, \theta^k) = \int_0^T \int_{\R^n} \rho_t(x)
  \left[\|f(x; \theta)\|_2^2 - 2 f^\intercal(x; \theta)
  \left.\frac{\ud}{\ud\tau}\right|_{\tau=0}
  \E[X_{t + \tau} | X_t = x, y_{0:T}; \theta^k]\right] \ud x \ud t, 
\]
and from (\ref{eq:dif_state}) we can calculate that
\[
  \left.\frac{\ud}{\ud\tau}\right|_{\tau=0}
  \E\left[X_{t + \tau} | X_t = x, y_{0:T}\right]
  = f(x) + \frac{\sigma^2 \nabla\beta_t(x)}{\beta_t(x)}.
\]
Therefore, (\ref{eq:dif_q}) holds in Algorithm~\ref{alg:diffusion}.

\section{Monte Carlo method}

For some simple cases like the linear Gaussian dynamics
(Appendix~\ref{sec:dif_linear}), we may have explicit solutions of the SPDEs
(\ref{eq:dif_fwd}) and (\ref{eq:dif_bwd}). In general cases, however, we have to
resort to numerical solutions of the Stratonovich-Kushner and Zakai equations. A
direct approach is the \emph{finite-difference splitting}. One can take a fixed
non-random grid $\{x_{1},\dots,x_{N}\}$ in the state space, and calculate
$\pi_{t}(x_{i})$ to approximate the evolving measure. See the line of work
\cite{bensoussan1990approximation, bensoussan1992approximation,
florchinger1991time, jentzen2009numerical} for the theoretical proof of the
convergence of this kind of numerical schemes. However, for a $d$-dimensional
state space with uniform grid size $h$, the number of grid points will be
$O\left(h^{-d}\right)$, which is unaffordable for high-dimensional problems.
Note that for the particle filter, the convergence rate is $C N^{-1/2}$. Though
$C$ might have exponential dependence in $d$, this is different from the
situation when the rate itself depends on $d$. In practice the Monte Carlo
method makes it possible to solve some high-dimensional problems like
(\ref{eq:cubic_scalar}) with much fewer samples.

In the following we propose a Monte Carlo sampling method to combine with the
parameter update (\ref{eq:dif_q}). For state estimation, notice that $\pi_t(x_t)
\propto p(X_t = x_t | y_{0:t})$ and $\rho_t(x_t) = p(X_t = x_t | y_{0:T})$, we
can modify the particle filter and smoother to generate samples to describe
those posterior distributions. For the objective (\ref{eq:dif_q}) in the
parameter estimation, the integral over $x$ can be calculated by summing over
the samples. Note that we take the vanilla particle filter as illustration. The
continuous formulation also gives us more freedom to adaptively choose proper
discretization and sampling scheme.

First, we solve the filtering problem $\tilde\pi_t(x) = p(X_t = x | y_{0:t})$.
Discretize $[0, T]$ with time step $\Delta t$. For each $t = 0, \Delta t, 2
\Delta t, \dots, T$, we will have $N$ samples $\xi_t^i$, $i = 1, \dots, N$, such
that the filtered $\tilde\pi_t(x)$ can be approximated by
\begin{equation}
  \hat\pi_t(x) = \frac{1}{N} \sum_{i=1}^N \delta_{\xi_t^i}(x).
  \label{eq:particle}
\end{equation}
Let $\xi_0^i \sim \pi_0$ be i.i.d.\ samples of the initial distribution. Assume
that we already have $\xi_t^i$. Resample
\[
  \tilde\xi_t^i \sim \sum_{i=1}^N
  \frac{p\left(\Delta y_t | X_t = \xi_t^i\right)}
  {\sum_{i'=1}^N p\left(\Delta y_t | X_t = \xi_t^{i'}\right)}
  \delta_{\xi_t^i}(x)
\]
independently, then the distribution of $\tilde\xi_t^i$ satisfies
$p(X_t | y_{0:t + \Delta t})$. Thus we can sample
\[
  \xi_{t + \Delta t}^i
  = \tilde\xi_t^i + f\left(\tilde\xi_t^i\right) \Delta t + \sigma\Delta W_t^i, 
  \quad \Delta W_t^i \sim \cN(0, \Delta t)\ \textrm{i.i.d.},
\]
and $\xi_{t + \Delta t}^i$ is a set of samples for $t + \Delta t$.

Next, we solve the smoothing problem $\rho(x) = p(X_t = x|y_{0:T})$,
approximating it by weighting the samples
\begin{equation}
  \hat\rho(x) = \sum_{i=1}^N w_t^i \delta_{\xi_t^i}(x).
\end{equation}
We calculate the weight $w_t^i$ backwards for $t = T, T - \Delta t, T -2 \Delta
t, \dots, 0$. Since $\tilde\pi_T = \rho_T$, we have $w_T^i = 1/N$. Notice that
\begin{align*}
  \rho_t(x_t)
  & = p(X_t = x_t | y_{0:T}) \\
  & = \int p(X_t = x_t | X_{t + \Delta t} = x_{t + \Delta t}, y_{0:T})
  p(X_{t + \Delta t} = x_{t + \Delta t} | y_{0:T}) \ud x_{t + \Delta t} \\
  & = \int p(x_t | x_{t + \Delta t}, y_{0:t + \Delta t})
  p(x_{t + \Delta t} | y_{0:T}) \ud x_{t + \Delta t} \\
  & = \int \frac{p(x_{t + \Delta t}, \Delta y_t | x_t)p(x_t | y_{0:t})}
  {\int p(x_{t + \Delta t}, \Delta y_t | x_t') p(x_t' | y_{0:t}) \ud x_t'}
  p(x_{t + \Delta t} | y_{0:T}) \ud x_{t + \Delta t} \\
  & = \tilde\pi_t(x_t) \int \frac{p(x_{t + \Delta t} | x_t)p(\Delta y_t | x_t)}
  {\int p(x_{t + \Delta t} | x_t') p(\Delta y_t | x_t')
  \tilde\pi_t(x_t') \ud x_t'}
  \rho_{t + \Delta t}(x_{t + \Delta t}) \ud x_{t + \Delta t}.
\end{align*}
Replace $\tilde\pi_t$ and $\rho_{t + \Delta t}$ by $\hat\pi_t$
and $\hat\rho_{t + \Delta t}$ respectively, we have
\begin{align*}
  \hat\rho_t(x_t)
  \hat{\rho}_{t}(x_{t})
  & = \frac{1}{N} \sum_{i=1}^N \delta_{\xi_t^i}(x_t)
  \sum_{j=1}^N
  \frac{p(\xi_{t + \Delta t}^j | x_t) p(\Delta y_t | x_t) w_{t + \Delta t}^j}%
  {\frac{1}{N} \sum_{i'=1}^N p(\xi_{t + \Delta t}^j | \xi_t^{i'})
  p(\Delta y_t | \xi_t^{i'})} \\
  & = \sum_{i=1}^N \sum_{j=1}^N
  \frac{p(\xi_{t + \Delta t}^j | \xi_t^i) p(\Delta y_t | \xi_t^i)
  w_{t + \Delta t}^j}
  {\sum_{i'=1}^N p(\xi_{t + \Delta t}^j | \xi_t^{i'})
  p(\Delta y_t | \xi_t^{i'})} \delta_{\xi_t^i}(x_t)
\end{align*}
Therefore, the weights are calculated as
\begin{equation}
  w_t^i = \sum_{j=1}^N
  \frac{p(\xi_{t + \Delta t}^j | \xi_t^i) p(\Delta y_t | \xi_t^i)
  w_{t + \Delta t}^j}
  {\sum_{i'=1}^N p(\xi_{t + \Delta t}^j | \xi_t^{i'})
  p(\Delta y_t | \xi_t^{i'})}.
\end{equation}
In addition, since
\[
  \beta_t(x_t) \propto \frac{\rho_t(x_t)}{\tilde\pi_t(x_t)}
  = \int \frac{p(x_{t + \Delta t} | x_t) p(\Delta y_t |x_t)}%
  {\int p(x_{t + \Delta t} | x'_t) p(\Delta y_t |x'_t)
  \tilde\pi_t(x'_t) \ud x'_t}
  \rho_{t + \Delta t}(x_{t + \Delta t}) \ud x_{t + \Delta t},
\]
let
\begin{equation}
  \hat\beta_t(x_t) = \sum_{j=1}^N
  \frac{p(\xi_{t + \Delta t}^j | x_t)p(\Delta y_t | x_t) w_{t + \Delta t}^j}
  {\sum_{i'=1}^N
  p(\xi_{t + \Delta t}^j | \xi_t^{i'}) p(\Delta y_t|\xi_t^{i'})}, 
\end{equation}
then $\beta_t$ can be approximated by $\hat\beta_t$ (times a constant factor,
which will be eliminated in the $\nabla\beta_t / \beta_t$ term in
(\ref{eq:dif_q})). Since both $p(\xi_{t + \Delta t}^j | x_t)$ and $p(\Delta y_t
| x_t)$ are Gaussian, the gradient $\nabla\hat\beta_t(x_t)$ can be calculated
analytically. Notice that $w_t^i = \hat\beta_t\left(\xi_t^i\right)$.

Now we can apply the Monte Carlo approach to the Algorithm~\ref{alg:diffusion}
for parameter estimation. In the E-step, the posterior distributions can be
sampled as above. In the M-step, the parameter update (\ref{eq:dif_q}) can be
modified as $\theta^{k+1} = \argmin_\theta \hat\cQ(\theta, \theta^k)$ where
\begin{equation}
  \hat\cQ(\theta, \theta^k) = \sum_{t=0}^T\sum_{i=1}^N w_t^i
  \left[\left\|f\left(\xi_t^i; \theta\right)\right\|_2^2
  - 2 f^\intercal\left(\xi_t^i; \theta\right)
  \left[f\left(\xi_t^i; \theta^k\right)
  + \frac{\sigma^2}{w_t^i} \nabla\hat\beta_t\left(\xi_t^i\right)\right]\right].
  \label{eq:dif_q_mc}
\end{equation}

If $f(x; \theta)$ is linear in $\theta$ (may be nonlinear in $x$): $f(x; \theta)
= A(x)\theta + b(x)$, then
\[
  \hat\cQ(\theta, \theta^{k})
  =\sum_{t=0}^T \sum_{i=1}^N w_t^i \bigg[
  \theta^\intercal A^\intercal\left(\xi_t^i\right) A\left(\xi_t^i\right) \theta
  - 2{\left[A\left(\xi_t^i\right) \theta
  + b\left(\xi_t^i\right)\right]}^\intercal 
  \left[A\left(\xi_t^i\right) \theta^k
  + \frac{\sigma^2}{w_t^i} \nabla \hat\beta_t\left(\xi_t^i\right)\right]
  - \left\|b\left(\xi_t^i\right)\right\|_2^2\bigg],
\]
and we have the explicit solution
\begin{align}
  \theta^{k+1}
  & = \argmin_\theta \hat\cQ(\theta, \theta^k) \nonumber \\
  & = \theta^k + \sigma^2
  {\left[\sum_{t=0}^T \sum_{i=1}^N
  w_t^i A^\intercal\left(\xi_t^i\right) A\left(\xi_t^i\right)\right]}^{-1}
  \left[\sum_{t=0}^T \sum_{i=1}^N
  A^\intercal\left(\xi_t^i\right) \nabla\hat\beta_t\left(\xi_t^i\right)\right].
\end{align}
In the general case when (\ref{eq:dif_q_mc}) has no explicit solution, we may
solve $\theta^{k+1}$ using gradient-based methods.

\paragraph{%
  Discussion about previous parameter estimation methods in applications%
}
Here we briefly discuss the previous related work on parameter estimation. In
\cite{ching2006bayesian, campillo2009convolution}, the authors regard the
parameter $\theta$ as part of the hidden state. They consider $\tilde X_t =
(X_t, \theta)$ with dynamics
\[
  \left[\begin{array}{c} X_{t+1} \\ \theta_{t+1} \end{array}\right]
  = \left[\begin{array}{c} f(X_t) \\ \theta_t \end{array}\right]
  + \left[\begin{array}{c} W_t \\ 0 \end{array}\right],
\]
and apply state estimation methods for $\{\tilde X_t\}$.

A more effective approach is to apply the EM algorithm with state estimation and
parameter update alternately. \cite{shumway1982approach} proposes an EM
algorithm for discrete-time Kalman filter. \cite{dembo1986parameter} considers
continuous diffusion process, and gives theoretical formulas for the EM
algorithm. \cite{campillo1989mle} further compares direct maximization of the
likelihood function with the EM algorithm, and shows that smoothing is necessary
for the EM approach. For practical problems, however, most previous work still
considers discrete-time settings. In particular, \cite{schon2011system,
zhao2013parameter, feng2016expectation, lucini2019model} all take the following
approach. Recall the parameter update $\theta^{k+1} = \argmax_\theta \cQ(\theta,
\theta^k)$ for $\cQ$ defined in (\ref{eq:dif_q0}). The E-step uses the particle
filter and smoother to sample from $p(X_{0:T} | y_{0:T}; \theta^k)$. Instead of
simplifying the objective, they calculate $\cQ(\theta, \theta^k)$ directly as a
sum over the discrete time grid and samples. Then the M-step maximizes over
$\theta$ using the gradients $\nabla_\theta\cQ(\theta, \theta^k)$. As a
comparison, we rewrite $\cQ(\theta, \theta^k)$ in the continuous form. The
explicit solution of the linear case comes as a straightforward byproduct. In
the nonlinear case, we use the standard particle filter and smoother. But note
that the continuous formulation gives us more freedom to adaptively choose
proper discretization and sampling scheme (like the extensions of particle
filter) for better performance.

\section{Simulation results}%
\label{sec:dif_simulate}

\subsection{Generator of hidden jump process}%
\label{sec:jump_simulate}

\begin{figure}[tp]
  \centering
  \includegraphics[width=0.45\textwidth]{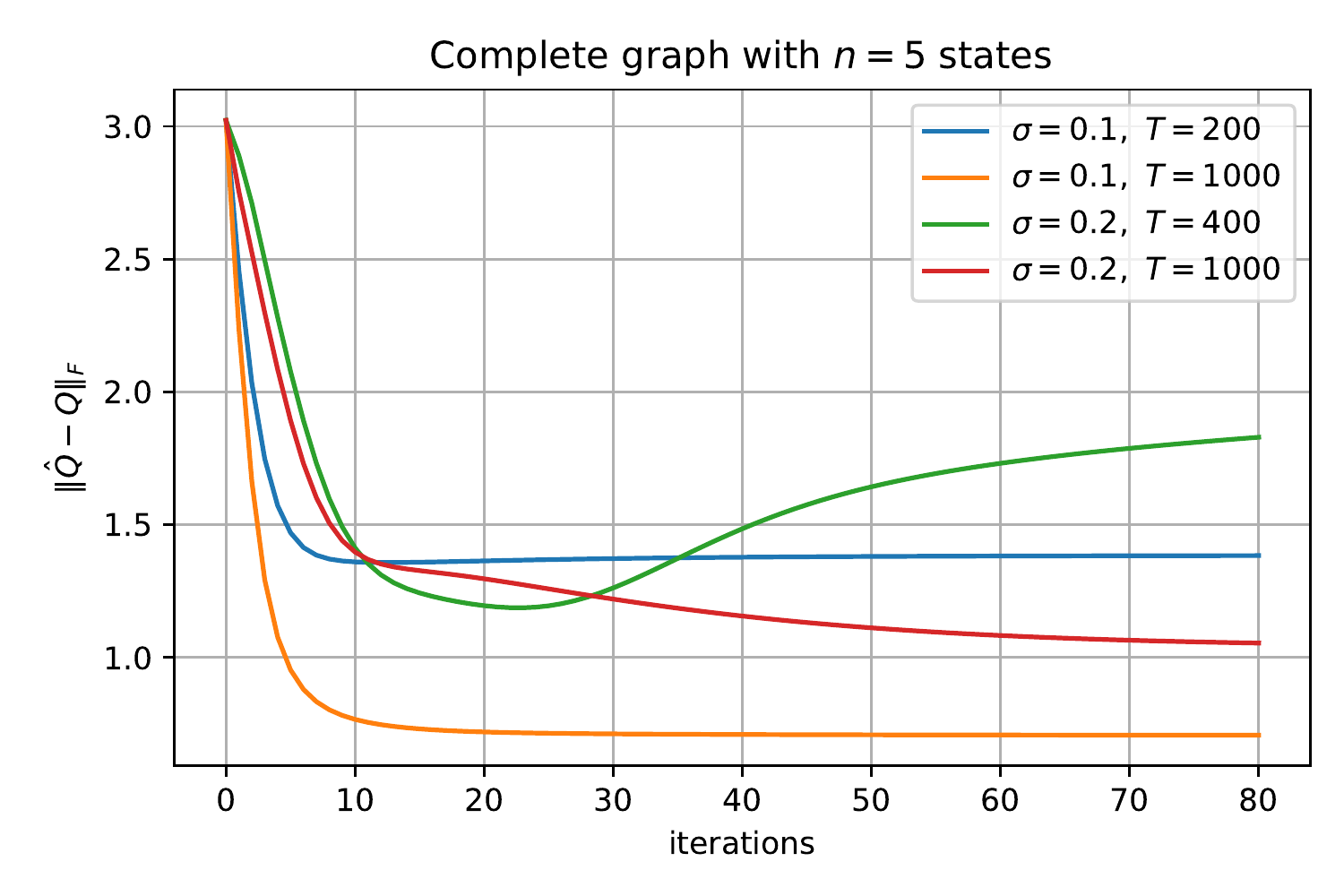}
  \includegraphics[width=0.45\textwidth]{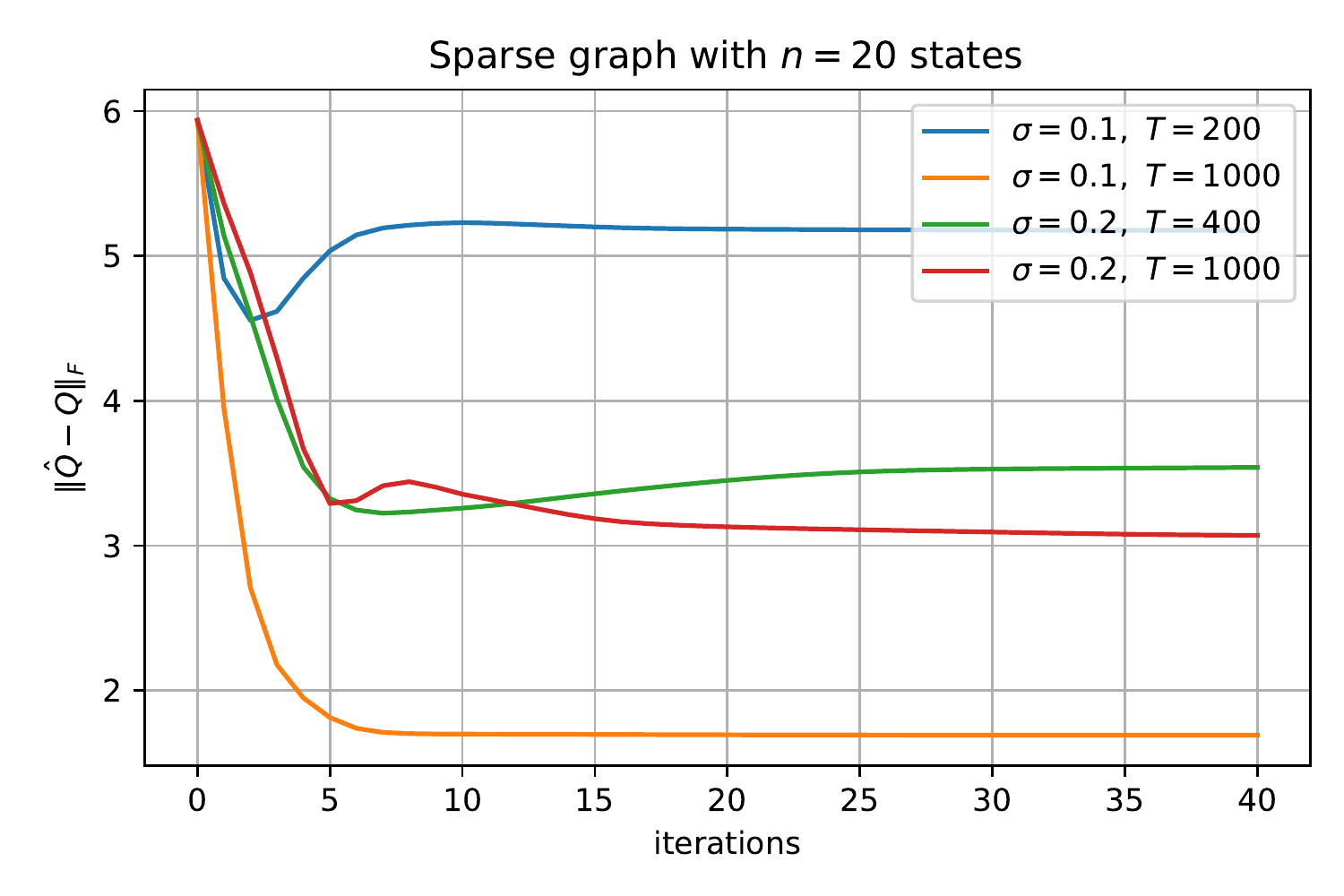}
  \includegraphics[width=0.45\textwidth]{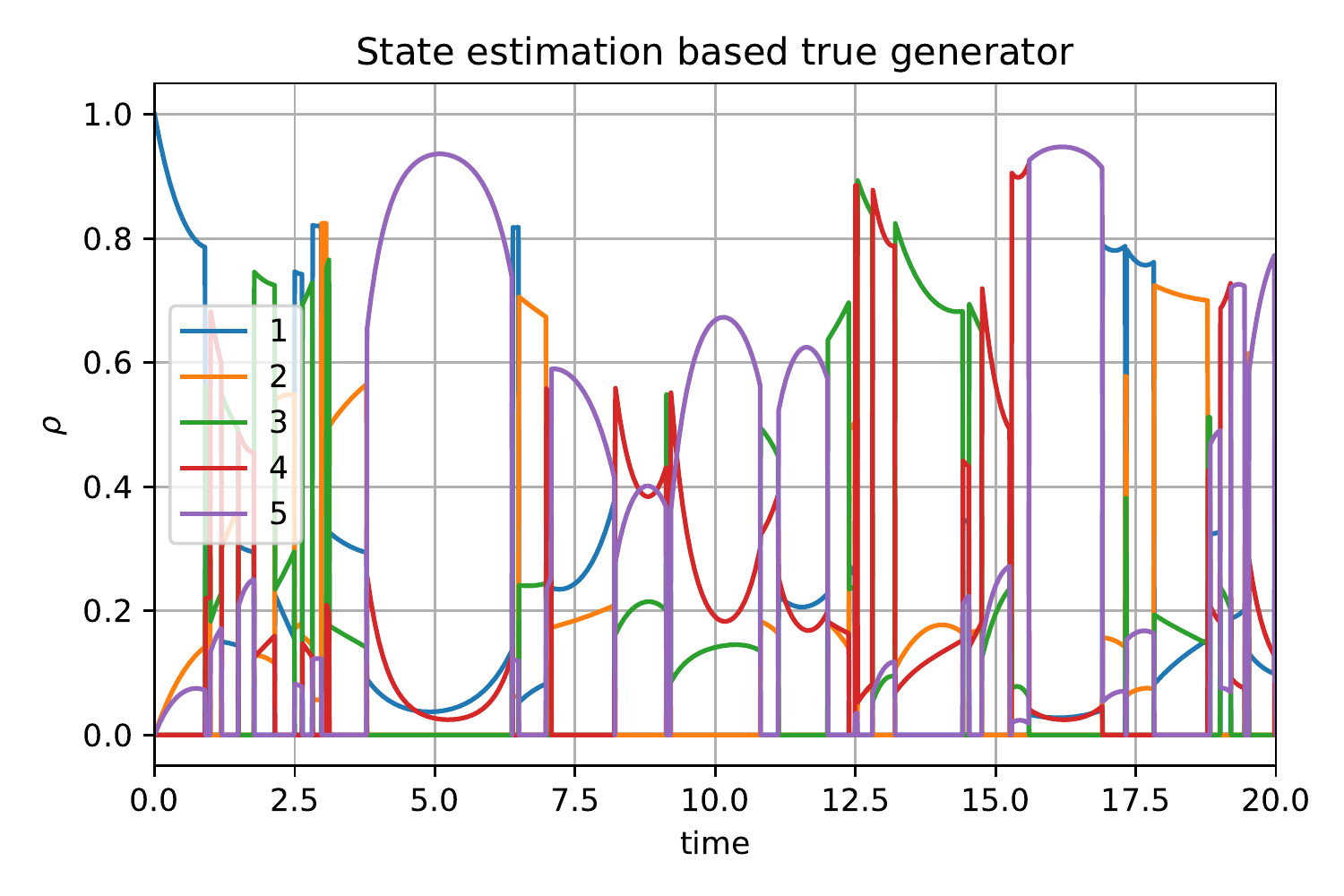}
  \includegraphics[width=0.45\textwidth]{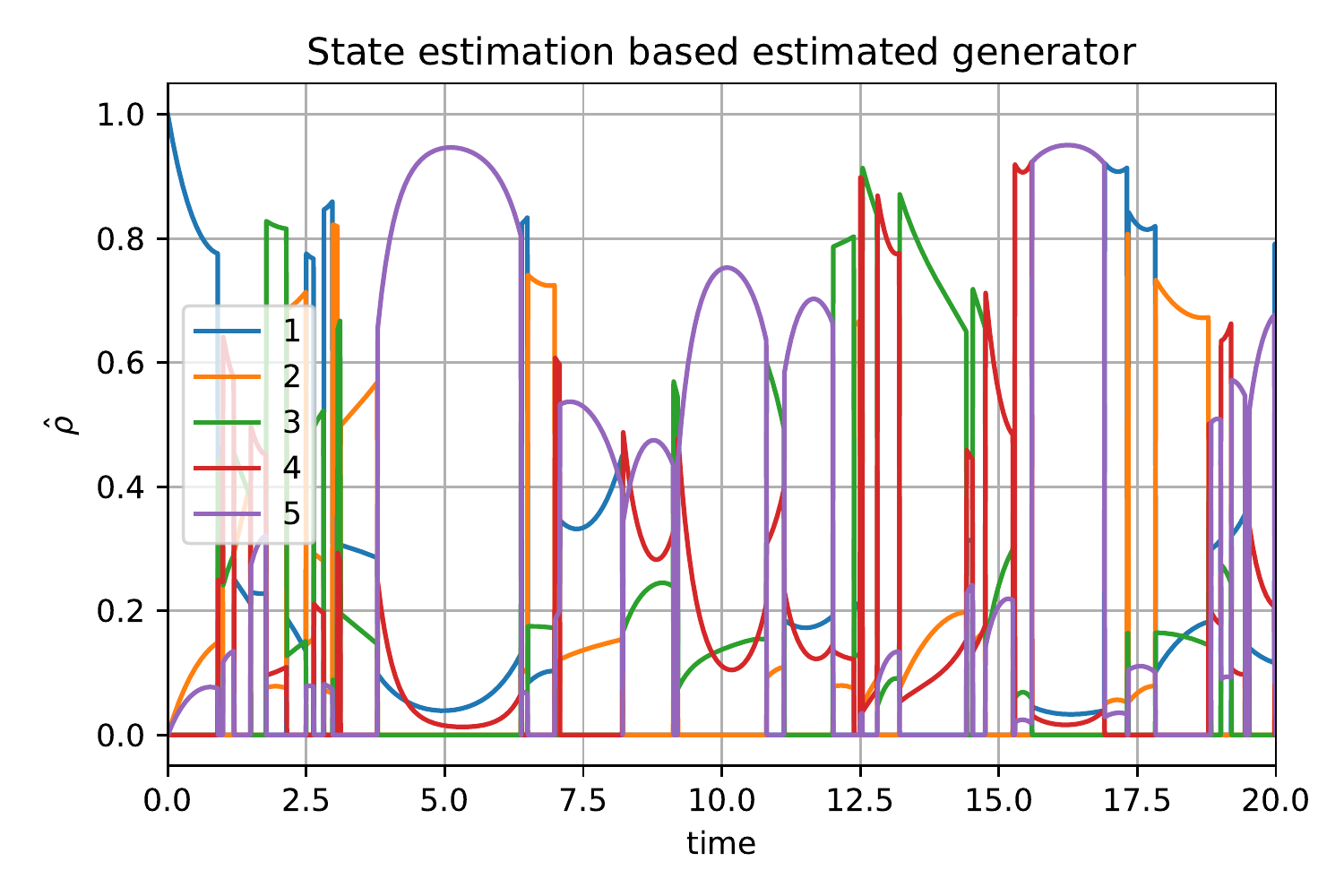}
  \caption[Simulation results for CT-HMM with hidden jump process]{%
    The simulation results for the CT-HMM with hidden jump process. Top: the
    distance $\|Q^k - Q\|_F$ under the Frobenius norm during the EM iterations
    for different noise levels $\sigma$ and total time $T$. Complex models may
    require larger $T$ to avoid overfitting, and the convergence is faster for
    smaller noise level $\sigma$. Bottom: the posteriors distribution
    $\rho_t(i)$ for each $i$ based on the true generator $Q$ and estimated
    $\hat{Q}$. The EM algorithm indeed converges. $\rho_t$ jumps at each
    $\tau_s$, but may still have significant changes between $(\tau_s,
    \tau_{s+1})$. Here $n = 5$, $\sigma = 0.2$, $T = 1000$ and we only show the
    results for $t \in [0, 20]$.
  }\label{fig:jump}
\end{figure}

We first implement our algorithm for the CT-HMM with hidden jump process in
Section~\ref{sec:hmm_jump}. Notice that the piecewise ODEs (\ref{eq:jump_fwd})
and (\ref{eq:jump_bwd}) for state estimation, if solved directly, may lead to
overflow or underflow of $\alpha_t$ and $\beta_t$ for large $T$, and we modify
the implementation as follows: at each $\tau_s$, adaptively choose $\kappa_s \in
\R_+$, then update
\[
  \alpha_{\tau_s} = \kappa_s \alpha_{\tau_s^-} (Q - D) R(y_{\tau_s}), \quad
  \beta_{\tau_s^-} = (Q - D) R(y_{\tau_s}) \beta_{\tau_s} / \kappa_s.
\]
The dynamics between $(\tau_s, \tau_{s+1})$ remains the same.

Here we test two simulation models:
\begin{enumerate}
  \item A complete graph with $n = 5$ states as \cite{liu2015efficient}. The
    generator $Q$ is randomly drawn as $-Q_{ii} \sim \cU[1, 5]$, $Q'_{ij} \sim
    \cU[0,1]$ and $Q_{ij} = \frac{Q'_{ij}}{\sum_{j' \ne i} Q'_{ij'}}(-Q_{ii})$
    for $i, j \in \{1, \dots, 5\}$, $i \ne j$, where $\cU$ is the uniform
    distribution. We initialize $Q_{ii}^0 = -3$ and $Q_{ij}^0 = 3/4$ for the EM
    algorithm.
  \item A sparse graph with $n = 20$ states. For each state $i = 1, \dots, 20$,
    randomly choose another 5 states $V_i \subset \{1, \dots, 20\} \setminus
    \{i\}$, $|V_i| = 5$, and let $Q_{ij} \sim \cU[0, 1]$ for $j \in V_i$,
    $Q_{ii} = -\sum_{j \in V_i} Q_{ij}$ and $Q_{ij} = 0$ for $j \notin V_i \cup
    \{i\}$. The initialization $Q_{ii}^0 = -19/8$ and $Q_{ij}^0 = 1/8$.
\end{enumerate}

For both of the models, the observations $Y_t \in \{1, \dots, n\}$ are given by
$r_i(i) = 1 - 2 \sigma$ and $r_i(i-1) = r_i(i+1) = \sigma$ (let $r_1(0) =
r_1(n)$ and $r_n(n+1) = r_n(1)$). Assume that the state starts from $X_0 = 1$,
i.e., $\pi_0 = [1, 0, \dots, 0]$. We test different noise level $\sigma$ and
total time $T$.

Figure~\ref{fig:jump} shows the simulation results. The top two figures show
the convergence of $Q^k$ for both models. Complex models may require larger $T$
to avoid overfitting, and the convergence is faster for smaller noise level
$\sigma$. The bottom two figures compare the state estimation $\rho_t(i)$ for
each $i$ based on the true generator $Q$ and estimated $\hat Q$. We can see that
the EM algorithm indeed converges. In addition, $\rho_t$ jumps at each $\tau_s$,
but may still have significant changes between $(\tau_s, \tau_{s+1})$.

\subsection{Bearings-only tracking}

\begin{figure}[tp]
  \centering
  \includegraphics[width=0.45\textwidth]{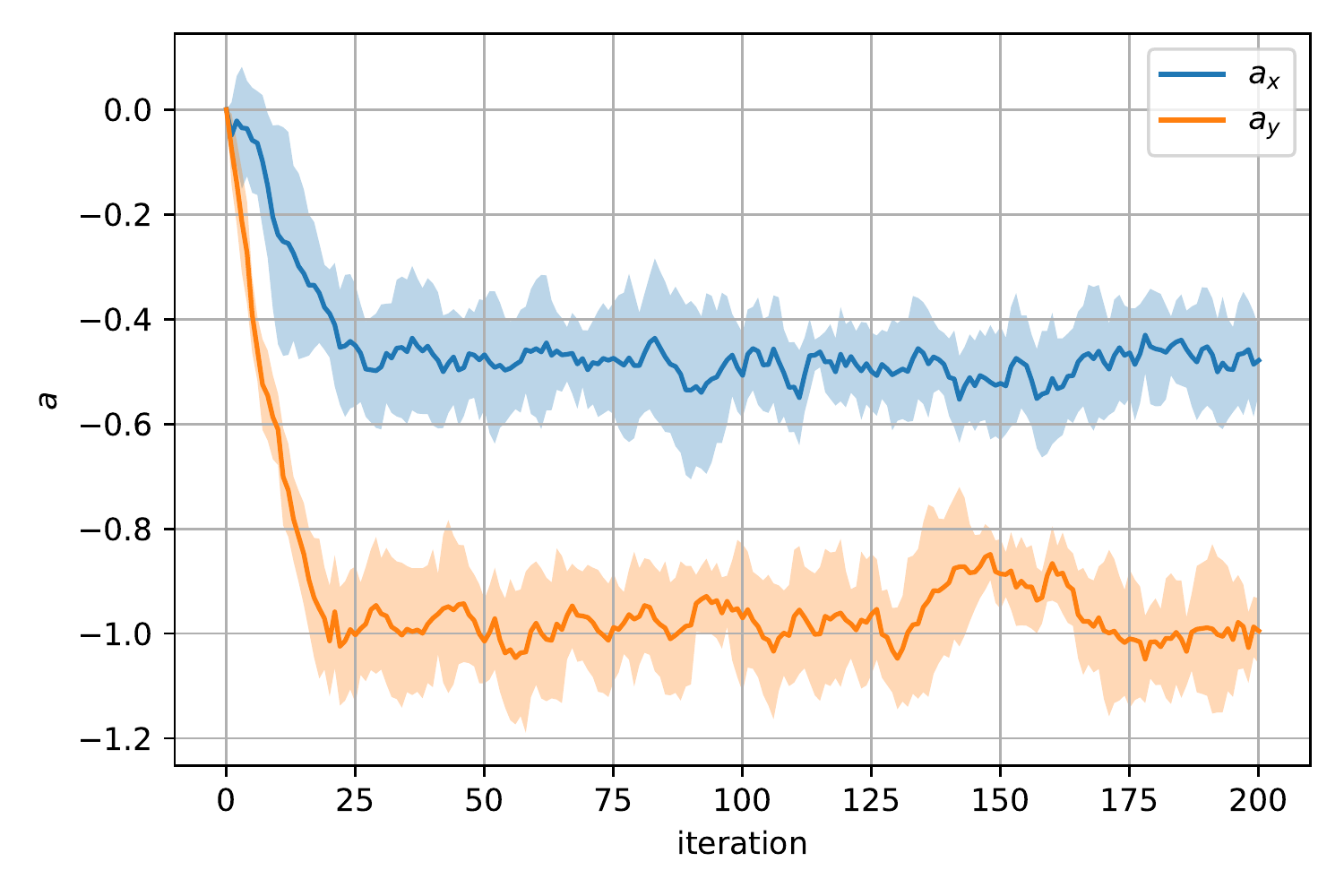} \\
  \includegraphics[width=0.25\textwidth]{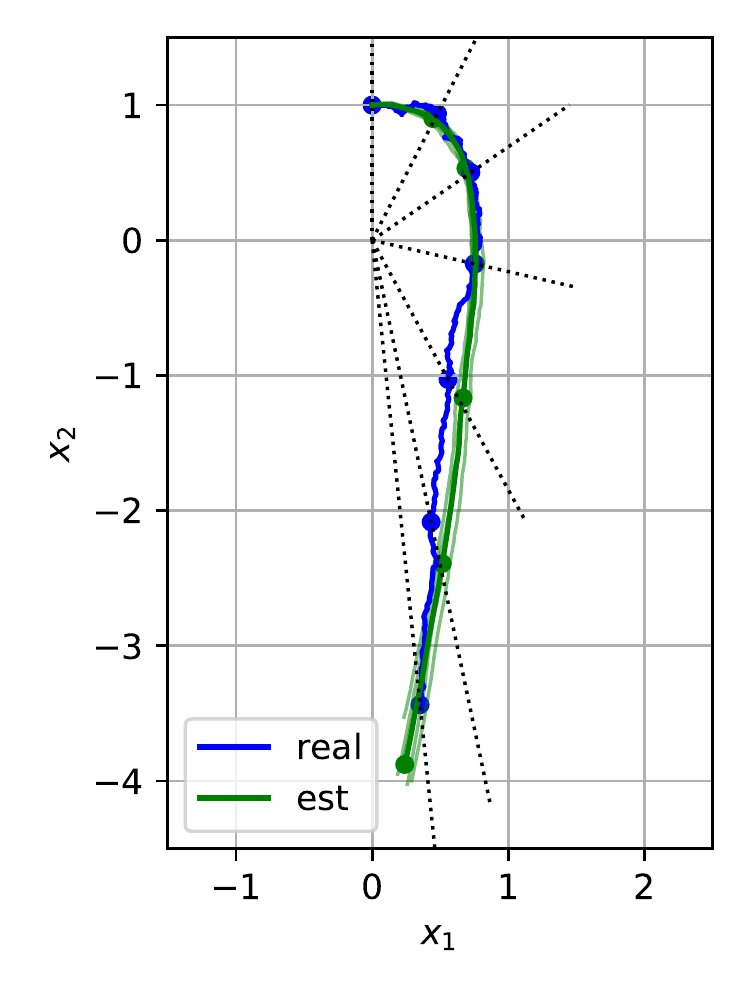}
  \includegraphics[width=0.25\textwidth]{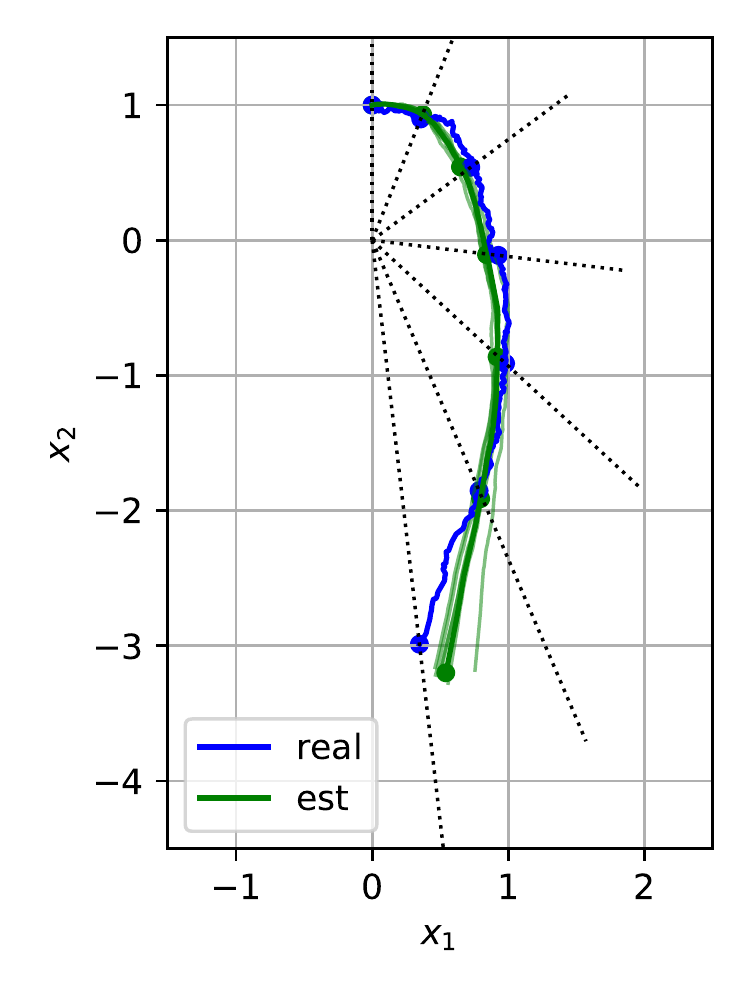}
  \includegraphics[width=0.25\textwidth]{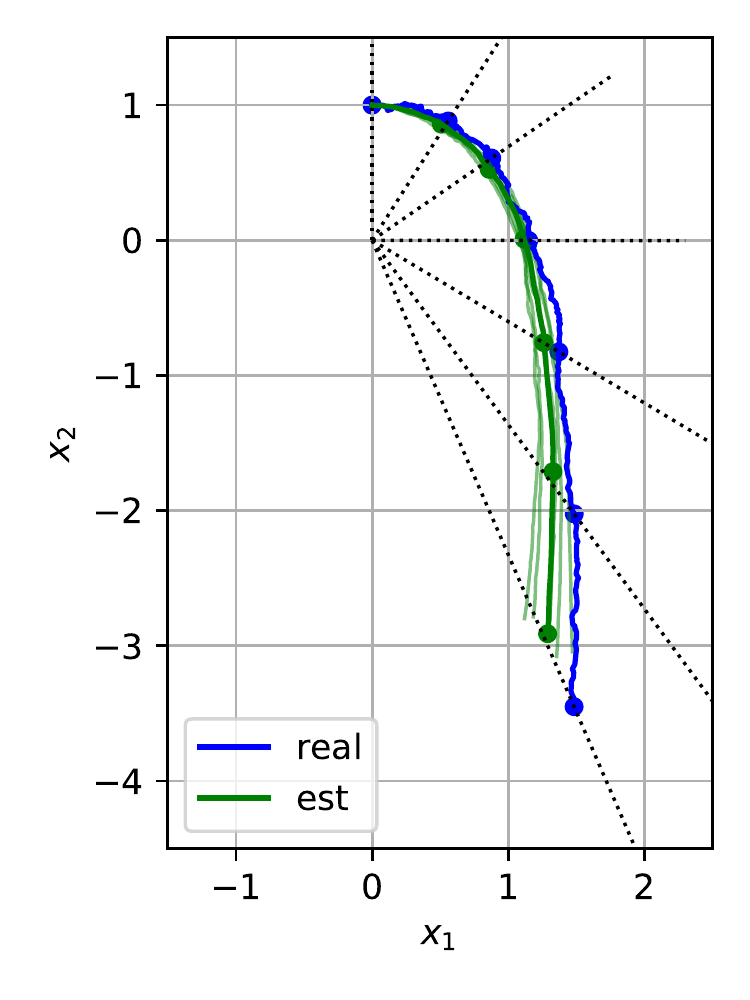}
  \caption[The simulation results for bearings-only tracking]{%
    The simulation results for bearings-only tracking (\ref{eq:bearings}). Top:
    the estimated acceleration during the EM iterations. Here we repeat the
    simulations 5 times, and plot the mean and standard deviation. Bottom: the
    true trajectories and the estimation based on estimated $\left(a_x^k,
    a_y^k\right)$. The 3 figures show 3 of the 5 simulations. The light green
    lines are the trajectories for $k = 175, 180, 185, \dots, 200$, and the dark
    green line is their average. We also mark the positions at time points $t =
    0, 0.5, 1, \dots, 3$ to compare the bearings.
  }\label{fig:bearings}
\end{figure}

The bearings-only tracking is a widely used test problem for filtering methods.
The goal is to track an object moving in the $x$-$y$ plane, and a fixed observer
at the origin takes noisy measurements of the target bearings. The problem can
be tracked back to \cite{aidala1983utilization}.  \cite{gordon1993novel,
pitt1999filtering} analyze its linear discrete-time dynamics, and
\cite{kushner2000nonlinear, kushner2008numerical} consider the continuous-time
hidden process with discrete-time observation. Here we assume that both the
hidden states and observation occur in continuous-time as
\begin{gather}
  \ud \left[\begin{array}{c}
  x \\ \dot x \\ y \\ \dot y \end{array}\right]_t
  = \left[\begin{array}{c}
  \dot x \\ a_x \\ \dot y \\ a_y \end{array}\right]_t \ud t
  + \sigma \ud W_t, \nonumber \\
  \ud \varphi_t = \arctan(y_t / x_t) \ud t + \eta \ud B_t.
  \label{eq:bearings}
\end{gather}
Here the position and volatility $\left(x, \dot x, y, \dot y\right)_t$ is the
hidden state, the acceleration $(a_x, a_y)$ is the unknown parameter, and the
angle $\varphi_t$ is the observation. (Unlike our general notations, in this
problem we sometimes use lower case letters as random variables).

Assume that the initial position $(x_0, y_0) = (0, 1)$, the initial volatility
$\left(\dot x_0, \dot y_0\right) = (1, 0)$, the noise $\sigma = 0.1$, $\eta =
0.02$, and the total time $T = 3$. The true value of the acceleration is $(a_x,
a_y) = (-0.5, -1)$, and we initialize $a_x^0 = a_y^0 = 0$. The time
discretization $\Delta t = 0.01$ and the number of samples $N = 128$. We
simulate the true trajectory and observations 5 times, and run the EM algorithm
for each of them. Figure~\ref{fig:bearings} (top) shows the mean and standard
deviation of estimated $\left(a_x^k, a_y^k\right)$ during the 5 runs. The
estimations converge to near the true values, and we can expect that more
repeated simulations and longer $T$ may improve the accuracy.

We are also interested in recovering the true trajectory. Let $\left(\hat a_x,
\hat a_y\right)$ be the estimation from the EM algorithm. Taking the Monte Carlo
state estimation under $(a_x, a_y)=\left(\hat a_x, \hat a_y\right)$, we have
samples $\left\{\xi_t^i\right\}$ for the filter and weights
$\left\{w_t^i\right\}$ for the smoother. Then the true states $x_t$ can be
estimated by the simple or weighted average as
\begin{equation}
  \hat x^\textrm{filter}_t = \frac{1}{N} \sum_{i=1}^N \xi_t^i, \quad
  \hat x^\textrm{smooth}_t = \sum_{i=1}^N w_t^i \xi_t^i, \quad
  t = 0, \Delta t, \dots, T.
  \label{eq:dif_state_mc}
\end{equation}
For the bearings-only tracking problem, since we only observe the angle but not
the radius, a single estimation path may not exactly coincide with the true
trajectory. Here we take the last few iterations $k = 175, 180, 185, \dots,
200$. For each $k$, repeat the Monte Carlo smoother estimation
(\ref{eq:dif_state_mc}) 5 times, and calculate the average of the 5 estimated
trajectories. In Figure~\ref{fig:bearings} (bottom), the light green lines are
the average trajectories for each $k$, and the dark green line is the average of
the light green lines. We also mark the positions for time points $t=0, 0.5, 1,
\dots, 3$ to compare the bearings. We can see that after the averaging, the dark
green line is close to the true trajectory (blue), and shares similar bearing at
the same time points.

\subsection{Cubic sensor problems for matrix estimation}

\begin{figure}[tp]
  \centering 
  \includegraphics[width=0.45\textwidth]{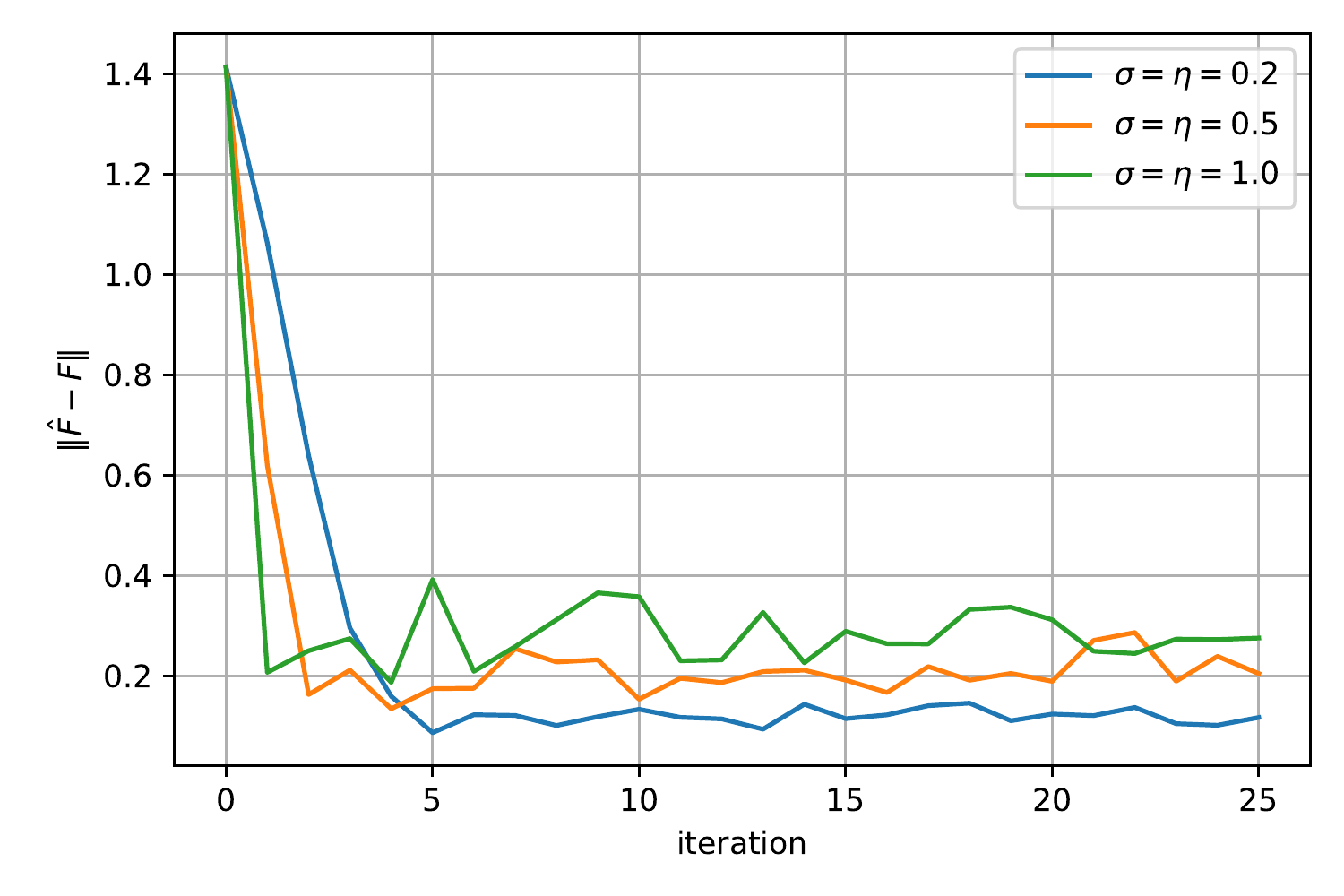}
  \includegraphics[width=0.45\textwidth]{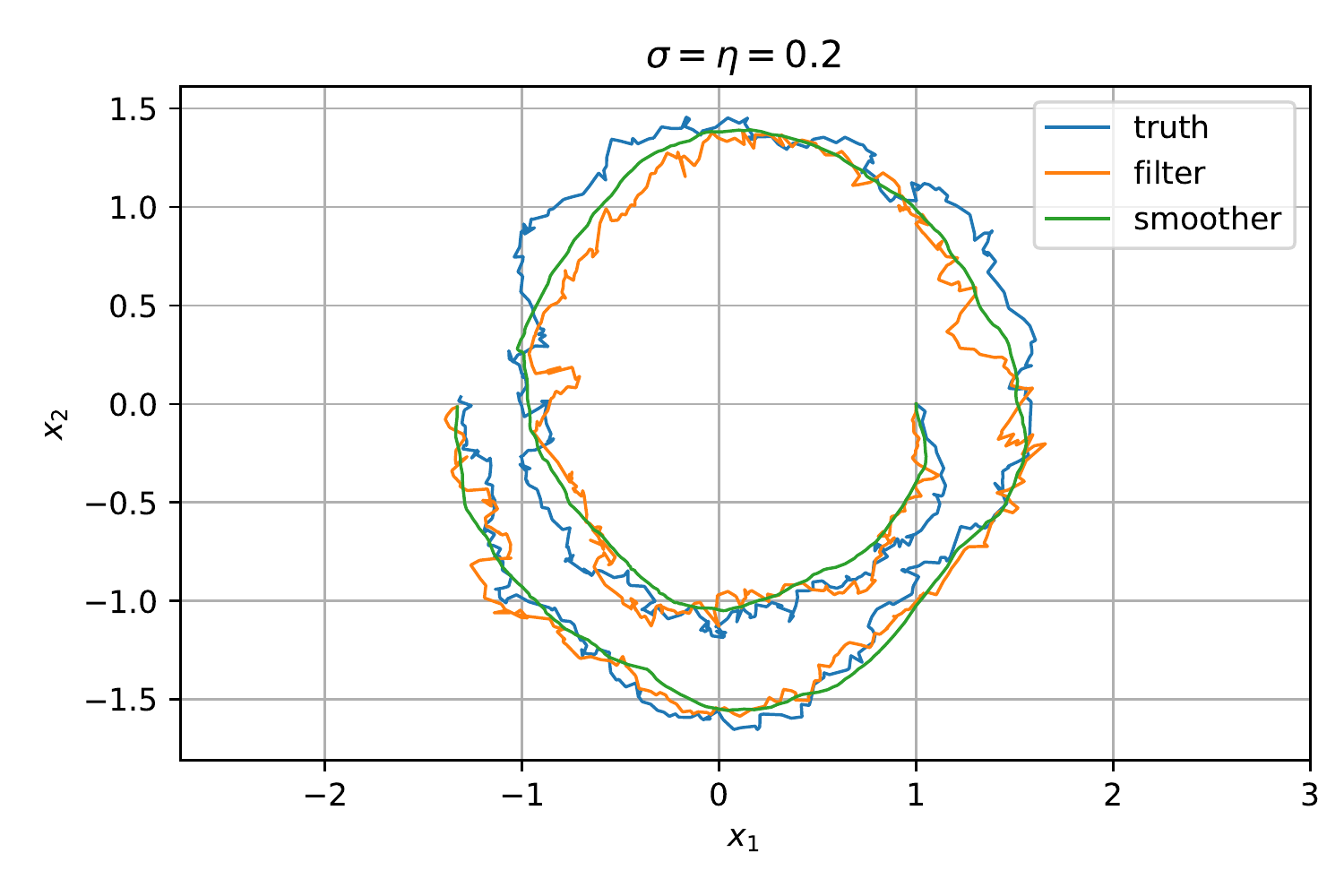} \\
  \includegraphics[width=0.45\textwidth]{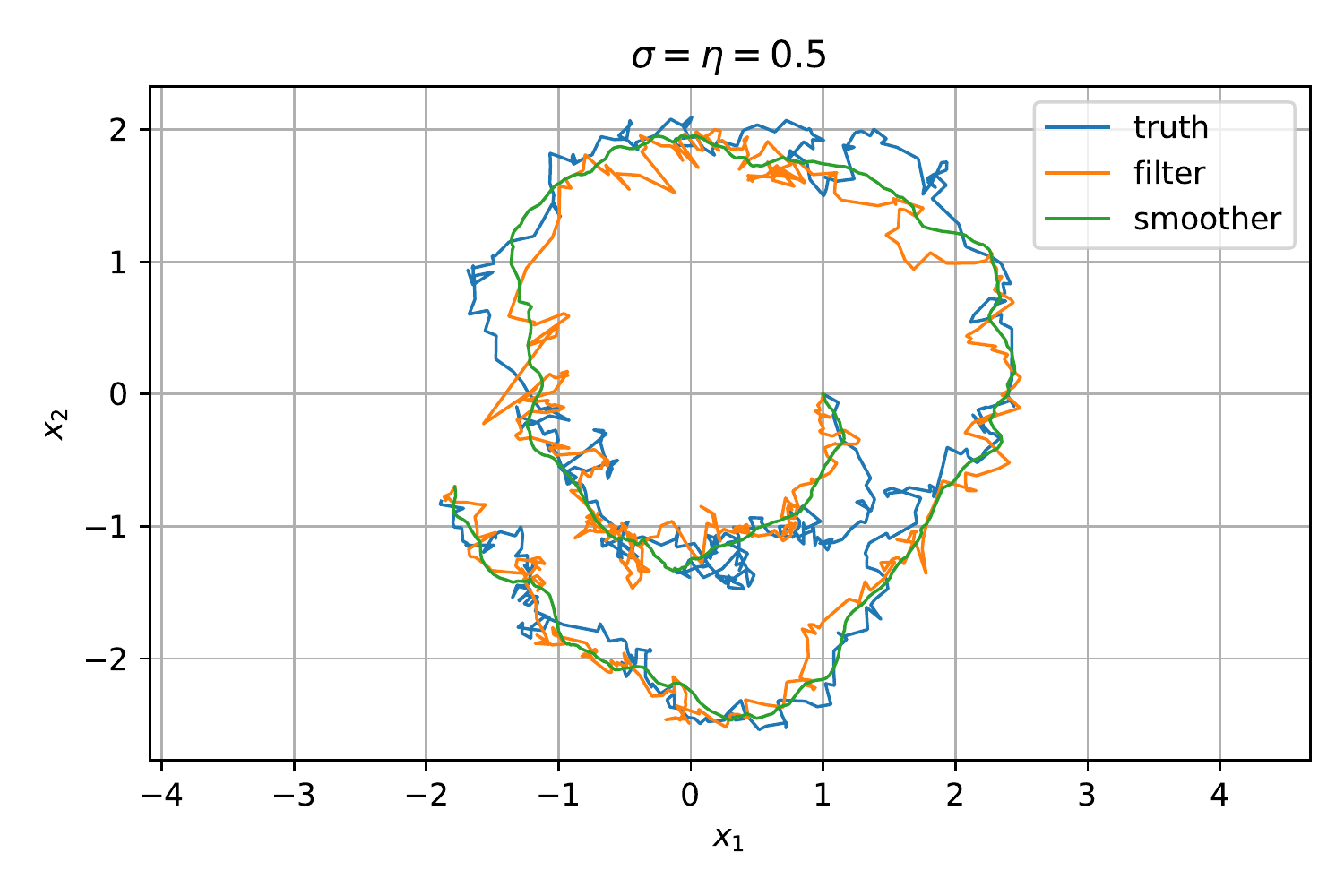}
  \includegraphics[width=0.45\textwidth]{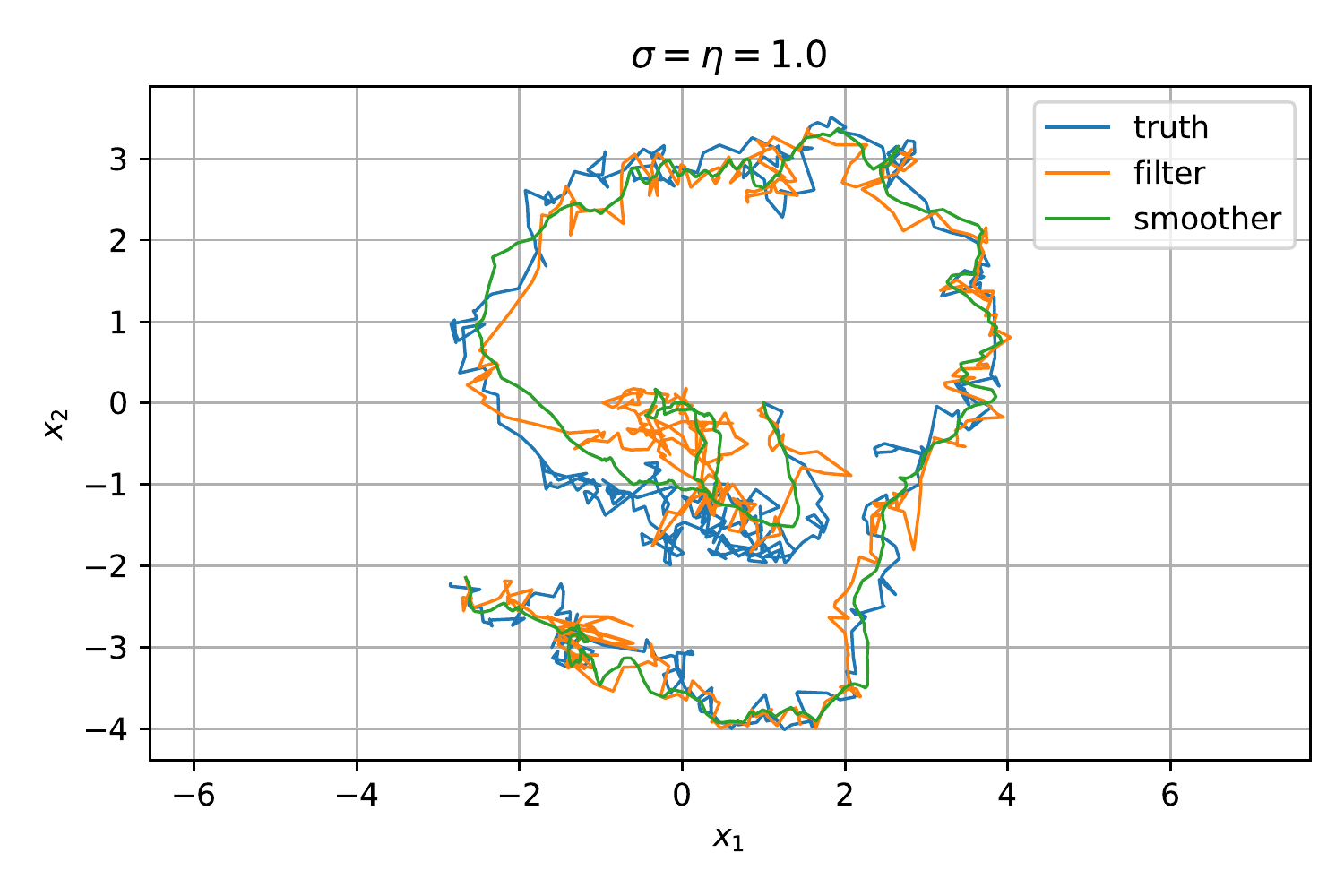}
  \caption[The cubic sensor problem for the matrix parameter estimation]{%
    The cubic sensor problem for the matrix parameter estimation
    (\ref{eq:cubic}). The upper left figure shows the distance $\left\|F^k -
    F\right\|_F$ under the Frobenius norm during the iterations. The EM
    algorithm converges fast for all the noise levels, while the larger noise
    level has faster convergence rate but lower accuracy. Then we calculate the
    solutions of the filter and smoother based on $\hat F$. The other three
    figures compare the solutions and the true trajectories under different
    noise levels. The solution of the smoothing is indeed smoother and closer to
    the true trajectory.
  }\label{fig:cubic_mat}
\end{figure}

In the cubic sensor problems \cite{steinberg1988optimal,
katayama2013equivalent}, the observation has the same dimension as the hidden
state, and is given by $h([x_1, \dots, x_d]) = \left[x_1^3, \dots,
x_d^3\right]$. Here we assume that $f$ is linear, then the dynamics is given by
\begin{align}
  \ud X_t & = F X_t \ud t + \sigma \ud W_t, \nonumber \\
  \ud Y_t & = X_t^3 \ud t + \eta \ud B_t.
  \label{eq:cubic}
\end{align}

First consider the case that the whole matrix $F$ is the unknown parameter to be
estimated, i.e., $\theta = F$. Then the parameter update (\ref{eq:dif_q_mc}) in
our Monte Carlo method becomes
\begin{equation}
  F^{k+1} = \argmin_F\hat\cQ(F, F^k) = F^k + \sigma^2
  \Big[\sum_{t, i} \nabla\hat\beta\left(\xi_t^i\right)\xi_t^{i\intercal}\Big]
  \Big[\sum_{t, i} w_t^i \xi_t^i \xi_t^{i\intercal}\Big]^{-1}.
\end{equation}

Now let the dimension $d = 2$, the true matrix
$F = \left[\begin{array}{rr} 0 & 1 \\ -1 & 0 \end{array}\right]$, 
the initial condition $X_0 = [1, 0]$ at $t = 0$, and $T = 10$. We test different
noise levels $\sigma = \eta = 0.2, 0.5, 1.0$. For each level, take the
discretization $\Delta t = 0.02$, the number of samples $N = 128$, and the
initialization $F^0 = 0$. Figure~\ref{fig:cubic_mat} shows that of $F^k$
converges fast for all the noise levels, while larger noise level has faster
convergence rate but lower accuracy. In the end we get estimation
\begin{align*}
  \hat F =
  \left[\begin{array}{rr} 0.00 & 0.88 \\-0.98 &-0.01 \end{array}\right], \quad
  & \sigma = \eta = 0.2, \\
  \hat F =
  \left[\begin{array}{rr}-0.10 & 0.83 \\ 1.02 &-0.01 \end{array}\right], \quad
  & \sigma = \eta = 0.5, \\
  \hat F =
  \left[\begin{array}{rr}-0.19 & \phantom{-} 0.87 \\
  -1.12 & 0.10 \end{array}\right], \quad
  & \sigma = \eta = 1.
\end{align*}
We further calculate the solutions of the filter and smoother
(\ref{eq:dif_state_mc}) based on $\hat F$. Instead of taking average of
trajectories as the previous problem, here we only generate a single estimated
trajectory for each noise level. Comparing with the true trajectory of $X_t$.
We can see that the solution of the smoothing is indeed smoother and closer to
the true trajectory.

\subsection{High dimensional linear dynamics}

\begin{figure}[p]
  \centering
  \includegraphics[width=0.45\textwidth]{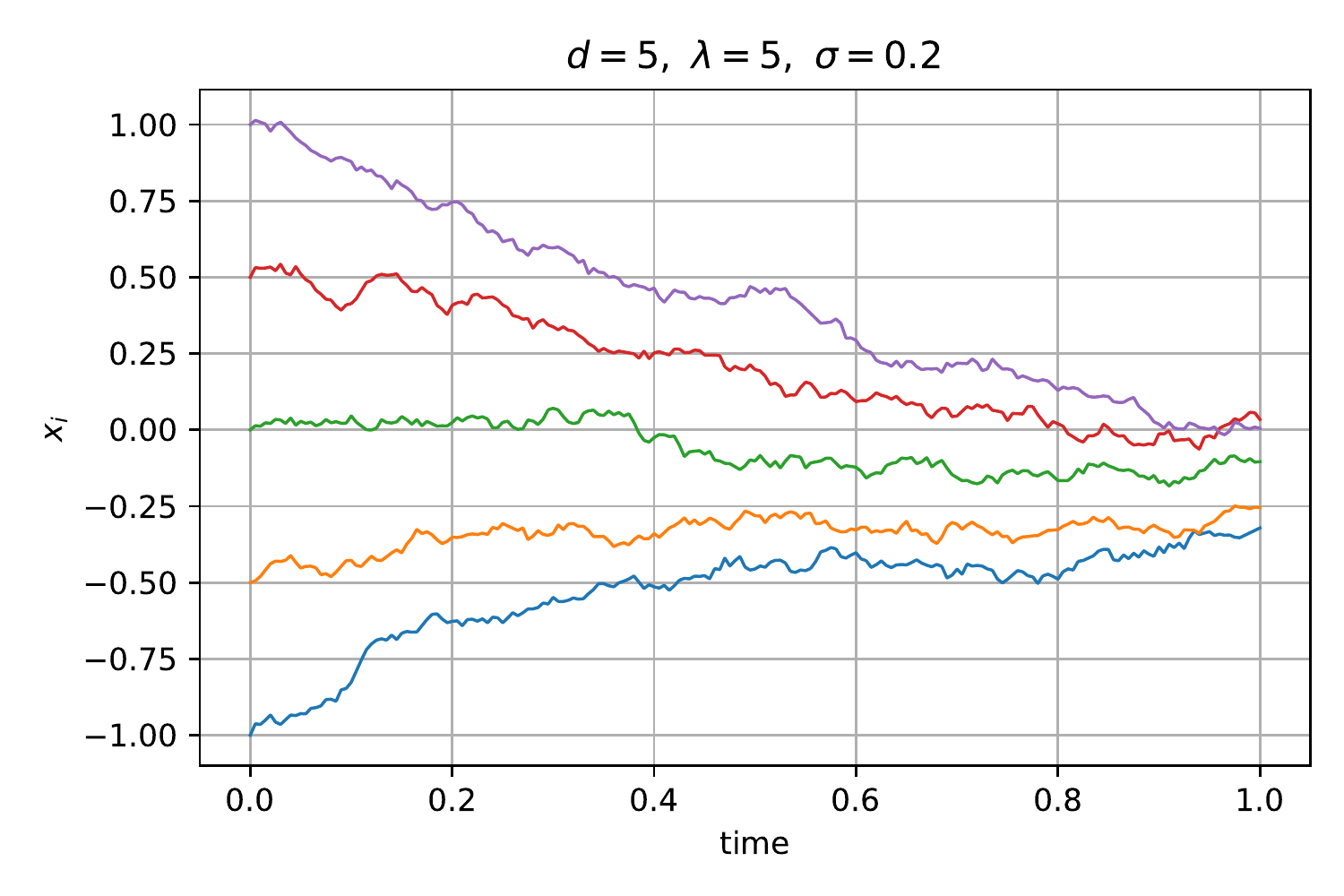}
  \includegraphics[width=0.45\textwidth]{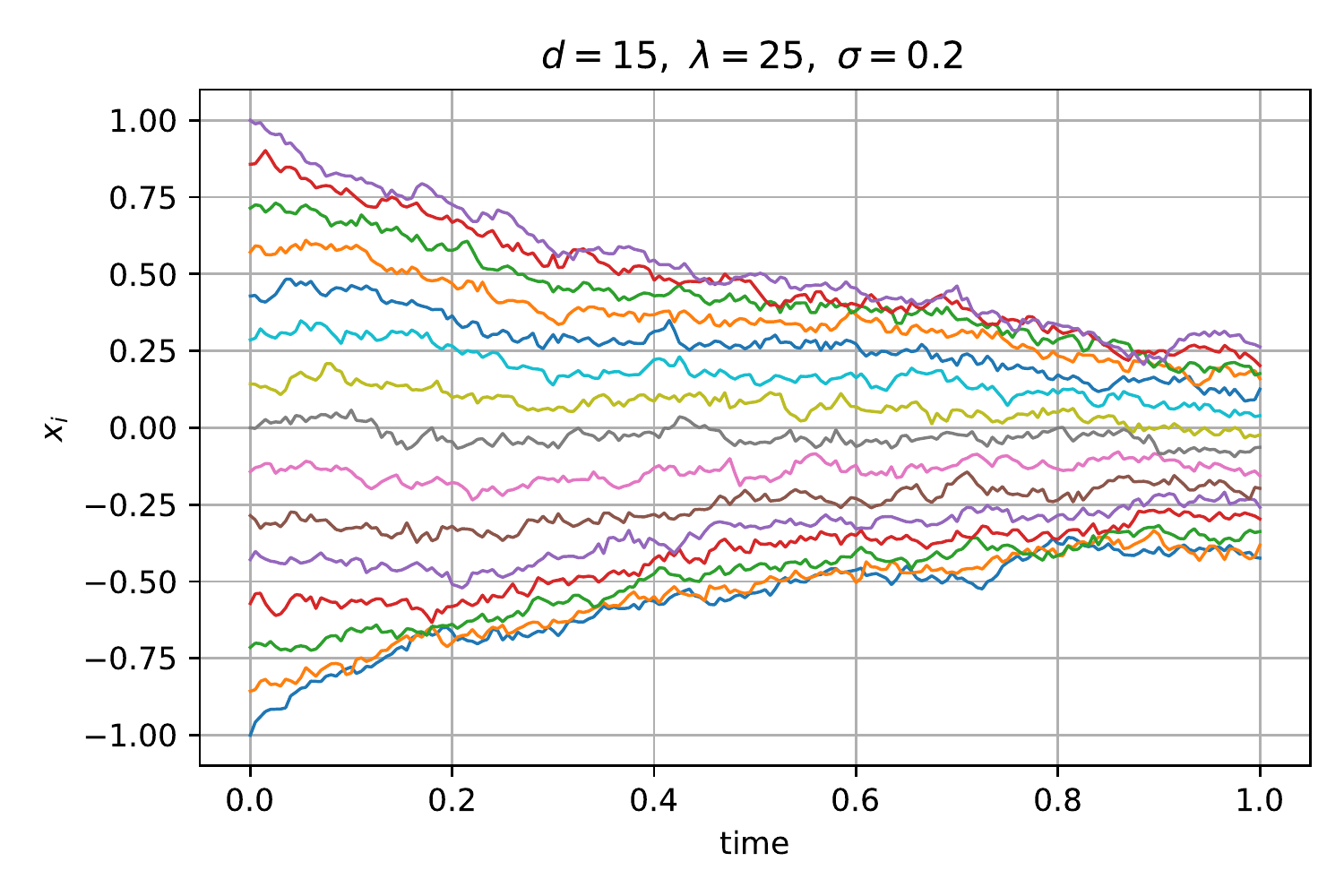} \\
  \includegraphics[width=0.45\textwidth]{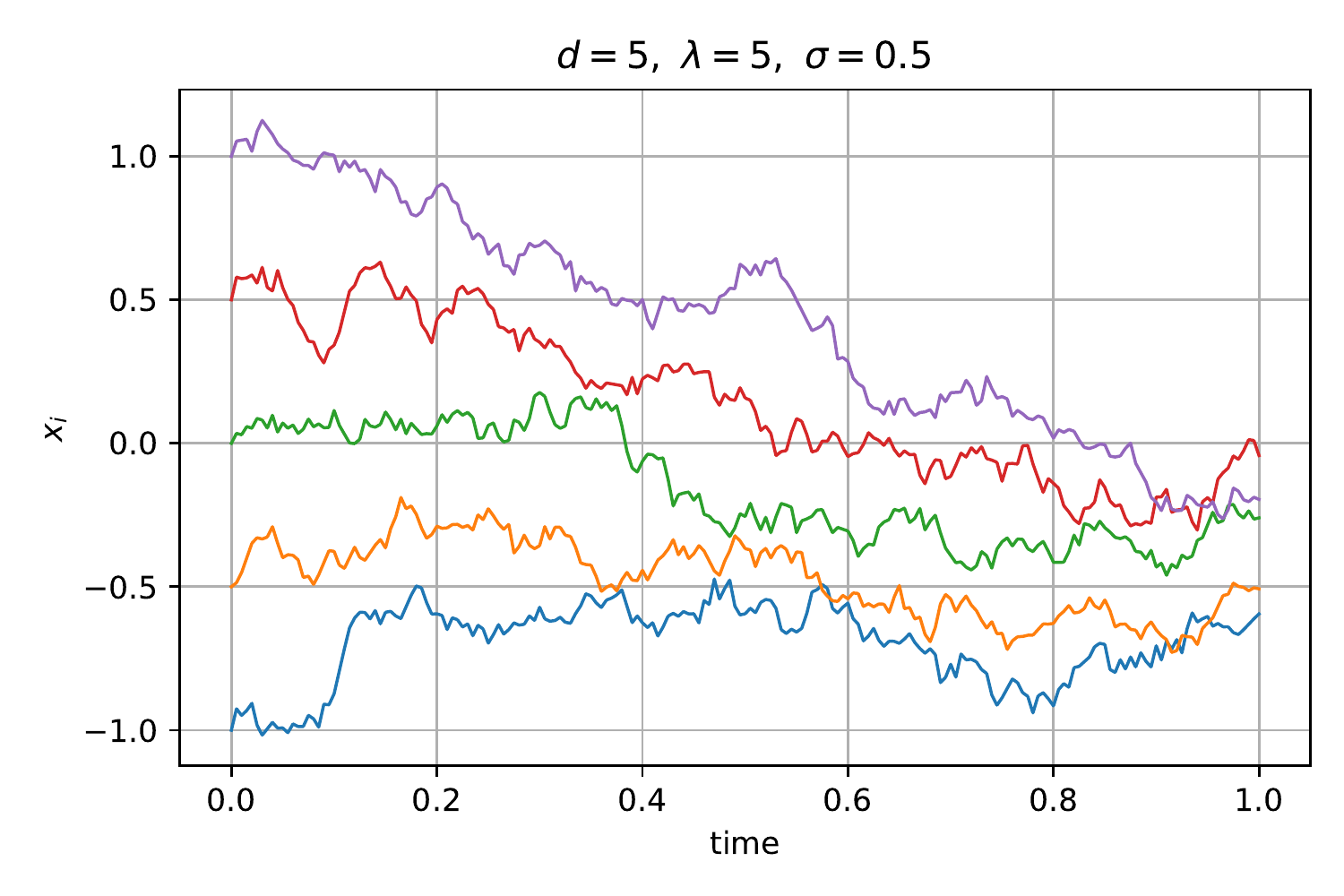}
  \includegraphics[width=0.45\textwidth]{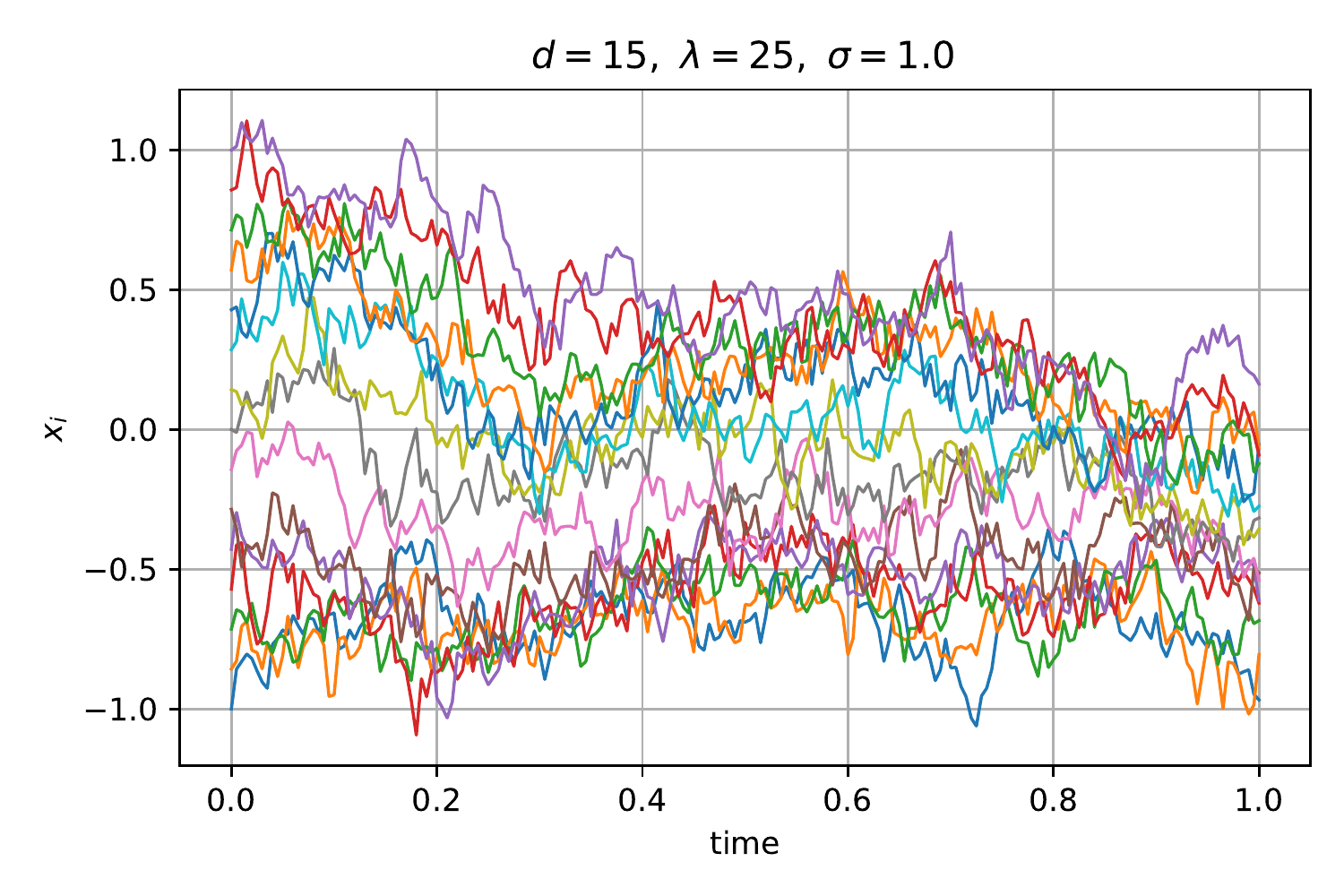} \\
  \includegraphics[width=0.45\textwidth]{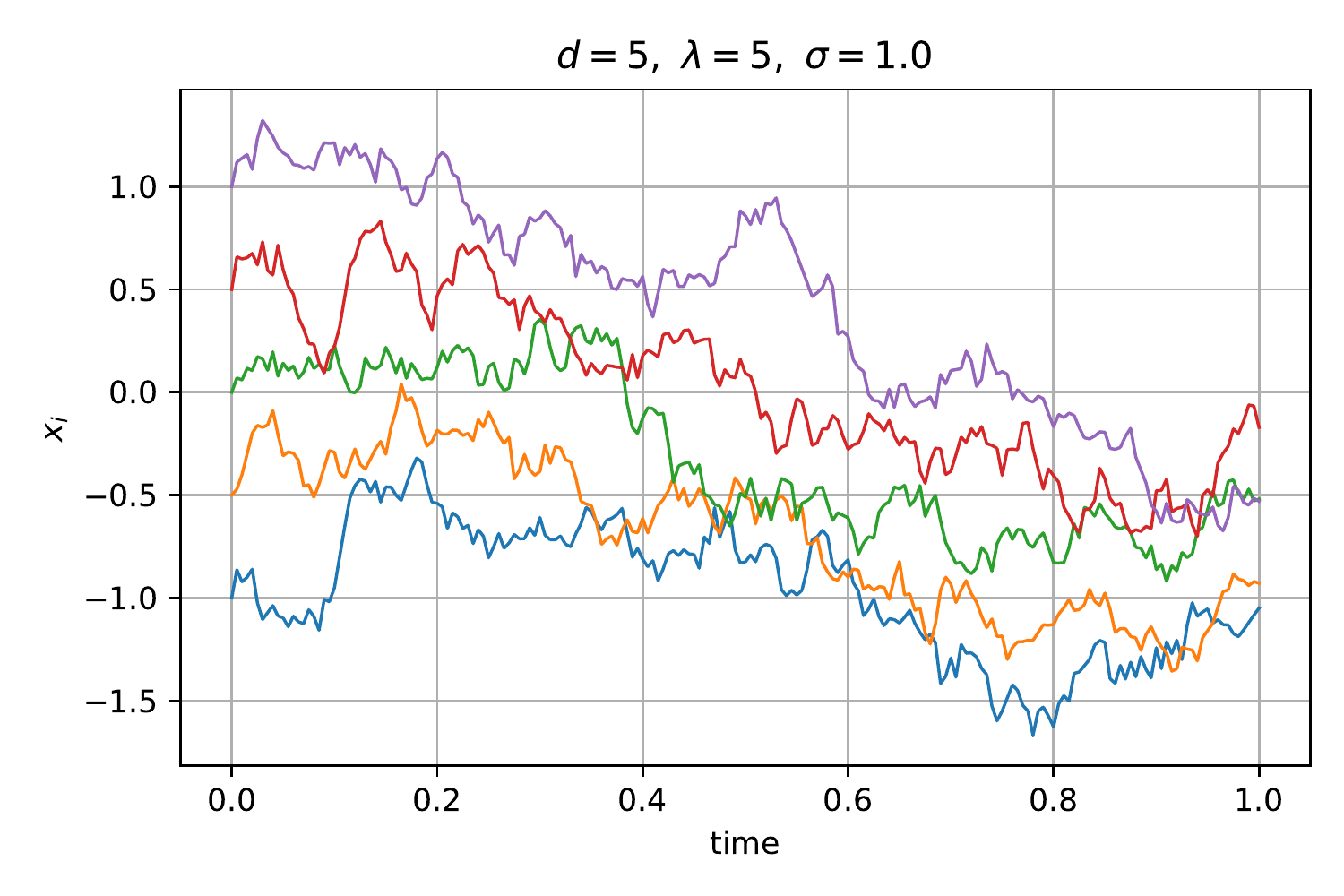}
  \includegraphics[width=0.45\textwidth]{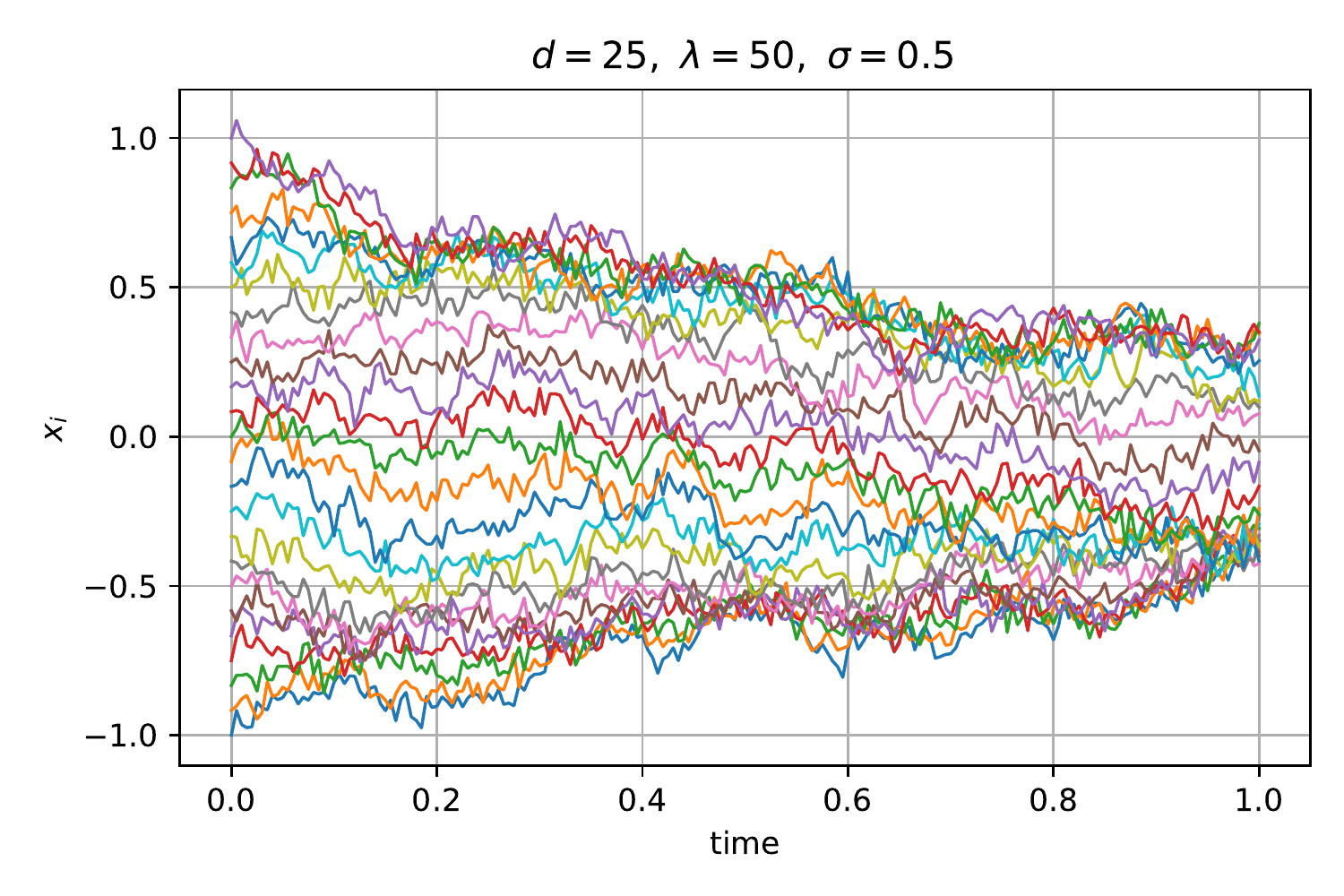}
  \caption[The hidden process for the scalar parameter estimation]{%
    The hidden process $\{X_{t}\}$ for the scalar parameter estimation problem
    (\ref{eq:cubic_scalar}). Each line shows the dynamics of one component of
    $X_{t}$. Here we take different dimension $d$, parameter $\lambda$ and noise
    $\sigma$.
  }\label{fig:cubic_scalar_x}
\end{figure}

\begin{figure}[p]
  \centering
  \includegraphics[width=0.45\textwidth]{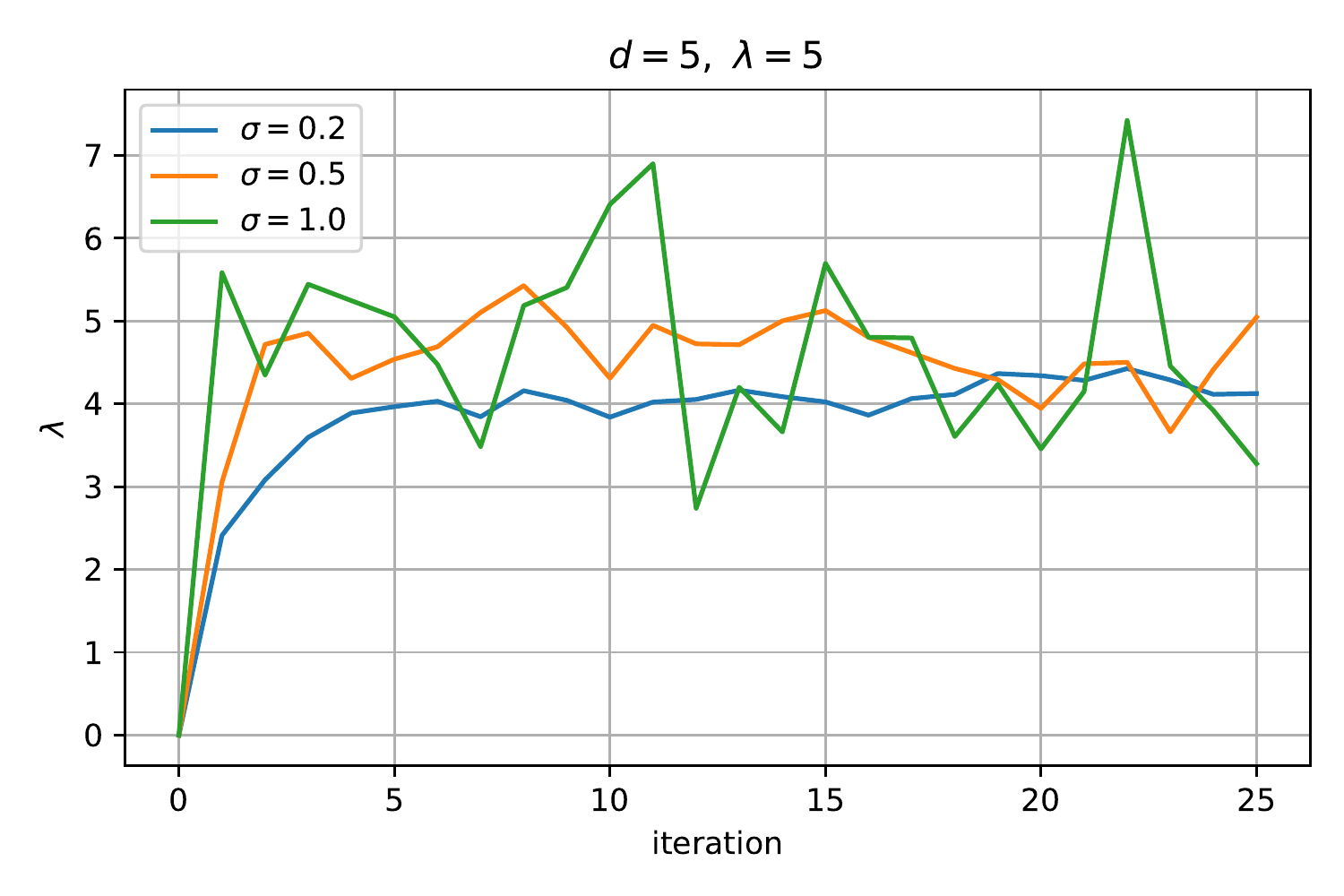}
  \includegraphics[width=0.45\textwidth]{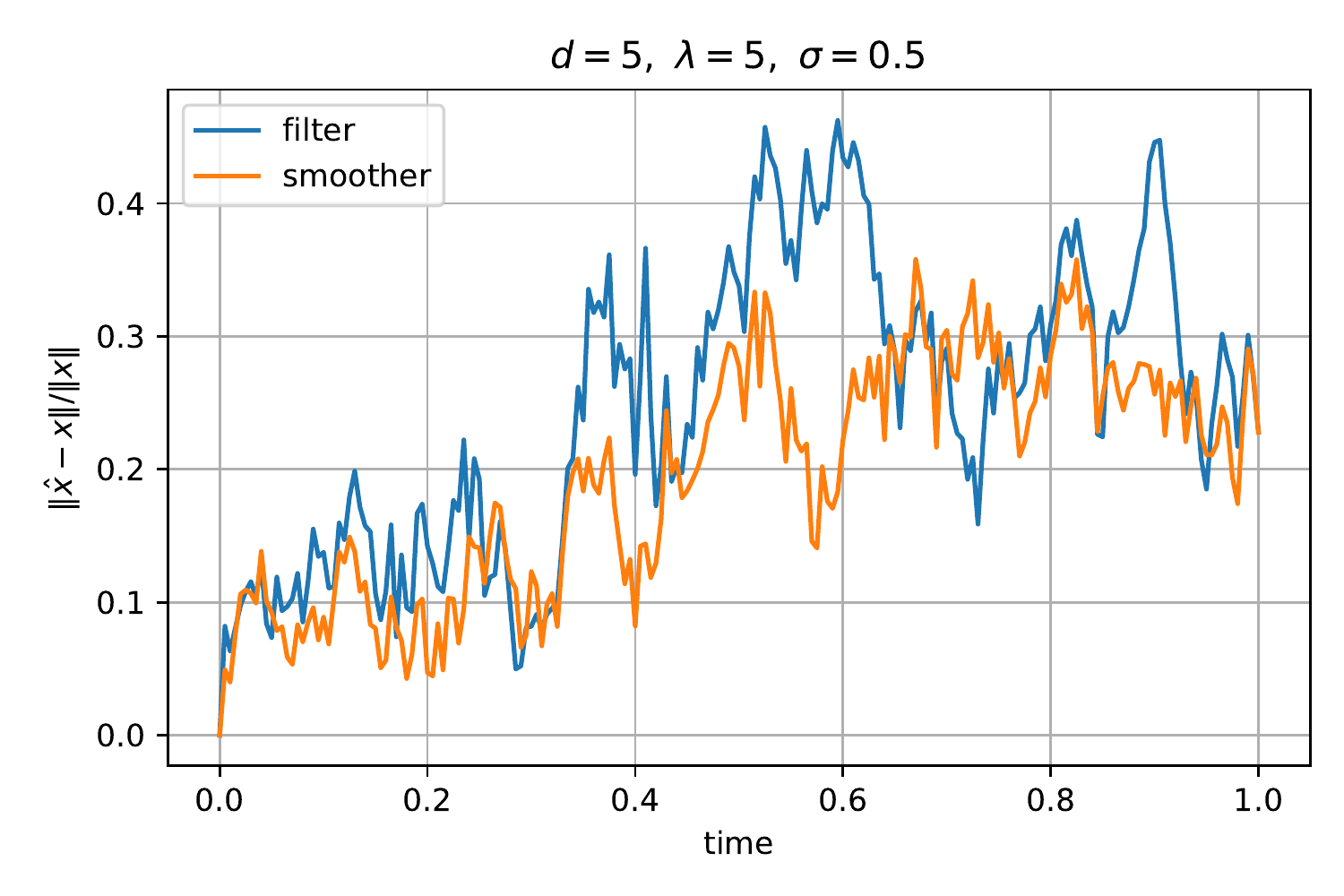} \\
  \includegraphics[width=0.45\textwidth]{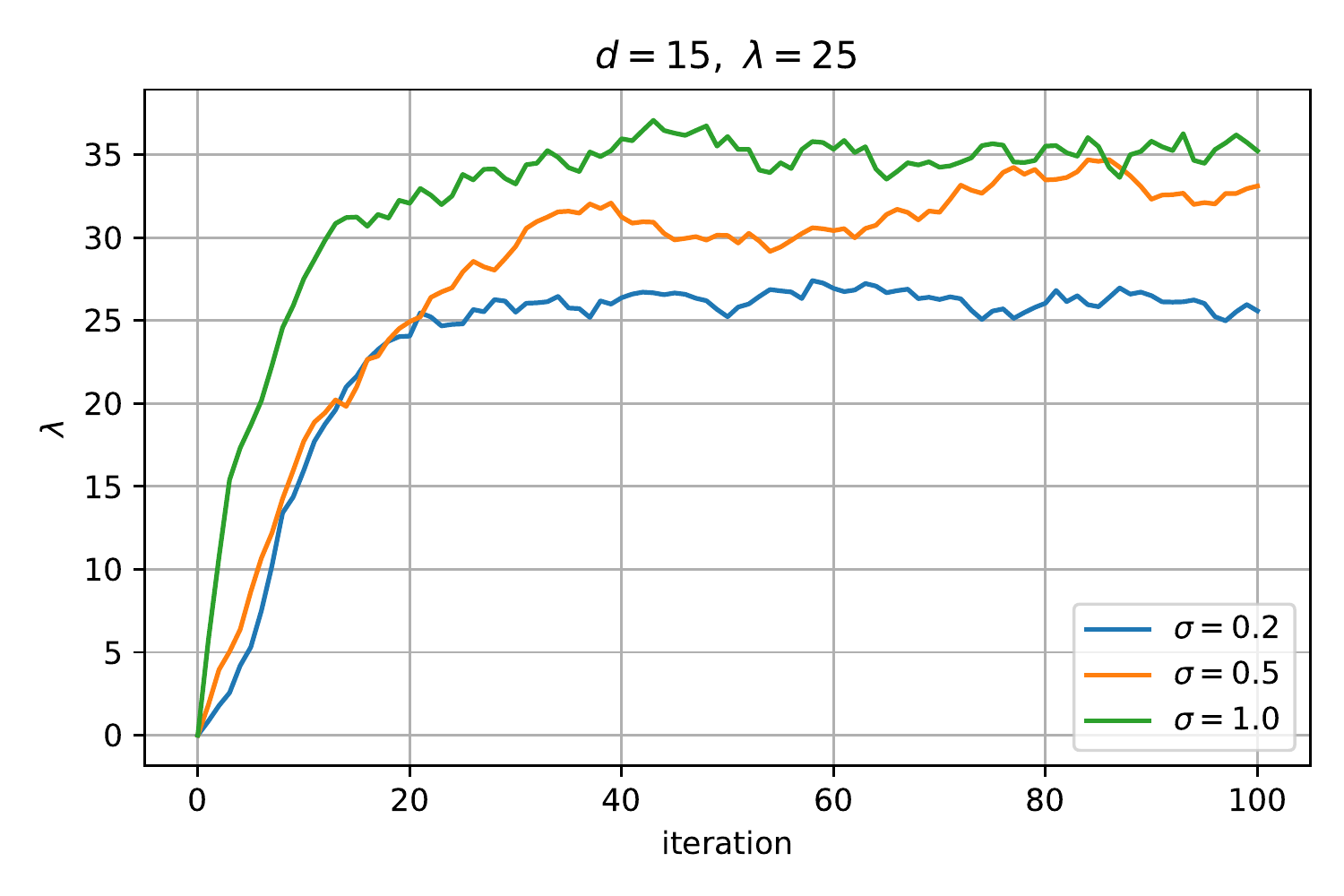}
  \includegraphics[width=0.45\textwidth]{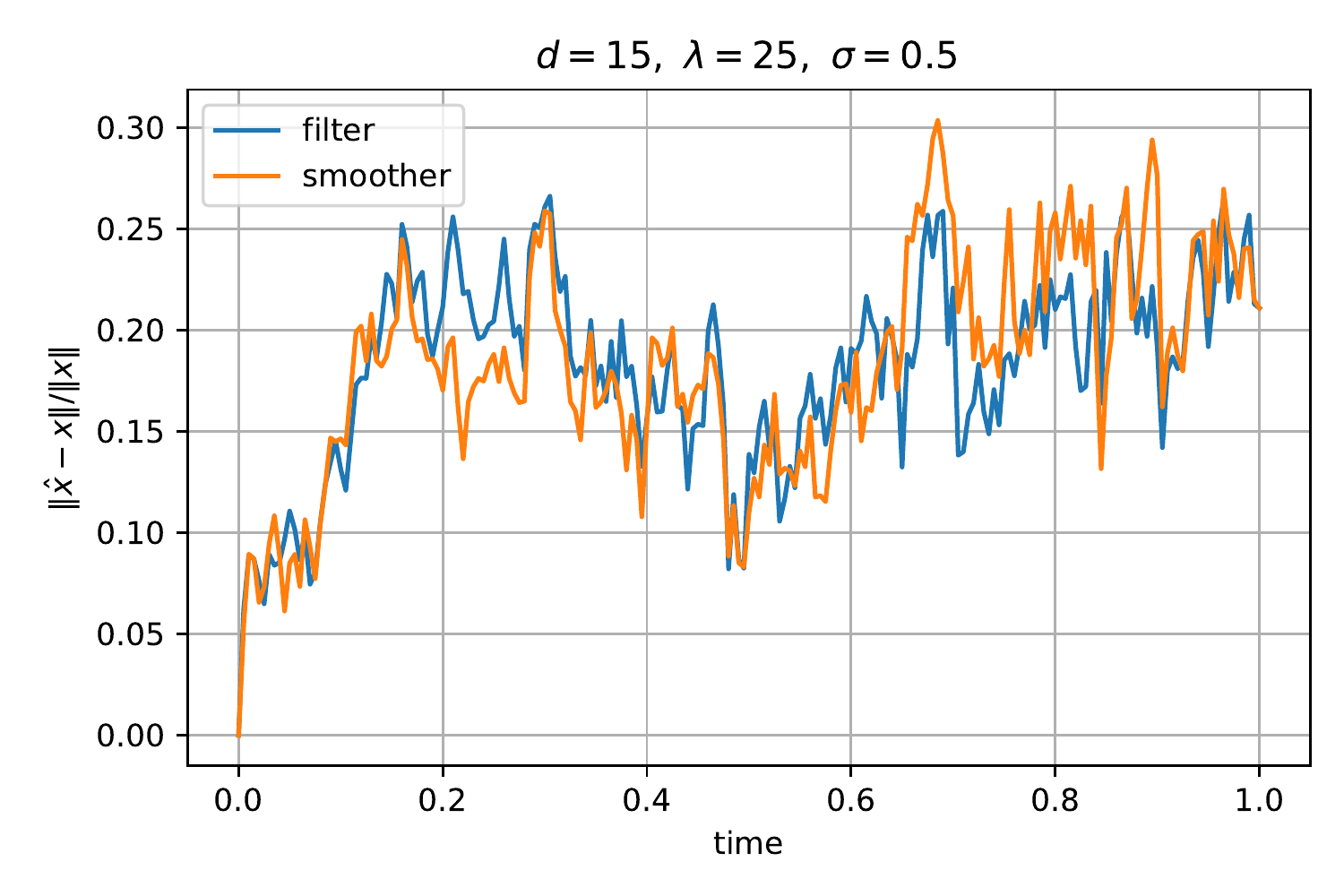} \\
  \includegraphics[width=0.45\textwidth]{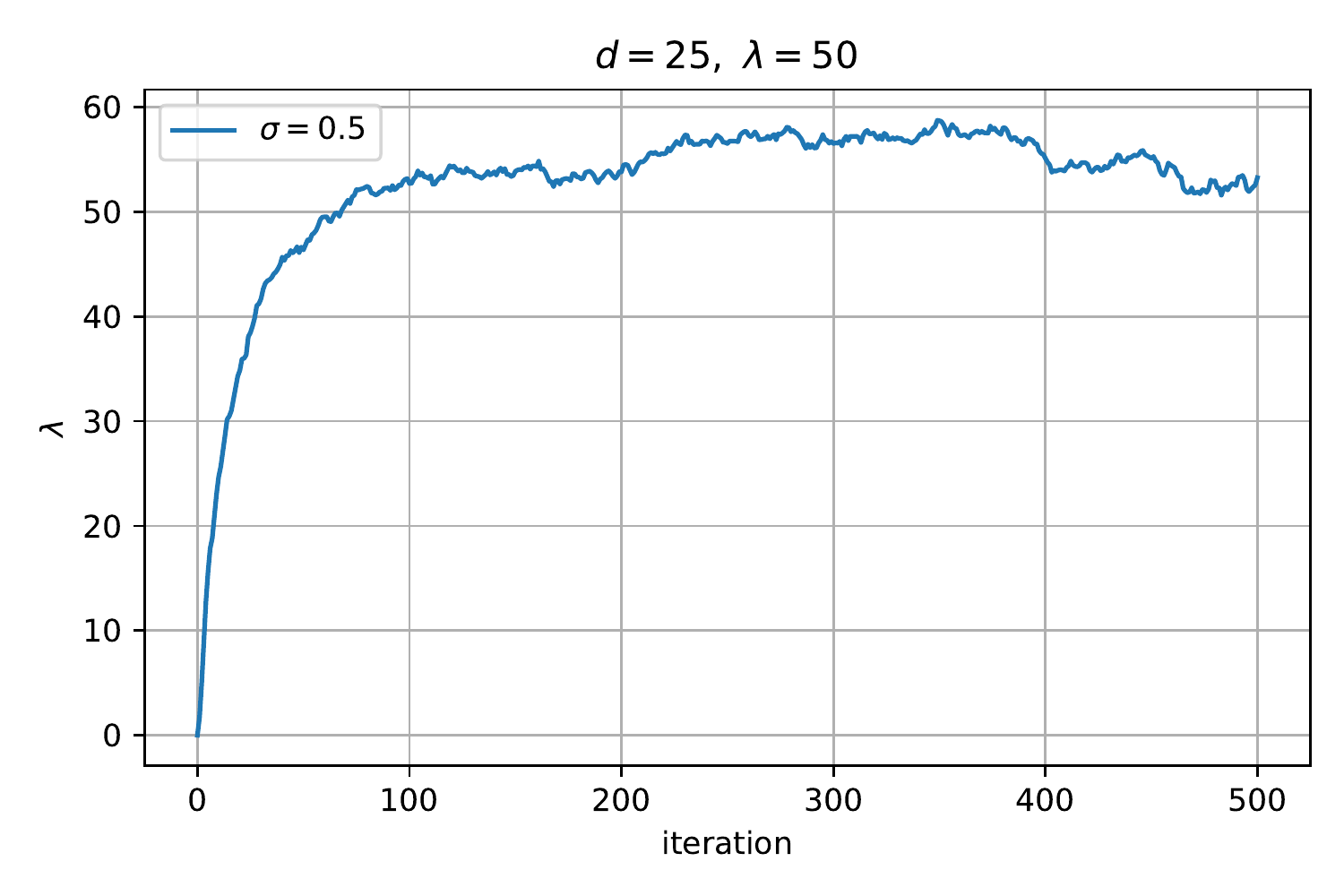}
  \includegraphics[width=0.45\textwidth]{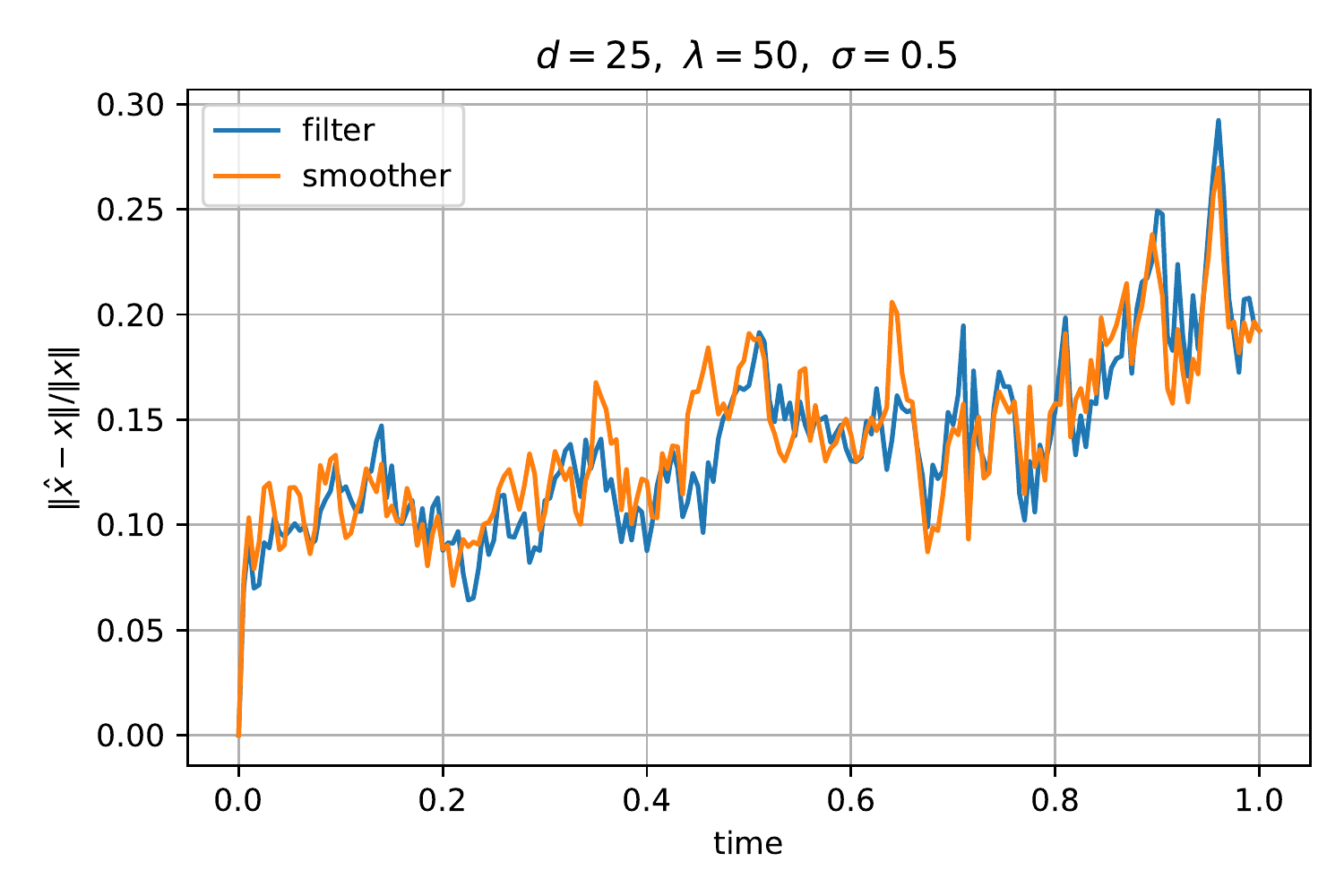}
  \caption[The cubic sensor problem with high dimensional linear dynamics]{%
    The cubic sensor problem (\ref{eq:cubic}) for the scalar parameter
    estimation (\ref{eq:cubic_scalar}). Top: $d = 5$, $\lambda = 5$; middle: $d
    = 15$, $\lambda = 25$; bottom: $d = 25$, $\lambda = 50$. The first row shows
    the update of $\lambda^k$ during the iterations. The solutions of the filter
    and smoother calculated using $\hat\lambda$, and the relative error
    $\left\|\hat x_t - x_t\right\|_2 / \|x_t\|_2$ is shown in the second column.
  }\label{fig:cubic_scalar}
\end{figure}

Here we apply our Monte Carlo method to a problem with high dimensional state
space. In the cubic sensor dynamics (\ref{eq:cubic}), let $F \in \R^{d \times
d}$ be the tridiagonal matrix
\begin{equation}
  F = \lambda
  \left[\begin{array}{cccccc}
    -1 &  1 \\
     1 & -2 & 1 \\
       &  1 & -2 & 1 \\
       &    & \cdots & \cdots & \cdots \\
       &    &  &  1 & -2 &  1 \\
       &    &  &    &  1 & -1
  \end{array}\right],%
  \label{eq:cubic_scalar}
\end{equation}
where the scalar $\lambda$ is the unknown parameter to be estimated. We consider
different dimensions $d = 5, 15, 25$, and take $\lambda = 5, 25, 50$
respectively. For each dimension, consider different noise levels $\sigma = 0.2,
0.5, 1.0$ and fix $\eta = 0.01$. Let the initial condition
\[
  X_0 = \left[-1, -1 + \frac{2}{d-1}, -1 + \frac{4}{d-1}, \dots,
  1 - \frac{2}{d-1}, 1\right],
\]
and $T = 1$. In Figure~\ref{fig:cubic_scalar_x} we generate hidden processes
$\{X_t\}$ for different $d$, $\lambda$ and $\sigma$.

For each experiment, take $\Delta t = 0.005$, the number of samples $N = 128$,
and the initialization $\lambda^0 = 0$. Figure~\ref{fig:cubic_scalar} shows the
simulation results. The EM algorithm with Monte Carlo method is efficient even
for high dimensional problem. The higher the dimension, the more iterations are
required. We also see that at larger noise level the convergence is faster at
the beginning but is more volatile afterwards. The solutions of the filter and
smoother (\ref{eq:dif_state_mc}) based on the estimated $\hat\lambda$ are also
close to the true path of $\{X_t\}$.

\subsection{Lorenz 96 model}

\begin{figure}[tp]
  \centering
  \includegraphics[width=0.45\textwidth]{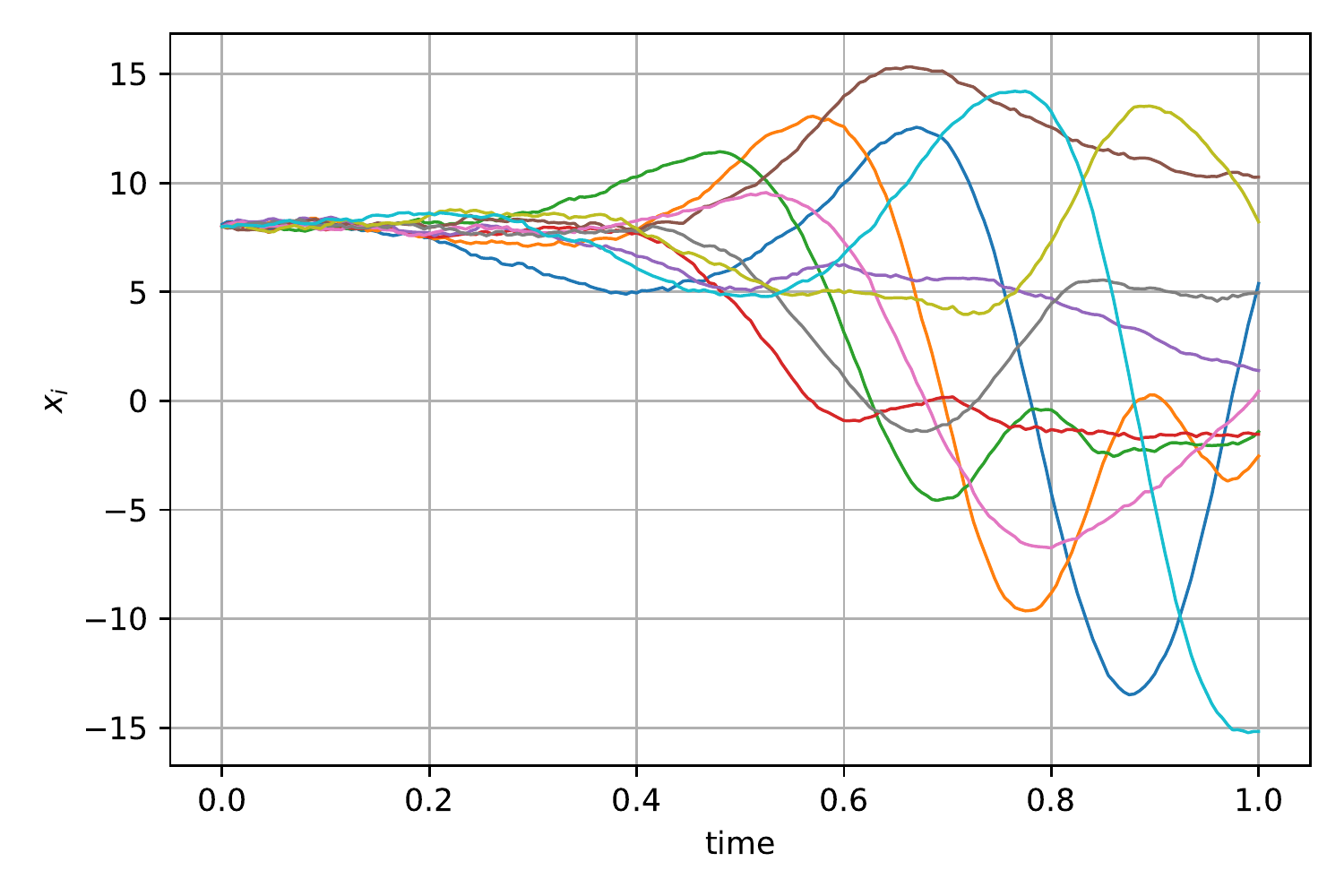}
  \includegraphics[width=0.45\textwidth]{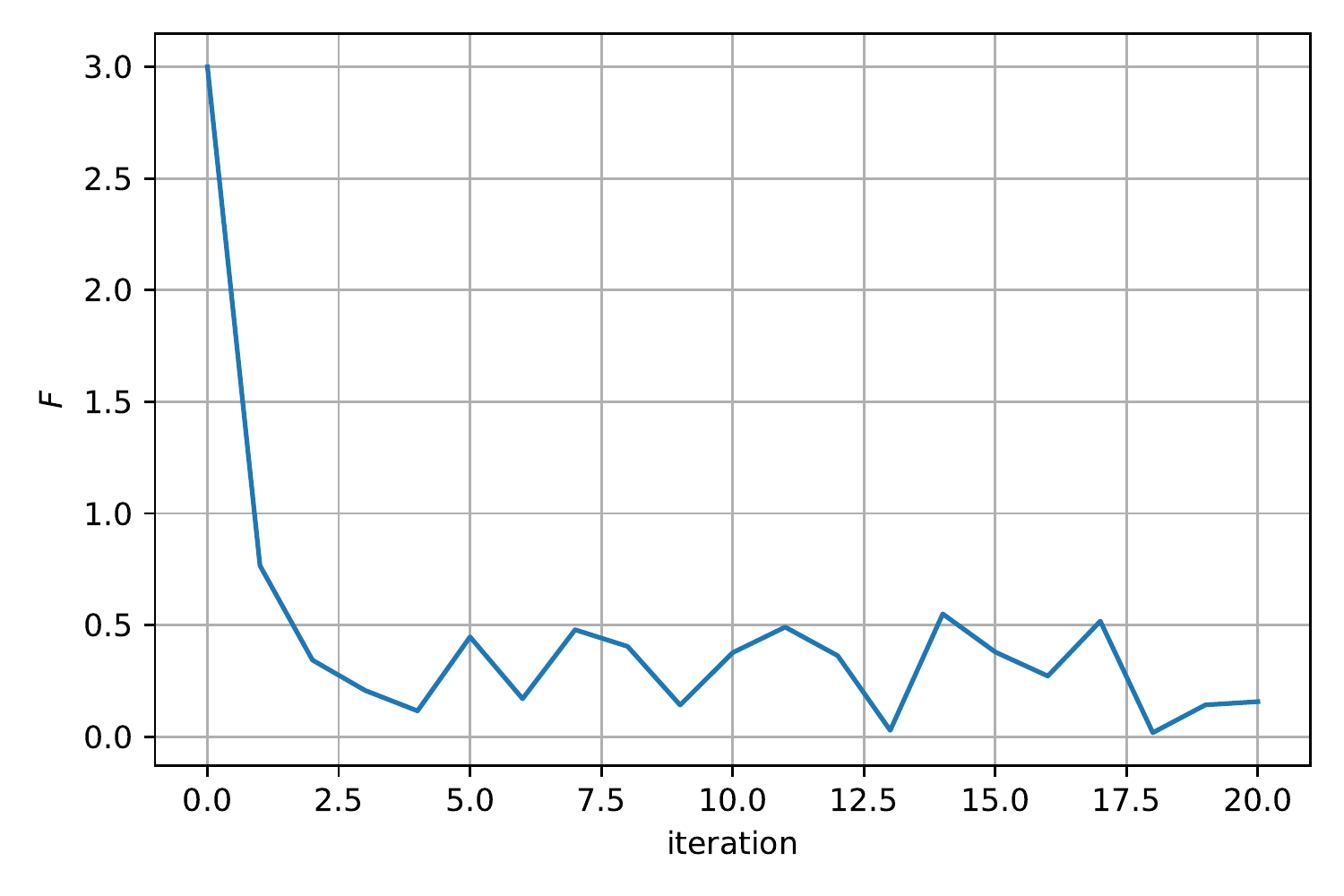} \\
  \includegraphics[width=0.45\textwidth]{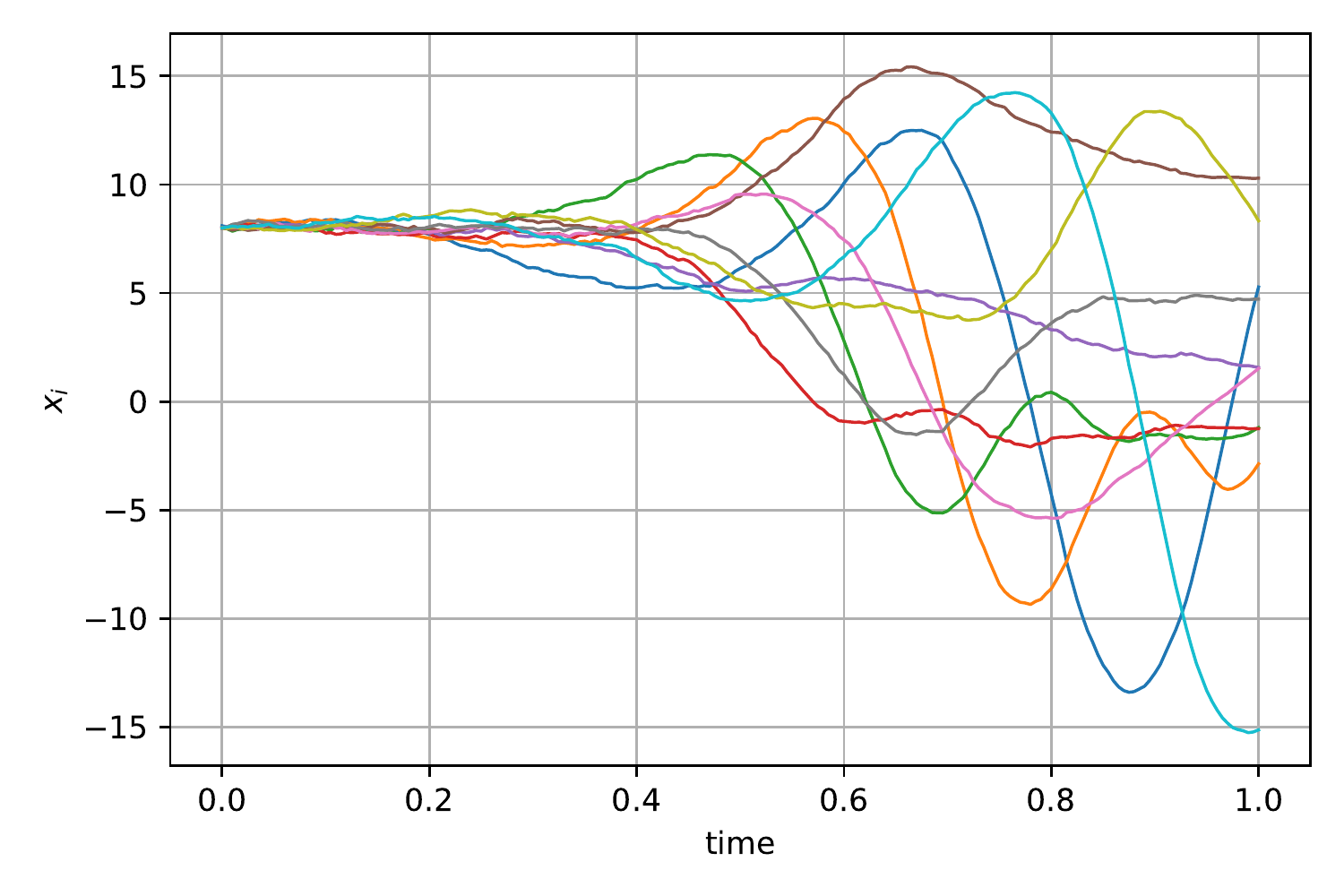}
  \includegraphics[width=0.45\textwidth]{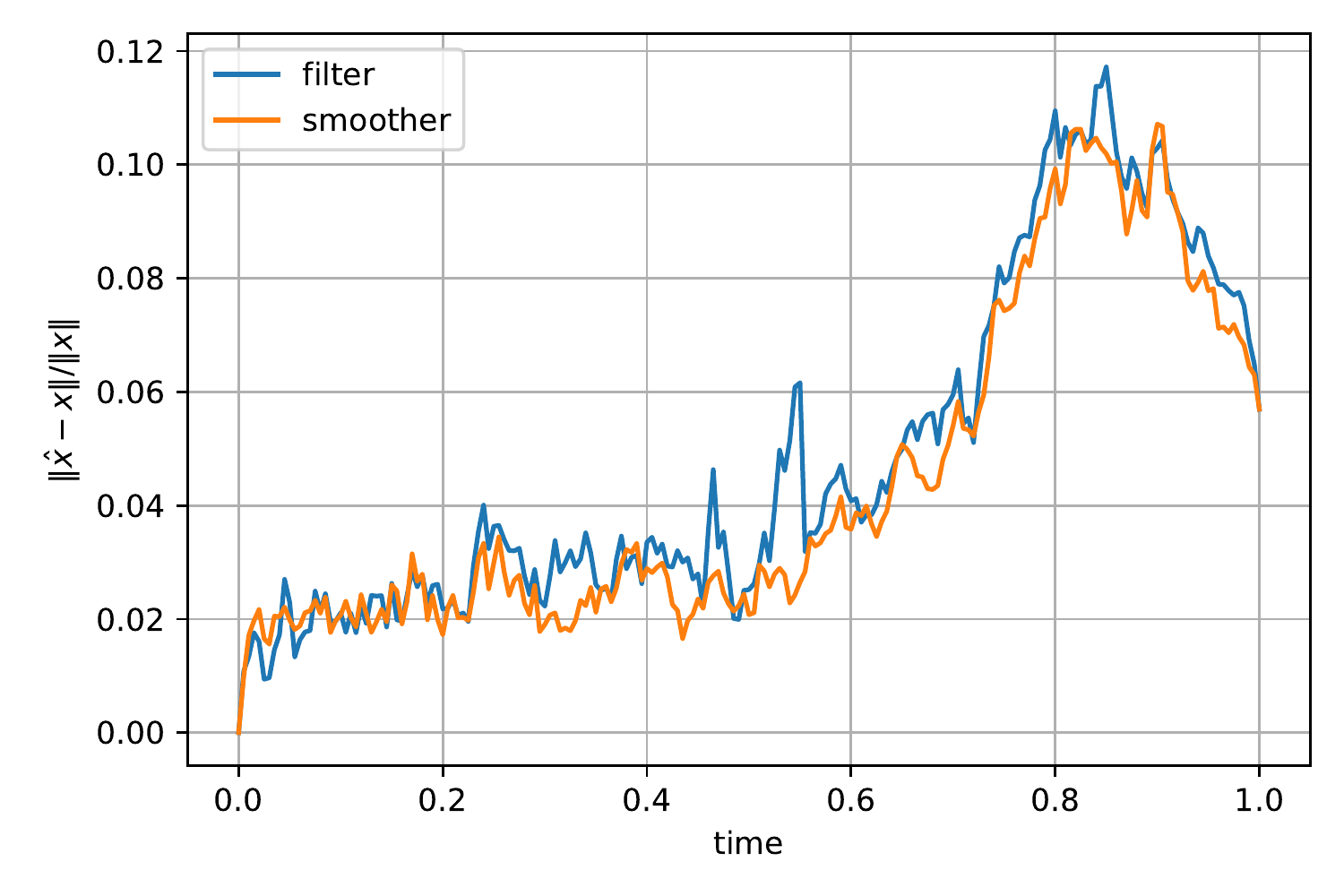}
  \caption[The Lorenz 96 model with cubic sensor observations]{%
    The Lorenz 96 model (\ref{eq:lorenz}) with cubic sensor observations
    (\ref{eq:cubic}). The upper left figure is the dynamics of the components of
    $X_t$ under the force $F = 8$ and initialization $X_{i,0} = F$, which shows
    the chaotic behavior after starting from the equilibrium. The upper right
    figure shows that the distance $|F^k - F|$ converges fast during the
    iterations. Then we calculate the solutions of the filter and smoother based
    on $\hat F$. The bottom left figure shows that the smoother solution is
    close to the true trajectory, and indeed captures the chaotic behavior. The
    bottom right figure gives the relative error $\|\hat x_t - x_t \|_2 /
    \|x_t\|_2$.
  }\label{fig:lorenz}
\end{figure}

The Lorenz 96 model is a dynamical system proposed by Edward Lorenz in 1996
\cite{lorenz1996predictability}, and is commonly used in data assimilation.
Denote the components of the hidden state as $X_t = [X_{1,t}, X_{2,t}, \dots,
X_{d,t}]$. Consider the following dynamics
\begin{equation}
  \frac{\ud X_{i,t}}{\ud t}
  = (X_{i+1,t} - X_{i-2,t}) X_{i-1,t} - X_{i,t} + F + \sigma W_t.
  \quad i = 1, 2, \dots, d,
  \label{eq:lorenz}
\end{equation}
where $X_{-1,t} = X_{d-1,t}$, $X_{0,t} = X_{d,t}$ and $X_{d+1,t} = X_{1,t}$. The
scalar $F$ is a forcing constant. The observations still follow the cubic sensor
(\ref{eq:cubic}).

Assume that the force $F$ is the unknown parameter to be estimated. Following
\cite{lucini2019model}, set the true value $F = 8$ that is commonly known to
cause chaotic behavior. Let the dimension $d = 10$ and initialize the state from
the equilibrium $X_{i,0} = F$, $i = 1, \dots, d$. Let the noise $\sigma = 1$,
$\eta = 5$ and the total time $T = 1$. Take the discretization $\Delta t =
0.05$ and the number of samples $N = 128$.

Figure~\ref{fig:lorenz} shows the simulation results. We can see that the hidden
states show chaotic behavior after starting from the equilibrium. The parameter
estimation $\hat F$ nevertheless converges fast to the true value. We also
calculated the solution of the filter and smoother (\ref{eq:dif_state_mc}) based
on the estimated $\hat F$. The relative error is small, and the state estimation
indeed captures the chaotic behavior.

\section{Conclusion}

In this paper, we propose a unified framework to formally obtain the state and
parameter estimation for CT-HMM by taking the continuous-time limit of the
Baum-Welch algorithm. We also propose a Monte Carlo approach for numerically
handling the continuous formulation based on the standard particle filter and
smoother, which may be further improved by using adaptive filtering methods. For
the CT-HMM problems in practice, instead of discretizing time from the
beginning, now one can first derive the continuous equations following our
framework, and then choose the proper discretization and sampling scheme for
better performance. Our numerical results demonstrate the effectiveness of the
proposed algorithms.

\paragraph{Acknowledgements}
We are grateful to Prof.~Ramon van Handel for his valuable suggestions on a
previous draft of this paper. This work is supported by a gift to Princeton
University from iFlytek.

\bibliographystyle{plainnat}
\bibliography{ct_hmm}

\appendix

\section{CT-HMM with discrete-time observations}%
\label{sec:hmm_disc_obs}

While this paper mainly focuses on the continuous-time states and observations,
in some applications like disease progression, we may only have observations at
some discrete time points and they are assumed to be independent (see
Section~\ref{sec:jump_setting}). The Baum-Welch framework can also be applied
to this setting as we now show.

\subsection{Hidden jump process}

Here we modify the CT-HMM $(Q, r, \pi_0)$ with hidden jump process in
Section~\ref{sec:hmm_jump}. The hidden states $\{X_t\}$ remains to be a jump
process with generator $Q$ and initial probability $\pi_0$. Instead of
continuous-time observations, here we assume that we only can observe at
discrete time points $0 \le \tau_1 < \cdots < \tau_S \le T$, and $Y_{\tau_s} \in
\{1, \dots, m\}$ are independently generated from
\begin{equation}
  r_i(y) = \Pr\{Y_{\tau_s} = y | X_{\tau_s} = i\}.
\end{equation}
The following are the results for state estimation and parameter estimation
given observations $Y_{\tau_s} = y_{\tau_s}$, $s = 1, \dots, S$.

\paragraph{State estimation}
Let $\alpha_t = [\alpha_t(1)\ \cdots\ \alpha_t(n)] \in \R^{1 \times n}$ and
$\beta_t = [\beta_t(1)\ \cdots\ \beta_t(n)]^\intercal \in \R^{n \times 1}$ be
the solutions of the forward and backward piecewise ODEs respectively
\begin{align}
  & \begin{cases}
    \alpha_0 = \pi_0, \\
    \dot\alpha_t = \alpha_t Q, & t \in [\tau_s, \tau_{s+1}), \\
    \alpha_{\tau_s} = \alpha_{\tau_s^-} R(y_{\tau_s}), 
  \end{cases} \\
  & \begin{cases}
    \beta_T = 1, \\
    \dot\beta_t = -Q \beta_t, & t \in [\tau_s, \tau_{s+1}), \\
    \beta_{\tau_s^-} = R(y_{\tau_s}) \beta_{\tau_s}, 
  \end{cases}
\end{align}
where the left limit $\alpha_{\tau_s^-} = \lim_{t \uparrow \tau_s} \alpha_t$,
and $R(y) = \diag(r_1(y), \dots, r_n(y))$. Then the posterior distribution of
the hidden states satisfies
\begin{equation}
  \rho_t(i) = \Pr\{X_t = i | Y_{\tau_s} = y_{\tau_s}, s = 1, \dots, S\}
  = \frac{\alpha_t(i) \beta_t(i)}{\alpha_0 \cdot \beta_0}.
\end{equation}

\paragraph{Parameter estimation}
Assume that the generator $Q$ is the unknown parameter to be estimated. The EM
algorithm can be applied: in the E-step, using the current estimate $Q = Q^k$,
we solve the state estimation for $\alpha_t$, $\beta_t$ and $\rho_t$; in the
M-step, update the generator using
\begin{equation}
  Q_{ij}^{k+1} = 
  \frac{\int_0^T \alpha_t(i) Q_{ij}^k \beta_t(j) \ud t}
  {(\alpha_0 \cdot \beta_0) \int_0^T \rho_t(i) \ud t}
\end{equation}
for $j \ne i$, and $Q_{ii}^{k+1} = -\sum_{j \ne i} Q_{ij}^{k+1}$.

\subsection{Hidden diffusion process}

For the CT-HMM with hidden diffusion process in
Section~\ref{sec:hmm_diffusion}, take the same hidden process
\[
  \ud X_t = f(X_t) \ud t + \sigma \ud W_t, 
\]
with initial distribution $\pi_0$, and assume discrete-tine observations
$Y_{\tau_s}$, $s = 1, \dots, S$ from
\begin{equation}
  r(y|x) = p(Y_{\tau_s} = y | X_{\tau_s} = x).
\end{equation}

\paragraph{State estimation}

For the filtering problem, the unnormalized distribution $\pi_t(x) = p(X_t = x |
Y_{\tau_s} = y_{\tau_s}, \tau_s < t) Z_t$ (for some constant $Z_t$) satisfies
the forward piecewise PDE
\begin{equation}
  \begin{cases}
  \pi_0(x) = p(X_0 = x), \\
  \ud\pi_t(x) = \cL^*\pi_t(x) \ud t, & t \in [\tau_s, \tau_{s+1}), \\
  \pi_{\tau_s}(x) = \pi_{\tau_s^-}(x) r(y_{\tau_s} | x), 
  \end{cases}
\end{equation}
where $\pi_{\tau_s^-}(x) = \lim_{t \uparrow \tau_s} \pi_t(x)$. For the smoothing
problem, let $\beta_t(x)$ be the solution of the backward piecewise PDE
\begin{equation}
  \begin{cases}
  \beta_T(x) = 1, \\
  \ud\beta_t(x) = -\cL\pi_t(x) \ud t, & t \in [\tau_s, \tau_{s+1}), \\
  \beta_{\tau_s^-}(x) = \beta_{\tau_s}(x) r(y_{\tau_s} | x), 
  \end{cases}
\end{equation}
then the posterior distribution of the hidden states is given by
\begin{equation}
  \rho_t(x) = p(X_t = x | Y_{\tau_s} = y_{\tau_s}, s = 1, \dots, S)
  = \frac{1}{Z_T} \pi_t(x) \beta_t(x).
\end{equation}

\paragraph{Parameter estimation}

Assume $f(x) = f(x; \theta)$ where $\theta$ is the unknown parameter to be
estimated. The EM algorithm is similar to the continuous-time observations. In
the E-step, using the current estimate $\theta = \theta^k$, we solve the state
estimation for $\pi_t$, $\beta_t$ and $\rho_t$; in the M-step, update the
parameter using $\theta^{k+1} = \argmin_\theta\tilde\cQ(\theta, \theta^k)$ where
\begin{equation}
  \tilde\cQ(\theta, \theta^k) = \int_0^T \int_{\R^d} \rho_t(x)
  \left[\|f(x; \theta)\|_2^2 - 2 f^\intercal(x; \theta)
  \left[f(x; \theta^k) + \frac{\sigma^2 \nabla\beta_t(x)}{\beta_t(x)}\right]
  \right] \ud x \ud t.
\end{equation}

\section{Hidden diffusion process with linear Gaussian dynamics}%
\label{sec:dif_linear}

In this section, we apply our algorithm to the linear Gaussian dynamics for the
CT-HMM with hidden diffusion process in Section~\ref{sec:hmm_diffusion}. Here
both $f$ and $h$ are linear and the initial distribution $\pi_0$ is Gaussian,
thus the posterior distribution of the hidden state is always Gaussian. The
continuous-time Kalman filter and associated parameter estimation has been given
in many literatures \cite{rauch1965maximum, dembo1986parameter,
elliott1997exact}, and here we would like to show the derivation under the
Baum-Welch framework. We will see that the SPDEs (\ref{eq:dif_fwd})
(\ref{eq:dif_bwd}) in state estimation and the optimization (\ref{eq:dif_q}) in
parameter estimation have explicit solutions, and are consistent with the
discrete-time Kalman filter and smoother as well as the EM algorithm under the
limit $\Delta t \to 0$.

Let $f(x) = Fx$ and $h(x) = Hx$ where the matrices $F \in \R^{n \times n}$ and
$H \in \R^{m \times n}$. We also consider $\sigma \in \R^{n \times n}$ and $\eta
\in \R^{m \times m}$ to be matrices. The dynamics of the hidden process
$\{X_t\}$ and the observation $\{Y_t\}$ becomes
\begin{align}
  \ud X_t & = FX_t \ud t + \sigma \ud W_t, \nonumber \\
  \ud Y_t & = HX_t \ud t + \eta \ud B_t.
  \label{eq:lin_model}
\end{align}
Assume that the initial distribution is Gaussian $X_0 \sim \cN(\mu_0, P_0)$. So
the posterior distribution of $X_t$ is always Gaussian.

\subsection{Continuous-time Kalman filter and smoother}

Given the Gaussian property in the linear case, to describe the posterior
distribution of $X_t$, we only need to find the mean and variance from the SDEs
(\ref{eq:dif_fwd}) and (\ref{eq:dif_bwd}). Here the SPDEs become
\begin{align*}
  \ud \pi_t(x) & = \cL^*\pi_t(x) \ud t
  + \pi_t(x) (Hx)^\intercal \eta^{-\intercal} \eta^{-1} \ud Y_t, \\
  \ud \beta_t(x) & = -\cL\beta_t(x) \ud t
  - \beta_t(x) (Hx)^\intercal \eta^{-\intercal} \eta^{-1} (\ud Y_t - Hx \ud t).
\end{align*}
One can prove that the solutions $\pi_t$ and $\beta_t$ are unnormalized
Gaussian, and $\rho_t(x) = \pi_t(x) \beta_t(x) / Z_T$ is Gaussian, 
i.e., for each $t$ there exist constants $Z_t^\pi$ and $Z_t^\beta$, 
such that
\begin{equation}
  \pi_t(x) Z_t^\pi \sim \cN(\mu_t^\pi, P_t^\pi), \quad
  \beta_t(x) Z_t^\beta \sim \cN(\mu_t^\beta, P_t^\beta), \quad
  \rho_t(x) \sim \cN(\mu_t^\rho, P_t^\rho).
\end{equation}
Solving the SPDEs, we have
\begin{align}
  \ud \mu_t^\pi & = F \mu_t^\pi \ud t
  + P_t^\pi H^\intercal \eta^{-\intercal} \eta^{-1}
  \left(\ud Y_t - H \mu_t^\pi \ud t\right), \nonumber \\
  \frac{\ud P_t^\pi}{\ud t} & = F P_t^\pi + P_t^\pi F^\intercal
  - P_t^\pi H^\intercal \eta^{-\intercal} \eta^{-1} H P_t^\pi
  + \sigma \sigma^\intercal, \nonumber \\
  \ud \mu_t^\beta & = F\mu_t^\beta \ud t
  - P_t^\beta H^\intercal \eta^{-\intercal} \eta^{-1}
  \left(\ud Y_t - H\mu_t^\beta \ud t\right), \nonumber \\
  \frac{\ud P_t^\beta}{\ud t} & = F P_t^\beta + P_t^\beta F^\intercal
  + P_t^\beta H^\intercal \eta^{-\intercal} \eta^{-1} H P_t^\beta
  - \sigma \sigma^\intercal,
  \label{eq:lin_sde}
\end{align}
and
\begin{align}
  \mu_t^\rho & = (P_t^\rho)^{-1}
  \left[(P_t^\pi)^{-1} \mu_t^\pi + (P_t^\beta)^{-1} \mu_t^\beta\right],
  \nonumber \\
  P_t^\rho & = \left[(P_t^\pi)^{-1} + (P_t^\beta)^{-1}\right]^{-1}.
  \label{eq:lin_state}
\end{align}
For the initial conditions, $X_0 \sim \cN(\mu_0, P_0)$ gives $\mu_0^\pi = \mu_0$
and $P_0^\pi = P_0$. However, since $\beta_T(x) = 1$, we have $P_T^\beta =
+\infty$ and $\mu_T^\beta$ is undefined. We can instead use $(P_T^\beta)^{-1}
\mu_T^\beta = 0$, $(P_T^\beta)^{-1} = 0$ and
\begin{align*}
  \ud \left[(P_t^\beta)^{-1} \mu_t^\beta\right]
  & = - F (P_t^\beta)^{-1} \mu_t^\beta \ud t
  + (P_t^\beta)^{-1} \sigma \sigma^\intercal (P_t^\beta)^{-1} \mu_t^\beta \ud t
  - H^\intercal \eta^{-\intercal} \eta^{-1} \ud Y_t, \\
  \frac{\ud(P_t^\beta)^{-1}}{\ud t}
  & = - F (P_t^\beta)^{-1} - (P_t^\beta)^{-1} F^\intercal
  + (P_t^\beta)^{-1} \sigma \sigma^\intercal (P_t^\beta)^{-1}
  - H^\intercal \eta^{-\intercal} \eta^{-1} H.
\end{align*}

As a comparison, in the following we consider Kalman filter and smoother under
the limit $\Delta t \to 0$. We will see that the result is consistent with the
differential equations above.

The discretization of the model gives
\begin{align*}
  X_{t + \Delta t} & = (I + F \Delta t) X_t + \sigma \Delta W_t, \\
  \Delta Y_t & = H X_t \Delta t + \eta \Delta B_t, 
\end{align*}
where we omit the $o(\Delta t)$ term. Here we use the notations in Kalman
filter: for $t \le s$, let
\begin{equation}
  X_t | Y_{0:s} \sim \cN\left(\mu_{t|s}, P_{t|s}\right)
\end{equation}
be the distribution of $X_t$ condition on $\Delta Y_0, \Delta Y_1, \dots, \Delta
Y_{s - \Delta t}$ \emph{without} $\Delta Y_s$. We will prove that $\mu_{t|t} =
\mu_t^\pi$, $P_{t|t} = P_t^\pi$ and $\mu_{t|T} = \mu_t^\rho$, $P_{t|T} =
P_t^\rho$.

For the filtering problem, we calculate $\mu_{t|t}$ and $P_{t|t}$. The
discrete-time Kalman filter gives
\begin{align*}
  K_t & = P_{t|t} H^\intercal \Delta t
  \left(H P_{t|t} H^\intercal \Delta t^2
  + \eta \eta^\intercal \Delta t \right)^{-1}, \\
  \mu_{t | t + \Delta t} & = \mu_{t|t}
  + K_t \left(\Delta Y_t - H \mu_{t|t} \Delta t\right), \\
  P_{t | t + \Delta t} & = (I - K_t H \Delta t) P_{t|t},
\end{align*}
and
\begin{align*}
  \mu_{t + \Delta t | t + \Delta t}
  & = (I + F\Delta t) \mu_{t | t + \Delta t}, \\
  P_{t + \Delta t | t + \Delta t}
  & = (I + F\Delta t)P_{t|t + \Delta t}(I + F\Delta t)^\intercal
  + \sigma \sigma^\intercal \Delta t.
\end{align*}
We also have
\begin{align*}
  K_t & = P_{t|t} H^\intercal \eta^{-\intercal} \eta^{-1} + O(\Delta t), \\
  \mu_{t + \Delta t | t + \Delta t} - \mu_{t|t}
  & = F \mu_{t|t} \Delta t + K_t \left(\Delta Y_t - H \mu_{t|t} \Delta t\right)
  + o(\Delta t), \\
  P_{t + \Delta t | t + \Delta t} - P_{t|t}
  & = \left[F P_{t|t} + P_{t|t} F^\intercal - K_t H P_{t|t}
  + \sigma \sigma^\intercal\right] \Delta t + o(\Delta t).
\end{align*}
We get the continuous-time Kalman filter by taking $\Delta t \to 0$, where
$\mu_{t|t}$ and $P_{t|t}$ satisfy the same differential equations as $\mu_t^\pi$
and $P_t^\pi$ (\ref{eq:lin_sde}) respectively.

For the smoothing problem, we need to calculate $\mu_{t|T}$ and $P_{t|T}$.
According to the Rauch-Tung-Striebel smoother \cite{rauch1965maximum}, 
\begin{align*}
  C_t & = P_{t - \Delta t | t}(I + F \Delta t)^\intercal P_{t|t}^{-1}, \\
  \mu_{t - \Delta t | T}
  & = \mu_{t - \Delta t|t} + C_t \left(\mu_{t|T} - \mu_{t|t}\right), \\
  P_{t - \Delta t | T}
  & = P_{t - \Delta t | t} + C_t \left(P_{t|T} - P_{t|t}\right) C_t^T.
\end{align*}
Then
\begin{align*}
  C_t
  & = I - \left(F + \sigma \sigma^\intercal P_{t|t}^{-1}\right) \Delta t
  + o(\Delta t), \\
  \mu_{t - \Delta t | T} - \mu_{t|T}
  & = -\left[\left(F + \sigma \sigma^\intercal P_{t|t}^{-1}\right) \mu_{t|T}
  - \sigma \sigma^\intercal P_{t|t}^{-1} \mu_{t|t}\right] \Delta t
  + o(\Delta t), \\
  P_{t - \Delta t | T} - P_{t|T}
  & = -\left[\left(F + \sigma \sigma^\intercal P_{t|t}^{-1}\right) P_{t|T}
  + P_{t|T} \left(F + \sigma \sigma^\intercal P_{t|t}^{-1}\right)^\intercal
  - \sigma \sigma^\intercal\right] \Delta t + o(\Delta t).
\end{align*}
Taking $\Delta t \to 0$, we get the backward ODEs
\begin{align}
  \frac{\ud \mu_{t|T}}{\ud t}
  & = \left(F + \sigma \sigma^\intercal P_{t|t}^{-1}\right) \mu_{t|T}
  - \sigma \sigma^\intercal P_{t|t}^{-1} \mu_{t|t}, \nonumber \\
  \frac{\ud P_{t|T}}{\ud t}
  & = \left(F + \sigma \sigma^\intercal P_{t|t}^{-1}\right) P_{t|T}
  + P_{t|T} \left(F + \sigma \sigma^\intercal P_{t|t}^{-1}\right)^\intercal
  - \sigma \sigma^\intercal,
  \label{eq:lin_state_mix}
\end{align}
where the initial conditions $\mu_{T|T}$ and $P_{T|T}$ are given by the previous
filter.

Compare with our previous results, by plugging (\ref{eq:lin_sde}) in
(\ref{eq:lin_state}) we see that the ODEs for $\mu_t^\rho$ and $P_t^\rho$ are
exactly the same as (\ref{eq:lin_state_mix}). Notice that for Kalman smoother
(\ref{eq:lin_state_mix}), the ODEs for $\mu_{t|T}$ and $P_{t|T}$ depend on the
filter $\mu_{t|t}$ and $P_{t|t}$, while in our solution (\ref{eq:lin_sde}) and
(\ref{eq:lin_state}), the smoother $\rho_t \propto \pi_t \beta_t$, where the
differential equations for $\pi_t$ and $\beta_t$ are independent. Later we will
see that $\beta_t$ plays an important role in the EM algorithm.

\subsection{An EM algorithm for Kalman filter}

In the linear dynamics (\ref{eq:lin_model}), we assume that $H$, $\sigma$ and
$\eta$ are given and $F$ is the unknown parameter to be estimated. We further
assume that $\sigma$ and $\eta$ are scalars instead of matrices for simplicity.
With the current estimation $F = F^k$, let the solution of the E-step be
$\mu_t^\pi$, $P_t^\pi$, $\mu_t^\beta$, $P_t^\beta$ and $\mu_t^\rho$, $P_t^\rho$.
In the M-step, the parameter update (\ref{eq:dif_q}) becomes $F^{k+1} =
\argmin_F \tilde\cQ(F, F^k)$ where
\[
  \tilde\cQ(F, F^k) = \int_0^T \int_{\R^n} \rho_t(x)
  \left[x^\intercal F^\intercal F x - 2 x^\intercal F^\intercal
  \left(F^k x + \frac{\sigma^2 \nabla\beta_t(x)}{\beta_t(x)}\right)\right]
  \ud x \ud t.
\]
Since $\beta_t(x)$ is unnormalized Gaussian, i.e., $\beta_t(x) Z_t^\beta \sim
\cN(\mu_t^\beta, P_t^\beta)$, we have
\[
  \frac{\nabla\beta_t(x)}{\beta_t(x)}
  = - (P_t^\beta)^{-1} \left(x - \mu_t^\beta\right).
\]
Furthermore, 
\[
  \int_{\R^n} x \rho_t(x) \ud x = \mu_t^\rho, \quad
  \int_{\R^n} x x^\intercal \rho_t(x) \ud x
  = \mu_t^\rho \mu_t^{\rho\intercal} + P_t^\rho, 
\]
the objective function
\begin{align*}
  \tilde\cQ & (F, F^k)
  = \int_0^T \int_{\R^n} \rho_t(x) \left[x^\intercal F^\intercal F x
  - 2 x^\intercal F^\intercal \left(F^k x - \sigma^2 (P_t^\beta)^{-1}
  \left(x - \mu_t^\beta\right)\right)\right] \ud x \ud t \\
  & = \int_0^T \int_{\R^n} \rho_t(x)
  \left[x^\intercal \left(F^\intercal F - 2 F^\intercal F^k
  + 2 \sigma^2 F^\intercal (P_t^\beta)^{-1}\right) x
  - 2 \sigma^2 x^\intercal F^\intercal (P_t^\beta)^{-1} \mu_t^\beta\right]
  \ud x \ud t \\
  & = \int_0^T \left[\left(\mu_t^\rho \mu_t^{\rho\intercal} + P_t^\rho\right)
  : \left(F^\intercal F - 2 F^\intercal F^k
  + 2 \sigma^2 F^\intercal (P_t^\beta)^{-1}\right)
  - 2 \sigma^2 \mu^{\rho\intercal}F^\intercal (P_t^\beta)^{-1} \mu_t^\beta
  \right] \ud t\\
  & = (F^\intercal F)
  : \int_0^T \left(\mu_t^\rho \mu_t^{\rho\intercal} + P_t^\rho\right) \ud t \\
  & \qquad - 2 F : \left[
  F^k \int_0^T \left(\mu_t^\rho \mu_t^{\rho\intercal} + P_t^\rho\right) \ud t
  - \sigma^2 \int_0^T (P_t^\beta)^{-1}
  \left(\left(\mu_t^\rho - \mu_t^\beta\right) \mu_t^{\rho\intercal}
  + P_t^\rho\right) \ud t\right].
\end{align*}
Therefore, 
\begin{align}
  F^{k+1} & = \argmin_F\tilde\cQ(F, F^k) \nonumber \\
  & = F^k - \sigma^2 \left[\int_0^T (P_t^\beta)^{-1}
  \left(\left(\mu_t^\rho - \mu_t^\beta\right) \mu_t^{\rho\intercal}
  + P_t^\rho\right) \ud t\right]
  \left[\int_0^T \left(\mu_t^\rho \mu_t^{\rho\intercal}
  + P_t^\rho\right) \ud t\right]^{-1}.
\end{align}

In the following, we will show that our result is consistent with the
discrete-time EM algorithm for linear dynamics.

First, we calculate the joint distribution of $(X_t, X_{t + \Delta t})$
condition on the whole observation $Y_{0:T}$. We know that
\[
  \pi_t(x) \propto p(X_t = x, Y_{0:t}), \quad
  \beta_t(x) \propto p(Y_{t:T} | X_t = x), 
\]
thus
\begin{align*}
  p(X_t & = x, X_{t + \Delta t} = x' | Y_{0:T})
  \propto \pi_t(x) p(X_{t + \Delta t} = x', \Delta Y_t | X_t = x)
  \beta_{t + \Delta t}(x) \\
  & \propto \exp\Bigg[-\frac{1}{2} \left(x - \mu_t^\pi\right)^\intercal
  (P_t^\pi)^{-1} \left(x - \mu_t^\pi\right)
  - \frac{1}{2 \eta^2 \Delta t} \left\|\Delta Y_t - H x \Delta t\right\|^2 \\
  & \qquad
  -\frac{1}{2 \sigma^2 \Delta t} \left\|x' - (I + F \Delta t) x\right\|^2
  - \frac{1}{2} \left(x' - \mu_{t + \Delta t}^\beta\right)^\intercal
  (P_{t + \Delta t}^\beta)^{-1} \left(x' - \mu_{t + \Delta t}^\beta\right)
  \Bigg].
\end{align*}
So $(X_t, X_{t + \Delta t})$ are joint Gaussian. We can see that
\begin{align*}
  \E & [X_{t + \Delta t} | X_t = x, Y_{0:T}] =
  \left[\frac{1}{\sigma^2 \Delta t} + (P_{t + \Delta t}^\beta)^{-1}\right]^{-1}
  \left[\frac{1}{\sigma^2 \Delta t}(I + F \Delta t)x
  + (P_{t + \Delta t}^\beta)^{-1}\mu_{t + \Delta t}^\beta\right] \\
  & = x + F x \Delta t
  - \sigma^2 (P_t^\beta)^{-1} \left(x - \mu_t^\beta\right) \Delta t
  + o(\Delta t) \\
  & = x + F x \Delta t + \frac{\sigma^2 \nabla\beta_t(x)}{\beta_t(x)} \Delta t
  + o(\Delta t), 
\end{align*}
which is consistent with our result. On the other hand, consider the covariance
between $X_{t + \Delta t}$ and $X_t$, i.e., 
\[
  P_{t + \Delta t, t}
  = \E\left[\left(X_{t + \Delta} - \mu_{t + \Delta t}^\rho\right)
  \left(X_t - \mu_t^\rho\right)^\intercal |Y_{0:T}\right], 
\]
then
\[
  \E[X_{t + \Delta t} | X_t = x, Y_{0:T}]
  = \mu_{t + \Delta t}^\rho
  - P_{t + \Delta t, t} (P_t^\rho)^{-1} \left(x - \mu_t^\rho\right).
\]
Denote
\[
  \dot\mu_t^\rho = \frac{\ud \mu_t^\rho}{\ud t}, \quad
  S_t = \left.\frac{\ud P_{t + \tau, t}}{\ud \tau}\right|_{\tau = 0}.
\]
One can calculate that $S_t = \left(F - \sigma^2 (P_t^\beta)^{-1}\right)
P_t^\rho$. Now the objective function
\begin{align*}
  \tilde\cQ & (F, F^k)
  = \int_0^T \int_{\R^n} \rho_t(x)
  \left[x^\intercal F^\intercal F x - 2 x^\intercal F^\intercal
  \left.\frac{\ud}{\ud \tau}\right|_{\tau = 0}
  \E[X_{t + \tau} | X_t = x, Y_{0:T}; F^k]\right] \ud x \ud t\\
  & = \int_0^T \int_{\R^n} \rho_t(x)
  \left[x^\intercal F^\intercal F x - 2 x^\intercal F^\intercal
  \left[\dot\mu_t^\rho - S_t (P_t^\rho)^{-1} \left(x - \mu_t^\rho\right)\right]
  \right] \ud x \ud t \\
  & = \int_0^T \int_{\R^n} \rho_t(x) \left[
  x^\intercal \left(F^\intercal F + 2 F^\intercal S_t (P_t^\rho)^{-1}\right) x
  - 2 x^\intercal F^\intercal
  \left(\dot\mu_t^\rho - S_t (P_t^\rho)^{-1} \mu_t^\rho\right)
  \right] \ud x \ud t \\
  & = \int_0^T \left[\left(\mu_t^\rho \mu_t^{\rho\intercal} + P_t^\rho\right)
  : \left(F^\intercal F + 2 F^\intercal S_t (P_t^\rho)^{-1}\right)
  - 2 \mu_t^{\rho\intercal} F^\intercal
  \left(\dot\mu_t^\rho - S_t (P_t^\rho)^{-1} \mu_t^\rho\right)\right] \ud t\\
  & = \left(F^\intercal F\right)
  : \int_0^T \left(\mu_t^\rho \mu_t^{\rho\intercal} + P_t^\rho\right) \ud t
  - 2F : \int_0^T \left(\dot\mu_t^\rho \mu_t^{\rho\intercal} + S_t\right) \ud t.
\end{align*}
Therefore, 
\begin{align}
  F^{k+1}
  & = \argmin_F\tilde\cQ(F, F^k)
  = \left[\int_0^T \left(\dot\mu_t^\rho \mu_t^{\rho\intercal} + S_t\right)
  \ud t\right]
  \left[\int_0^T \left(\mu_t^\rho \mu_t^{\rho\intercal} + P_t^\rho\right)
  \ud t\right]^{-1} \nonumber \\
  & = \left[\int_0^T \left(\frac{\ud \mu_t^\rho}{\ud t} \mu_t^{\rho\intercal} +
  \left.\frac{\ud P_{t + \tau, t}}{\ud \tau}\right|_{\tau=0}\right) \ud t\right]
  \left[\int_0^T \left(\mu_t^\rho \mu_t^{\rho\intercal} + P_t^\rho\right)
  \ud t\right]^{-1}.
  \label{eq:lin_q}
\end{align}

Comparing with the EM algorithm for discrete-time linear dynamics
\cite{shumway1982approach}, the parameter update
\[
  I + F^{k+1} \Delta t
  = \left[\sum_{t=0}^T \left(\mu_{t + \Delta t}^\rho \mu_t^{\rho\intercal}
  + P_{t + \Delta t, t}\right)\right]
  \left[\sum_{t=0}^T \left(\mu_t^\rho \mu_t^{\rho\intercal}
  + P_t^\rho\right)\right]^{-1}.
\]
We can see that the formula (\ref{eq:lin_q}) is the limit of the discrete case.

\end{document}